\documentclass[manuscript,screen]{acmart}

\AtBeginDocument{%
	}

\setcopyright{rightsretained}
\acmJournal{TOPLAS}
\acmYear{2022} 
\acmVolume{1} 
\acmNumber{1} 
\acmArticle{1} 
\acmMonth{1} 
\acmPrice{}
\acmDOI{10.1145/3563943}

\newif\ifspacehack
\spacehackfalse

\newif\ifanon
\makeatletter
\if@ACM@anonymous\anontrue\else\anonfalse\fi
\makeatother

\usepackage{booktabs}
\usepackage{subcaption}

\usepackage{xspace}
\usepackage{float}
\usepackage{tikz}
\usepackage{multirow}
\usepackage{hhline}
\usepackage{multicol}
\usepackage{enumitem}
\usepackage{amsmath}
\usepackage{amsfonts}
\usepackage{mathtools}
\usepackage{todonotes}
\usepackage{tr/mathpartir}
\floatstyle{boxed}
\restylefloat{figure}
\definecolor{darkgreen}{rgb}{0.0,0.392,0.157}
\newcommand{\NEW}[1]{#1}
\newcommand{\NEWW}[1]{#1}
\newcommand{\NNEW}[1]{#1}
\newcommand{\CASENUM}[1]{{\bf\color{orange} #1}} 
\newcommand{\LINENUM}[1]{{\bf\color{cyan} #1}} 
\newcommand{\EVALNUM}[1]{{\bf\color{magenta} #1}} 
\renewcommand{\CASENUM}[1]{#1} 
\renewcommand{\LINENUM}[1]{#1} 
\renewcommand{\EVALNUM}[1]{#1} 

\newcommand{\REVISION}[1]{{\color{blue} #1}}
\renewcommand{\REVISION}[1]{#1}

\newcommand*\circled[1]{\tikz[baseline=(char.base)]{
		\node[shape=circle,draw,inner sep=1pt] (char) {#1};}}

\ifanon
  \newcommand{\aquarium}{Aviary\xspace}
  \newcommand{\aquariumX}{aviary\xspace}
  \newcommand{\ale}{Eagle\xspace}
  \newcommand{\casp}{Kiwi\xspace}
  \newcommand{\lowering}{low\-er\-ing\xspace}
  \newcommand{\lowerings}{low\-er\-ings\xspace}

  \newcommand{\lowertool}{Swoop\xspace}
  \newcommand{\synth}{Hatch\xspace}
\else
  \newcommand{\aquarium}{A\-quar\-ium\xspace}
  \newcommand{\aquariumX}{a\-quar\-ium\xspace}
  \newcommand{\ale}{Ale\-wife\xspace}
  \newcommand{\casp}{Cass\-io\-pea\xspace}
  \newcommand{\lowering}{low\-er\-ing\xspace}
  \newcommand{\lowerings}{low\-er\-ings\xspace}

  \newcommand{\lowertool}{\ale compiler\xspace}
  \newcommand{\synth}{synthesis engine\xspace}
\fi
  
\newcommand{\MD}{ma\-chine-de\-pen\-dent\xspace}
\newcommand{\MI}{ma\-chine-in\-de\-pen\-dent\xspace}

\newcommand{\ttint}{\ensuremath{\mathtt{int}}}
\newcommand{\ttbool}{\ensuremath{\mathtt{bool}}}

\usepackage{caption}
\usepackage{listings}
\usepackage{parcolumns}
\usepackage{algorithm,algpseudocode}

\lstdefinelanguage{alewife}{
  morekeywords = [1]{require,function,block,pre,post,type,value,let,in,region,lower,with,if,then,else}, 
  morekeywords = [2]{ptr,vec,bit,len,ref,int,bool,reg}, 
  morecomment=[n]{(*}{*)}  
}
\lstdefinelanguage{casp}{
	morekeywords = [1]{module,modify,lowering,invariant,defop,pre,post,frame,letstate,control,type,let,in,def,if,then,else,assert},
  morekeywords = [2]{bit,len,ref,memory,with,label,def,skip,reg,txt,sem,dec,hex,bin,lbl},
  morecomment=[n]{(*}{*)}
}

\begin{document}

\ifspacehack
\addtolength\leftmargini{-13.5pt}
\addtolength{\textfloatsep}{-5mm}
\addtolength{\dbltextfloatsep}{-5mm}
\addtolength{\floatsep}{-3mm}
\fi

\title{Towards Porting Operating Systems with Program Synthesis} 

\author{Jingmei Hu}
\affiliation{
  \department{John A. Paulson School of Engineering and Applied Sciences}
  \institution{Harvard University}
  \streetaddress{150 Western Ave.}
  \city{Boston}
  \state{MA}
  \postcode{02134}
  \country{United States of America}
}

\author{Eric Lu}
\affiliation{
  \department{John A. Paulson School of Engineering and Applied Sciences}
  \institution{Harvard University}
  \streetaddress{150 Western Ave.}
  \city{Boston}
  \state{MA}
  \postcode{02134}
  \country{United States of America}
}

\author{David A. Holland}
\affiliation{
  \department{John A. Paulson School of Engineering and Applied Sciences}
  \institution{Harvard University}
  \streetaddress{150 Western Ave.}
  \city{Boston}
  \state{MA}
  \postcode{02134}
  \country{United States of America}
}

\author{Ming Kawaguchi}
\affiliation{
  \department{John A. Paulson School of Engineering and Applied Sciences}
  \institution{Harvard University}
  \streetaddress{150 Western Ave.}
  \city{Boston}
  \state{MA}
  \postcode{02134}
  \country{United States of America}
}

\author{Stephen Chong}
\affiliation{
  \department{John A. Paulson School of Engineering and Applied Sciences}
  \institution{Harvard University}
  \streetaddress{150 Western Ave.}
  \city{Boston}
  \state{MA}
  \postcode{02134}
  \country{United States of America}
}

\author{Margo I. Seltzer}
\affiliation{
  \department{Computer Science}
  \institution{The University of British Columbia}
  \streetaddress{2366 Main Mall}
  \city{Vancouver}
  \state{BC}
  \postcode{V6S 0K3}
  \country{Canada}
}

\renewcommand{\shortauthors}{Hu and Lu, et al.}

\begin{abstract}
The end of Moore's Law has ushered in a diversity of hardware not
seen in decades.
Operating system (and system software) portability is accordingly
becoming increasingly critical.
Simultaneously, there has been tremendous progress in program synthesis.
We set out to explore the feasibility of using modern program synthesis
to generate the \MD parts of an operating system.
\REVISION{
Our ultimate goal is to generate new ports automatically from
descriptions of new machines.
}


\REVISION{
One of the issues involved is writing specifications,
both for \MD operating system functionality and for instruction set
architectures.
We designed two domain-specific languages: \ale for \MI specifications
of \MD operating system functionality and \casp
for describing instruction set architecture semantics.
Automated porting also requires an implementation.
}
We developed a toolchain that, given an \ale specification and
a \casp machine description, specializes the \MI specification to the target
instruction set architecture
and synthesizes an implementation in assembly language with a customized symbolic execution engine.
\REVISION{
Using this approach, we demonstrate successful synthesis of 
a total of 140 OS components from two pre-existing OSes for four real
hardware platforms.
}
We also developed several optimization methods for OS-related
assembly synthesis to improve scalability.

\REVISION{
The effectiveness of our languages 
and ability to synthesize code for all 140 specifications
is evidence of the feasibility of program synthesis for \MD OS code. 
}
However, many research challenges remain; we also discuss the benefits and limitations of our synthesis-based approach to automated OS porting.

\end{abstract}




\maketitle

\section{Introduction}

Porting an operating system (OS) to a new machine architecture is
expensive.\footnote{
For example, NetBSD's AArch64 (64-bit ARM) port took approximately 300
commits by approximately 20 people (just in the kernel, not including
user-level material or toolchain work); this was spread over two and a
half years between serious work beginning and first release in 2020.
See \url{https://anonhg.NetBSD.org/src/file/tip/sys/arch/aarch64}.
Note that bugs are still being found; see for example
\url{https://gnats.NetBSD.org/56264}.
}
This expense manifests in many ways: 1) direct monetary cost;
2) time to market penalties; and 3) opportunity cost for experienced
kernel developers who could be handling more important issues, such as
new technology or security~\cite{security1,security2,security3,security4}.
Moreover, the end of Moore's Law has ushered in an era of new
hardware~\cite{turing18}, which requires new system software.
This suggests that porting
system software is an increasingly important challenge.


Meanwhile, the convergence of two other trends creates an opportunity.
First, some machine architecture vendors are now producing formal
specifications of their instruction set architectures (ISAs) \cite{armstrong-popl19}.
\REVISION{
Second, program synthesis techniques 
\cite{jitsynth,barman2010,memsynth,feser2015,flashfill,polozov2015}, 
and especially verification-based synthesis such as Counterexample-Guided
Inductive Synthesis (CEGIS)~\cite{solarLezama06, sketching-structures,
6679385},
have reached a point where we can expect to synthesize small but useful
program chunks in reasonable time.
}

\begin{figure}
	\includegraphics[width=\linewidth]{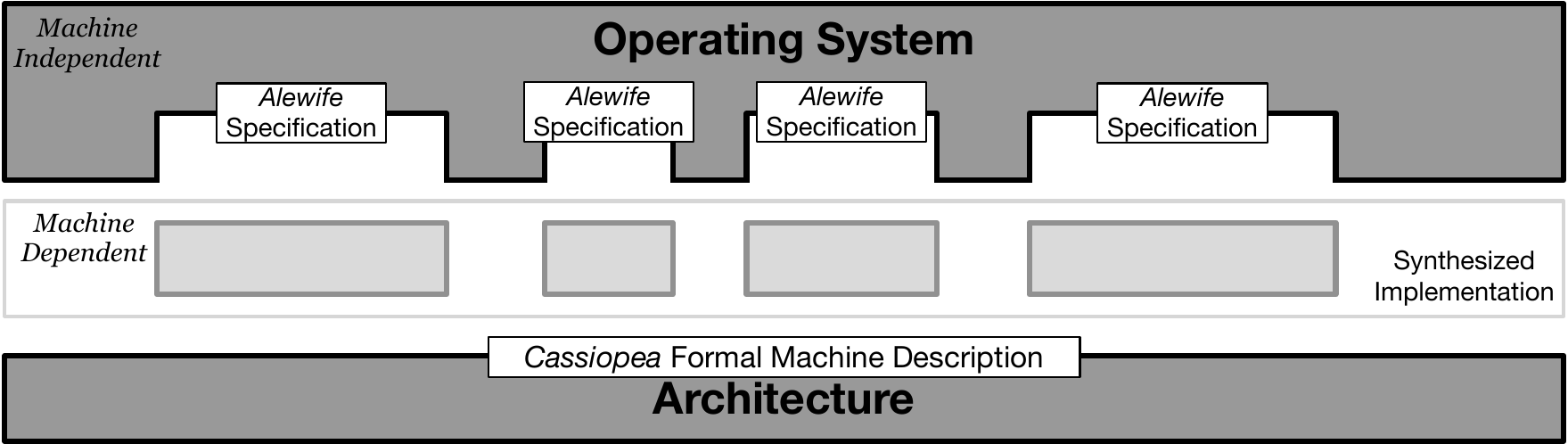}
	\caption{Porting an operating system (top box in dark gray) to an architecture (bottom box in dark gray)
		requires re-implementing the machine-dependent parts (middle boxes in light gray).}
	\label{fig:intro}
\end{figure}

The nature of OS architecture and of verification-based program synthesis suggests that a synthesis-based approach
might reduce porting costs and/or provide an efficient way to produce
verified operating system ports.
Most OSes are structured to be portable as shown in \autoref{fig:intro};
they have clearly delineated \MI (top box in dark gray) and \MD parts (middle boxes in light gray)~\cite{custer1992inside,LinuxKernel,BSDKernel,KernelArch,MachVM}.
Porting an OS requires re-implementing the \MD parts (light gray boxes in \autoref{fig:intro}), not
performing deep structural reorganization.
Further, some of the \MD parts are written in C; other parts are
written in assembler.
We focus on those parts written in assembler
(typically thousands of lines per port)\footnote{
\url{https://anonhg.NetBSD.org/src/file/tip/sys/arch/}
},
as those require in-depth knowledge of the assembly language for
the processor to which the system is being ported\footnote{
\REVISION{
Though the C code also requires knowledge of the processor, 
it does not, by definition, include operations that cannot be expressed in C, 
such as control register accesses, or code that must
not destroy registers that are ordinarily available to the compiler,
such as trap handlers.
The assembly code covers such material; it is therefore more challenging 
and is the proper first step for assessing feasibility.


}}.

The \MI parts of the OS use the \MD parts via a \MI interface whose functions
each implement a well-defined \MI operation,
such as turning on interrupts or saving a trapframe.
Overall, these assembly language \MD portions exhibit two important characteristics.
First, they are usually small and relatively isolated from one
another; they \emph{decompose readily into small independent pieces} of code,
which each exhibit simple control flow.
Each independent piece tends to be either implemented in a single function or
combined with related pieces into a sequence of semantically meaningful steps.
Second, each \MD portion implements a well-defined \MI operation, regardless of
the target architecture; while details may vary, \emph{each piece can be given a
\MI specification}.
This uniformity of specification is precisely the definition of the boundary
between \MI code and \MD code.

Taken together, these characteristics outline an approach for
synthesizing the \MD parts of an operating system. 
Operating system developers provide \REVISION{functional}
specifications of the \MD parts of
their operating system independent of any underlying hardware (white \emph{specification} boxes in \autoref{fig:intro}). 
Computer architects produce a \REVISION{functional} specification of
their platform, independent of any operating system 
(white \emph{formal machine description} box in \autoref{fig:intro}). By combining
\MI OS specifications with hardware specifications, a
synthesis engine produces the appropriate implementation (light gray
boxes in \autoref{fig:intro}). Thus, given
$M$ operating systems and $N$ hardware platforms, one writes $M+N$
specifications rather than implementing $M\times N$ ports.
\REVISION{
In our approach we use full functional specifications,
rather than, for example, partial specifications based on
input/output samples.
}

\REVISION{CEGIS-based} synthesis relies on verification to determine when an
implementation satisfies its specification, so components constructed in this
manner are correct by construction.
Such an approach offers an alternative to building verified operating system
ports from the ground up, which can take many person years~\cite{klein2009sel4}.

This approach presents myriad interesting research questions, spanning
many different areas of computer science.
\begin{enumerate}
\item What are the requirements for a language that expresses \MD functionality in a \MI way?
\item What are the requirements for a language that describes hardware
in a way amenable to synthesis? Must one model the entire ISA to synthesize
\MD OS code?
\item Some parts of an operating system are not only \MD but machine-specific, in that
they handle abstractions or structures that do not exist on other machines, such as segment tables on x86.
  How do we address parts of the operating system that are machine-specific and, thus, cannot be given a \MI specification? 
\item How well do modern synthesis techniques work in this domain?
\item What additional tooling or support is needed to integrate the results of program synthesis into a complete
working system?
\item How well does a synthesized system perform?
\item Does program synthesis make the porting process faster and/or easier?
\end{enumerate}

This paper proposes and evaluates solutions for the first two research questions and
touches upon the fourth.
We also designed and built an entire ecosystem in pursuit of the fifth
point.
A detailed description of that larger ecosystem is beyond the scope of
this paper, although we highlight several ways in which the work
discussed here leverages it.
The other research questions are either current research or intended
future work, but this area is
wide open, and we invite others to build upon our work and explore
alternative approaches.

\bigskip

To tackle these questions we developed a system to explore synthesis
of \MD parts of an OS.
Two domain specific languages are central to our approach.
\textit{\ale} is a language for \MI specifications of \MD OS
functionality (white \emph{specification} boxes in
\autoref{fig:intro}), and 
\textit{\casp}\footnote{
\casp is named for a jellyfish that features symbiotic
photosynthetic algae, which is for some reason spelled without 
the customary 'i'.
} is a register transfer language (RTL) style machine description
language (white \emph{formal machine description} box in \autoref{fig:intro}).

\ale expresses \MI specifications of \MD functionality by providing an
abstract machine to hide details.
%
In \ale, 
an OS designer
declares abstract predicates (or other functions) and abstract machine state.
These describe observable properties of a machine without providing an
exact implementation.
For example, one might declare the abstract predicate
\texttt{interrupts\_are\_on} to indicate whether interrupts are on or
\texttt{from\_usermode} to indicate whether a pending trap came from
user mode or kernel mode.
One might also declare \texttt{crt0\_argv\_reg} to be the register
that \texttt{argv} is placed in by the kernel at program startup.
These abstract declarations describe concepts that are independent of
any particular machine, but can be directly related to the exact
\MD state and functionality of each machine.
Then, using these abstractions, an OS designer can provide \MI pre-
and postcondition specifications.
For example, the postcondition of a block that is meant to turn interrupts on might require \texttt{interrupts\_are\_on()
== true}.

This formulation
leads to a beneficial separation of concerns.
OS developers can write \MI code in terms of the \MD functionality exposed
via \ale specifications.
Chip designers can write machine descriptions in \casp, possibly supplying
definitions of commonly used abstract declarations, such as interrupt state.
A programmer porting an OS to a new machine architecture need only specify how
the \ale abstract declarations used by the OS are defined in terms of
the state of the target machine.

Given an \ale input file, a \casp description of an architecture, and a mapping from \ale abstract declarations to the state of the target machine, our \lowertool turns the \ale \MI specifications into \MD specifications. Our \synth takes these \MD specifications and synthesizes assembly language programs that satisfy the specifications.


%
%

%

\medskip

\REVISION{
We evaluate our languages and tools via \CASENUM{35}
synthesis use cases from
eight complete \MD procedures 
in two pre-existing OSes: 
Barrelfish \cite{barrelfish} and the OS/161 instructional OS \cite{os161},
deployed on four real machine architectures:
32-bit MIPS, 32-bit ARM, 32-bit RISC-V and x86\_64.
Our goal in this evaluation is not (yet) to produce a complete port of
either OS, but instead, to show that synthesizing the code for such a
port is feasible.
For this reason we have selected a variety of code examples with a variety
of purposes.
Each example is taken from the \MD code for one port of one of the
OSes; we synthesize comparable code for each of the other target
architectures for which we have written machine specifications.
We successfully lower all \CASENUM{35} \MI specifications for each architecture with the \lowertool,
generating \CASENUM{140} \MD specifications in total for synthesis.
We validate the efficacy of the \synth by synthesizing and verifying assembly code for all \CASENUM{140} specifications;
all synthesis executions finished within a half-hour timeout with optimizations enabled in our \synth.
}


In summary, the contributions of this paper are:
\begin{itemize}
\item A novel approach to synthesizing the machine-dependent portions of an
OS.
\item The \ale language, which allows specification of \MD OS components in a
\MI fashion.
\item The \casp language, which allows describing real instruction
set architectures to enable synthesis of OS functionality.
\item An assembly language \synth and the \lowertool, which transforms
machine-independent specifications into machine-dependent specifications.
\item Several optimization techniques that improve scalability
for assembly language synthesis.
\item Identification and discussion of the challenges and payoffs of our
synthesis-based approach to automated OS porting.
\end{itemize}

In the next section, we begin with a brief overview of our approach to
program synthesis, named
\aquarium, after the ecosystem we built for producing synthesized
ports of \MD components of an operating system.
We then discuss our languages in detail
(Sections \ref{sec:ale} and \ref{sec:casp}) and describe the mechanism
for connecting them (\autoref{sec:lowering}).
In \autoref{sec:impl}, we describe the implementation of
the \lowertool and \synth.
Then, in \autoref{sec:validation}, we present use cases, validation,
and experimental results.
\autoref{sec:disc} contains further discussion, including when program synthesis
is not appropriate for implementation of \MD OS components (and a brief description of other tools we created to address these situations);
\autoref{sec:rw} discusses related work, and
\autoref{sec:conclusion} concludes.

\section{Overview: Program synthesis in \aquarium}\label{sec:overview}

\aquarium is a collection of tools to automatically construct the
\MD parts of an OS.
In this work, we focus on the languages we developed for 1) describing
machine-dependent OS functionality in a machine-independent fashion and
2) specifying a processor's semantics, and on using modern program synthesis
in the assembly language realm, where all state is global and data is untyped.

\begin{figure}
	\ifanon \includegraphics[width=\linewidth]{overview-horizontal} \else
	\includegraphics[width=\linewidth]{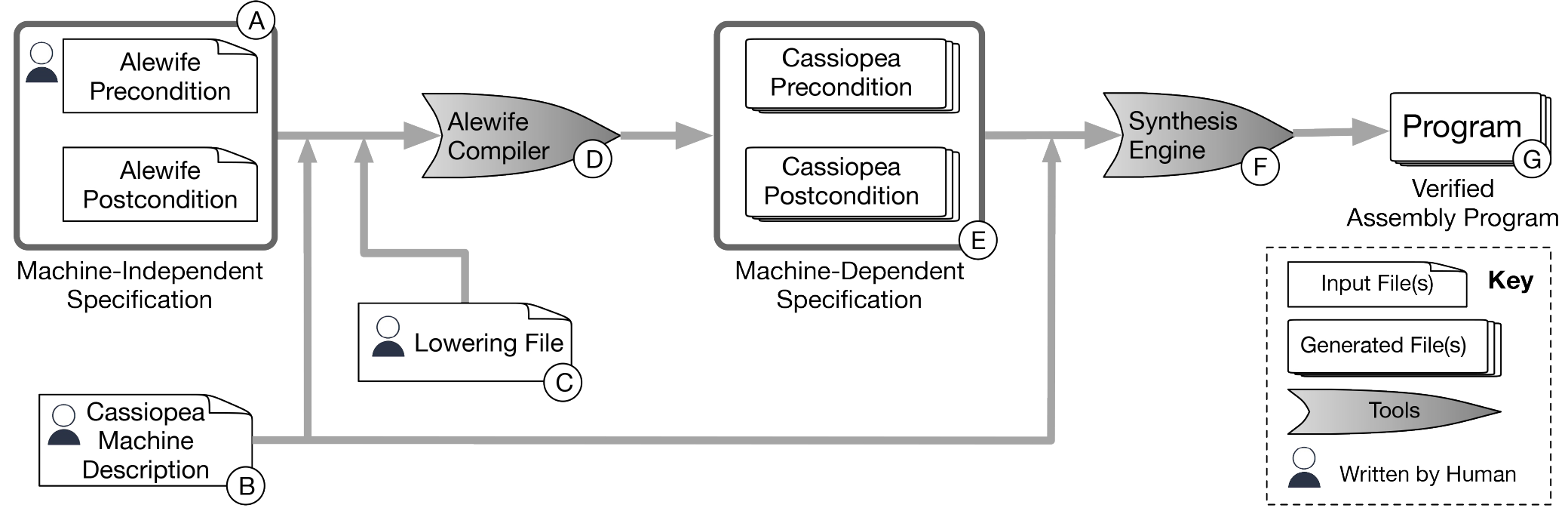} \fi
	\caption{\aquarium program synthesis workflow.
          Given a \MI \ale specification \protect\circled{A} (written by OS designers),
          a \casp machine description \protect\circled{B} for a specific architecture (written by chip designers),
          and an \ale-\casp \lowering file \protect\circled{C},          
          the \lowertool \protect\circled{D}
          generates a \MD \casp specification \protect\circled{E} for that architecture.
          The \casp \synth \protect\circled{F} uses the machine description and the \MD specification to synthesize a satisfying sequence of assembly instructions, verifies them against the specification, and translates them into real assembly programs \protect\circled{G}.}
	\label{fig:overview}
\end{figure}

\autoref{fig:overview} presents an overview of \aquarium's
four-step approach to OS synthesis:

\emph{Step 1: Writing a Machine Independent Specification.}
An OS developer writes an \ale specification for some piece of OS functionality
\circled{A}.
\ale reasons about \MD functionality in an abstract machine (\autoref{sec:ale}).

\emph{Step 2: Writing Machine Descriptions.}
A computer architect produces a machine model \circled{B} written
in \casp, which models the ISA semantics at the assembly language level (\autoref{sec:casp}).

\emph{Step 3: Lowering to Machine-Dependent Specifications.}
To specialize a \MD specification for synthesis, \aquarium
applies \emph{\lowering} files \circled{C} that
instantiate the \MI \ale abstractions for the target
architecture.
The \ale compiler \circled{D} performs the \lowering process, which takes a \casp machine description, an \ale specification, and the corresponding \lowering file to generate the \MD specification \circled{E} for that architecture (\autoref{sec:lowering}).

\emph{Step 4: Synthesis and Verification.}
Given the \casp machine description and the \MD specification, \aquarium uses the \casp \synth \circled{F} to synthesize a satisfying sequence of assembly instructions \circled{G} and verify it against the specification (\autoref{sec:impl}).

\subsection*{A Small Example of Aquarium Program Synthesis}
\begin{figure}
\begin{tabular}{c}
\begin{lstlisting}[language=alewife]
ldr r0, [r2, #88]
cmp lr, r0
movhs r0, #0x00000000
movlo r0, #0x00000001
\end{lstlisting}
\caption{ARMv7 \texttt{disp\_check} assembly program example from the Barrelfish operating system.}
\label{fig:stack-asm-mips}
\end{tabular}
\end{figure}
As a running example, we use an excerpt from the exception handling code
of the Barrelfish operating system~\cite{barrelfish}.
\autoref{fig:stack-asm-mips} shows the original ARMv7 Barrelfish code (which we call \verb|disp_check|).
The code sets the return value in \verb|r0| to 0 or 1
depending on the relationship between the value in the memory location
\verb|[r2,#88]| and the contents of the special register \verb|lr|.
The value in \verb|r0| is used in code that follows this block.
Since this code is written in assembly, its implementation is machine-dependent.
We discuss this code in more detail and present its
\MI specification in the sections that follow.

\newcommand{\ensm}[1]{\ensuremath{#1}}
\newcommand{\aprimitivetyp}{\atyp_{\mathit{base}}\xspace}
\newcommand{\bitsize}{\ensm{\mathit{bv_{len}}}\xspace}
\newcommand{\ttwptr}{\ensm{\mathtt{ptr}}\xspace}
\newcommand{\ttwvec}{\ensm{\mathtt{vec}}\xspace}
\newcommand{\ttwreg}{\ensm{\mathtt{reg}}\xspace}
\newcommand{\ttN}{\ensm{\mathit{N}}}
\newcommand{\ttmemid}{\ensm{\mathit{x_{mem}}}\xspace}
\newcommand{\ttmlen}{\ensm{\mathtt{len}}\xspace}
\newcommand{\ttbit}{\ensm{\mathtt{bit}}\xspace}
\newcommand{\memtyp}[3]{\ensm{#1\ \ttbit\ #2\ \ttmlen\ #3\ \texttt{ref}}}
\newcommand{\ptrform}[2]{\ensm{\ttinparen{#1, #2}}}
\newcommand{\ttinparen}[1]{\ensm{\mathit{(}#1\mathit{)}}\xspace}
\newcommand{\ttmemidi}{\ensm{\mathit{x_{mem}}_1}\xspace}
\newcommand{\ttmemidj}{\ensm{\mathit{x_{mem}}_2}\xspace}

\section{Writing \MI specifications in \ale}
\label{sec:ale}

Each \ale file (\circled{A} in \autoref{fig:overview}) 
provides a \MI specification for a single piece of \MD
functionality in terms of its effects on an abstract machine.
This allows specifications to reason about the underlying machine
state that is relevant to \MI code without having to know \MD
details.
We assume that these specifications are written by OS designers
with general knowledge of machines, but not in-depth knowledge of all
the architectures to which
their system will eventually be ported.

We call the unit of \MD functionality in an \ale file a \emph{block}.
A block is a sequence of assembly instructions
that has a single entry point,
no backward branches to code in the block,
and optionally a single branch to code outside the block.
The specification consists of four parts:
1) preconditions, 2) postconditions, 3) let-bindings, and 4) frame conditions.
The precondition describes the \emph{initial state} of the
abstract machine immediately before the block executes.
The generated code may assume that the precondition is true
and can behave in arbitrary or undefined ways if the precondition is false.
The postcondition describes the \emph{final state} of the 
abstract machine immediately after the block executes.
Provided that the precondition was true, correct generated code
(in our case a sequence of assembly instructions)
must guarantee that the postcondition is true after it executes.
Let-bindings are evaluated against the initial state and allow
the postcondition to refer to the initial state.
Frame conditions specify what machine state may be modified.
An \ale file may include other files, enabling reuse of declarations shared across multiple OSes.

The rest of this section describes the mechanisms by which \ale permits \MI
specification of \MD functionality;
\ale syntax and semantics appear in Appendixes
\ref{sec:ale_lang} and \ref{sec:ale_check}.

\autoref{fig:stack-alewife} shows the \ale specification for Barrelfish's \verb|disp_check|.
This block compares the saved program counter where an
exception occurred (previously placed in \verb|pc_reg|) to the
upper bound of the critical region in which the exception is
re-entrant and requires special handling.
That bound is ordinarily stored in the dispatcher structure.\footnote{
Barrelfish decomposes its kernel into a \MD CPU driver
and a machine independent, user-space Monitor.
The CPU driver ``performs dispatch and fast local messaging between
processes on the core''~\cite{barrelfish}.
The \texttt{dispatcher} structure belongs to the CPU driver and is
analogous to a process structure or process control block; among other
things it contains register save areas.
Thus, it is both OS-specific, as it is particular
to the Barrelfish architecture, and machine-dependent, as the size of the
structure varies with the pointer size, word size, and number of registers
of the architecture.
}
The dispatcher structure resides in \verb|DispMem| and
contains space to save registers; thus, both the
number of fields in it and their sizes are \MD and must be
abstracted.
\verb|get_crit_ptr| takes the base address of the dispatcher structure
as an argument and adds in the offset of the proper field in the structure.
We make \verb|get_crit_ptr| a function, rather than adding in an
abstract offset directly, for illustrative purposes.

The precondition (line 13) indicates that 
register \verb|disp_reg| (from line 4) must contain a pointer to the
base of the region \verb|DispMem|, which is defined on line 8.
The postcondition (line 14-15) ensures that the
register \verb|disp_area_reg| contains a bitvector whose value is \verb|1|
if the variable \verb|crit| (line 12) is \verb|true| or 
a bitvector whose value is \verb|0| if \verb|crit| is \verb|false|.
The variable \verb|crit| is a boolean condition that stores the comparison result 
between a register (\verb|pc_reg|) and the contents of the memory
location computed by the function \verb|get_crit_ptr|.

\subsubsection*{Memory model}
\ale memory regions, $\ttmemid$, have finite lengths 
with types $\memtyp{\ttN_1}{\ttN_2}{\ttN_3}$;
such a region contains $\ttN_2$ elements, each of which is an
$\ttN_1$-bit value; pointers into the region are $\ttN_3$-bit values. 
$\ttN_1$, $\ttN_2$, and $\ttN_3$ here can be 
either concrete lengths or abstract symbolic constants.
For example in \autoref{fig:stack-alewife}, the type of the \verb|DispMem| region has three symbolic constants:
it has \verb|DISP_MAX| locations (\verb|DISP_MAX len|),
each containing a bitvector with length \verb|wordsize| (\verb|wordsize bit|),
whose pointers are \verb|wordsize|-bit values (\verb|wordsize ref|).
We discuss the concretization of these abstract values in \autoref{sec:lowering}.

Memory is addressed by \textit{pointers}, which are pairs of a memory region
and offset; a pointer to the $n$th element of memory region $\ttmemid$ is represented
by the pair $\ptrform{\ttmemid}{n}$.
Memory regions do not overlap; that is,
given two memory regions $\ttmemidi$ and $\ttmemidj$, a pointer
$\ptrform{\ttmemidi}{\_}$ can never alias a pointer $\ptrform{\ttmemidj}{\_}$.
This ``swiss-cheese'' memory model is inspired by DieHard's
miniheaps~\cite{berger06} and array separation logic~\cite{asl} and
reflects the C standard.
Memory regions are second-class; pointers may be passed around, but
memory regions may not and every pointer must explicitly name a
specific memory region.

A memory region can optionally have a \emph{label}, which is the
symbol used to refer to its base address in assembly language text.
Note that while this symbol refers to a constant, the value of the
constant is not known until the code is assembled and linked into the
final output program; or in the case of typical shared libraries or
position-independent executables, not until program execution.
In general, position-independent code and ordinary code must use
different instructions to refer to labels.
For this reason, and because Barrelfish links its kernel as
position-independent code, we attach an access type to each label.

It is allowable to declare a memory region with \verb|0 len|, 
which means that we model only the existence of the region but not its
internal workings.
OS designers generally do not need to know the synthesis algorithm 
when writing these \ale specifications. However, limiting memory 
regions to their minimum size improves performance, since adding extra state slows down synthesis. 

\begin{figure}
\begin{tabular}{c}
\begin{lstlisting}[language=alewife]
require type word
require value wordsize: int
require value pc_reg: wordsize reg
require value disp_reg: wordsize reg
require value disp_area_reg: wordsize reg
require value DISP_MAX: int
require function get_crit_ptr: (word) word
region DispMem: wordsize bit DISP_MAX len wordsize ref
lower-with: disp_defs may_use_flags disp_scratch

let crit_ptr: wordsize ptr = get_crit_ptr([DispMem, 0])
let crit : bool = *pc_reg b< fetch(crit_ptr, wordsize)
pre: *disp_reg == [DispMem, 0]
post: if crit then *disp_area_reg == (1: wordsize vec)
else *disp_area_reg == (0: wordsize vec)
\end{lstlisting}
\caption{\ale specification for \texttt{disp\_check}.}
\label{fig:stack-alewife}
\end{tabular}
\end{figure}

\subsection*{Abstract Declarations and Abstract Functions}

Values imported with \verb|require| (lines 2-6 in \autoref{fig:stack-alewife}), 
such as the register \verb|pc_reg| and integer \verb|wordsize|, 
are examples of \emph{abstract declarations},
allowing an \ale specification to omit
details of how an implementation represents and stores data.
This is the key to making \ale specifications \MI.
\REVISION{
Such identifiers must be defined elsewhere for each target machine;
we bind the abstract declarations and their corresponding \MD definitions
in a later step, before conducting synthesis. 
}
\ale types (the complete specification is in \autoref{sec:ale_check})
can contain 
symbolic bit lengths 
and abstract type identifiers,
which will be specialized to different bit lengths and \casp types by \casp machine descriptions and \MD lowerings
\REVISION{
before synthesis takes place.
}
For example, in \autoref{fig:stack-alewife}, the value \verb|wordsize|, imported with \verb|require| in line 2, 
is used to define the dispatcher memory region \verb|DispMem| on line 8.
A machine description, written in \casp, provides the definition for
\verb|wordsize| appropriate for a specific machine.
Similarly, the size of the (\MD) dispatcher structure
is specified using an abstract integer \verb|DISP_MAX|.
We discuss details of the concretization of these abstract states 
and functions in \autoref{sec:casp} and \autoref{sec:lowering}, 
with examples from two different architectures (MIPS and ARMv7) whose relevant 
\casp descriptions appear in \autoref{fig:casp-all} and 
the lowering files in \autoref{fig:stack-lowering}.

\ale specifications also use \emph{abstract functions}.
For example, line 7 of \autoref{fig:stack-alewife}
declares a function \verb|get_crit_ptr|.
This example does not refer to machine state, but others might.
The predicate \verb|interrupts_are_on| mentioned earlier, for example,
will.

The keyword \verb|lower-with| (e.g., line 9 in \autoref{fig:stack-alewife})
is used to name modules
from which to get the \MD instantiations of the abstract definitions.
In this case, there are three: definitions related to \verb|disp_check|, 
permission to change the processor flags on machines that have them,
and scratch register assignments for this block.
These are called \emph{lowering modules} and are discussed in more detail in \autoref{sec:lowering}.

\subsection*{Frame Conditions}

An \ale specification may optionally provide \emph{frame conditions} that
\REVISION{
permit a block to modify additional machine state (registers or
register fields and memory).
}
By default, synthesized code can modify only machine state
explicitly mentioned in the postcondition.
All other state elements' final values must equal their initial values.
The \verb|modify| frame condition identifies additional
(abstract) state elements that may be modified (e.g., scratch registers).
As we discuss in \autoref{sec:lowering}, lowering modules can also
add frame conditions.
\REVISION{
In practice this is where most frame conditions come from.
}
The frame conditions are more than synthesis heuristics that help
to prune the search space;
they are a useful shorthand to materially affect the specification.

The default \REVISION{behavior} is desirable: most blocks should
not modify state irrelevant to their postcondition.
However, some computations require scratch registers.
Additionally, it is
sometimes necessary to read a register and write the same value back.
For example, many control registers contain multiple fields but can
only be read and written as a unit;
changing one field requires reading the whole
register, updating the desired field, and writing the whole register
back.
It is important both to permit these accesses and to ensure that the
other fields are not arbitrarily modified.
On MIPS, for example, bit 0 of the status register controls the
interrupt state; but the register contains many other fields where
arbitrary changes can have unwanted consequences, such as switching
between 32-bit and 64-bit mode.
A correct MIPS implementation of code to enable or disable interrupts
must be able to write to the bit controlling 64-bit mode, but must not
change it.

Since we do not prohibit writes entirely,
in theory, the synthesizer can generate programs that change
registers in arbitrary ways and then restore their initial value.
In practice, this is not an issue, because our synthesis
technique preferentially produces smaller programs, excluding
programs with redundant operations.
Another potential problem is that the synthesizer is obliged to waste
time considering and ruling out programs that make unwanted changes.
We address this in the implementation by \emph{gating} registers
(\autoref{subsec:gating})
that are irrelevant to the specification; this removes them
entirely from consideration.

	

\subsection*{Limitations of \ale}

\ale specifications are based on the premise that the \MD code in an
operating system is made up of \MD instantiations of \MI constructs;
this allows the bulk of the specification to be \MI with relatively
minor parts substituted on a per-machine basis.
In practice, this is largely, but not entirely, true.
Some \MD code in operating systems is not just \MD but
machine-\emph{specific}:
it handles concepts that do not exist on other machines, e.g.,
segment tables on x86.
These concepts primarily arise in code that is called directly by the
machine (such as trap handlers and the kernel startup code) rather
than by the \MI part of the OS.
It is \emph{possible} to specify such code in \ale, but only in a
degenerate form, where the entire specification is abstracted into a
single pre- and postcondition pair in the lowering file.




\renewcommand{\ensm}[1]{\ensuremath{#1}}
\newcommand{\atyp}{\tau}
\newcommand{\ttid}{\ensm{\mathit{id}}}
\newcommand{\ttC}{\ensm{\mathit{C}}}
\renewcommand{\ttN}{\ensm{\mathit{N}}}
\newcommand{\bnfeq}{\ensm{\Coloneqq}}
\newcommand{\bnfor}{\ensm{\mid}}
\renewcommand{\ptrform}[2]{\ensm{\ttinparen{#1, #2}}}
\renewcommand{\ttinparen}[1]{\ensm{\mathit{(}#1\mathit{)}}\xspace}
\renewcommand{\ttmemid}{\ensm{\mathit{x_{mem}}}\xspace}
\renewcommand{\ttmemidi}{\ensm{\mathit{x_{mem}}_1}\xspace}
\renewcommand{\ttmemidj}{\ensm{\mathit{x_{mem}}_2}\xspace}
\newcommand{\ttloc}{\ensm{\texttt{reg}}\xspace}

\newcommand{\bvtyp}[1]{\ensm{#1\ \ttbit}}
\newcommand{\bvaltyp}[1]{\ensm{\bvtyp{#1}}}
\newcommand{\bvltyp}[1]{\ensm{#1\ \ttloc}}
\renewcommand{\ttmlen}{\ensm{\mathtt{len}}\xspace}
\renewcommand{\ttbit}{\ensm{\mathtt{bit}}\xspace}

\renewcommand{\memtyp}[3]{\ensm{#1\ \ttbit\ #2\ \ttmlen\ #3\ \texttt{ref}}}
\newcommand{\labeltyp}[1]{\ensm{#1\ \ttlabel}}
\newcommand{\nbitmemtyp}[1]{\memtyp{\_}{\_}{#1}}
\newcommand{\ptradd}[2]{\ensm{#1\ \mathtt{p+}\ #2}}

\section{Writing machine descriptions in \casp}
\label{sec:casp}
\begin{figure}
\noindent
\begin{minipage}[t]{.49\textwidth}
\begin{lstlisting}[title=Listing A. \casp MIPS machine description, language=casp,frame=r]
(* bitvector and register types *)
let wordsize: int = 32
type word = 32 bit
type register = 32 reg
(* general purpose registers *)
letstate r0: register
...
(* control flags: cp0 $12: Status *)
letstate control cp0_12_ie: 1 reg ...
(* assembly text representation *)
let r0.txt = "$0" ...
(* machine invariant *)
invariant: *r0 == 0x00000000
(* instructions *)
defop SLTU rd:register rs:register
  rt:register {
 txt = format("sltu {1}, {2}, {3}",
  rd.txt, rs.txt, rt.txt),
 sem = [ (* check the validity
 of each operand *)
   assert (valid_gpreg(rd));
   assert (valid_gpreg(rs));
   assert (valid_gpreg(rt));
   assert (rs != rt);
   assert (rt != r0);
   if rd == r0 then skip
   else if ( *rs b< *rt ) then
     *rd <- 0x00000001
   else *rd <- 0x00000000 ]}
defop LW rd:register rs:register
  imm:16 bit {
 txt = format("lw {1}, {2}({3})",
   rd.txt, imm.dec, rs.txt),
 sem = [ ... (* check validity *)
   if rd == r0 then skip
   else let addr: word =
     *rs b+ sign_extend(imm)
   in *rd <- fetch[addr, 32] ]}
 ...
\end{lstlisting}
\end{minipage}\hfill
\begin{minipage}[t]{.49\textwidth}
\begin{lstlisting}[title=Listing B. \casp ARMv7 machine description, language=casp]
(* bitvector and register types *)
let wordsize: int = 32
type word = 32 bit
type register = 32 reg
(* general purpose registers *)
letstate r0: register
...
(* control flags: CPSR *)
letstate control Z: 1 reg
...
(* assembly text representation *)
let r0.txt = "r0"
...
(* instructions *)
defop MOV_imm rd:register ri:4 bit
  vi:8 bit cd:4 bit {
 txt = let wi:word = rotimm(ri, vi) in
  format("mov{1} {2}, #{3}",
  armcc(cd), rd.txt, wi.hex),
 sem = [ (* check validity *)
   assert (valid_cond(cond));
   assert (valid_gpreg(rd));
   assert (valid_rotimm(ri, vi));
   if checkcond(cd) then
     *rd <- rotimm(ri, vi)
   else skip ]}
defop LDR_imm rd:register rn:register
  imm:12 bit cd:4 bit {
 txt = format("ldr{1} {2}, [{3},#{4}]",
  armcc(cd), rd.txt, rn.txt, imm.dec),
 sem = [
   ... (* check validity *)
   if checkcond(cd) then
     assert (is_ptr( *rn ));
     let addr: word =
       *rn b+ zero_extend(32, imm)
     in *rd <- fetch[addr, 32]
   else skip ]}
...
\end{lstlisting}
\end{minipage}
\caption{Excerpts from \casp machine descriptions. Lines of the form (* ... *) are comments.\label{fig:casp-all}}
\end{figure}

To verify or synthesize programs against a particular machine, \aquarium uses a
model of the ISA semantics written in the \casp language (\circled{B} in
\autoref{fig:overview}).
This section presents an overview of the language, highlighting those
design features that
improve the efficiency of symbolic execution and synthesis.
The complete syntax and semantics appear in Appendices~\ref{sec:lang} and
\ref{sec:check}.

A \casp machine description provides an executable model of an ISA, declaring
\emph{machine state}, such as registers and memory, and defining
\emph{operations} that describe its assembly instructions.
\casp is a typed, interpreted language.
In the following discussions we refer to conditions detected during
the initial typechecking pass as ``static'' and those that are not as
``runtime errors''.
The symbolic execution that occurs during synthesis (see
\autoref{sec:impl}) explores all execution paths and treats any
runtime errors as conditions to avoid; that is, no output program
should ever produce a runtime error.

Throughout the rest of this section, we use
excerpts from two of our \casp
machine descriptions
(\autoref{fig:casp-all}:
MIPS (Listing A) and ARMv7 (Listing B)) as running examples.
Unless otherwise stated, line numbers refer to lines in both listings
from this figure.

\subsubsection*{Register model}
Registers in \casp have types of the form $\bvltyp{\ttC}$, which indicates
that a register can hold $\ttC$-bit values (line 4).
Registers are first-class values and can be passed as
arguments to functions and operations.
Each register has the text form used in the machine's assembly language
associated with it.
The \synth uses this to produce the synthesized assembly code; it allows
working with non-identifier register names, such as \verb|%eax| on x86.

\subsubsection*{Memory model}
\casp uses the same ``swiss-cheese'' memory model as \ale does (described in \autoref{sec:ale}).
However, unlike \ale, \casp memory regions, $\ttmemid$, 
have finite, statically known lengths.
Unbounded or variable-length memory regions do not typically appear in
assembly-language \MD OS components.
\casp memory regions have types of the form $\memtyp{\ttC_1}{\ttC_2}{\ttC_3}$,
where $\ttC_1$, $\ttC_2$, and $\ttC_3$ must be concrete constants.
As in \ale, labels have an access type that indicates whether
references should be position-independent; instruction definitions that
use labels can query or assert about the access type, so that the \synth
produces the appropriate code.
Note that most memory regions appear in \ale specifications 
and not in machine descriptions: we provide a
programmer's view of the machine, so the memory regions
available are the ones associated with the code at hand and not the
machine's entire memory space.

\subsubsection*{Bitvectors and pointers}
In a typical ISA, all values are bitvectors and can be used by any operation.
In contrast, as in typed assembly language~\cite{tal} and the Compcert
backend~\cite{leroy2009formally}, \casp distinguishes
between finite bitvector constants and pointers into memory regions.
The distinction is not statically checked, however; the type
$\bvaltyp{\ttC}$ includes both $\ttC$-length bitvectors and pointers
into memory regions of type $\memtyp{\ttC_1}{\ttC_2}{\ttC}$.
It is a runtime error to apply bitvector operations to pointer values and vice
versa, though we overload some operations to allow limited pointer arithmetic.
We deliberately omit the ability to convert between pointer values and
bitvectors, since we do not model absolute memory addresses.
This model simplifies reasoning about the semantics of programs, since they
cannot create pointers to arbitrary regions.
It is also a runtime error to make unaligned or out-of-bounds accesses
to memory.

\subsubsection*{Machine state}
All machine state is either registers or memory.
Such state is global in scope and declared using
\textbf{\texttt{letstate}}.
Registers and memory can contain only bitvectors and pointers.
\casp code can manipulate integers, booleans, and registers to define
instruction semantics, but these values cannot be stored directly in machine
state.

We refer to the state of a machine $\mathbf{M}$ with the notation
$\Sigma_\mathbf{M}$, which
is a pair $(\rho, \sigma)$.
$\rho$ is a map from registers to
values of bitvector type, and $\sigma$ is a map from pointers to
values of bitvector type.

Our example in \autoref{fig:casp-all} declares several pieces of machine state.
The general-purpose register \verb|r0| for both MIPS and ARM appears on
line 6.
Its assembly output text is specified by binding \verb|r0.txt|,
which appears on line 11 in \autoref{fig:casp-all}A and line 12
in \autoref{fig:casp-all}B.
We also show one control register for each machine:
\verb.cp0_12_ie. in MIPS (line 9 in \autoref{fig:casp-all}A)
and the \verb|Z| flag in ARM (line 9 in \autoref{fig:casp-all}B).
(Control registers vary widely across machines.)
There are no memory regions in the example because, as discussed
previously, memory regions normally arise in program specifications
and not in the machine model itself (see \autoref{fig:stack-alewife} line 8).


\subsubsection*{Operations and programs}
Machine state in \casp is manipulated by \emph{operations}.
These are typically a single assembly instruction 
but may represent a group of instruction variants and/or a short
sequence.
In the examples, lines 15--29 and 30--38 of \autoref{fig:casp-all}A
and lines 15--26 and 27--38 of \autoref{fig:casp-all}B
define operations with \textbf{\texttt{defop}}.

Each operation definition has two parts: \verb.txt. is an expression that
constructs the assembly text representation of the operation, and \verb.sem. is
a \casp statement defining the operation's full concrete semantics in terms of the machine
state.
In the definition, most operations check the validity of each operand first.
The MIPS instructions shown are 
\verb.SLTU. (unsigned set less than) and \verb|LW| (load word).
The ARM instructions shown are \verb|MOV_imm| (move immediate to register)
and \verb|LDR_imm| (load with immediate value offset).

Borrowing from JitSynth~\cite{jitsynth},
an \textit{operation} is a pair $(op, \mathbf{T})$, where $op$ is an
\textit{opcode}, and $\mathbf{T}$ is a map from argument name to type, which
can include registers, booleans, concrete bitvectors, and labels.
(When used as an argument to an instruction that accepts labels, a label
provides a pointer to the first entry in its memory region.)
Each operation defines a set of \textit{instructions} $(op,
\mathbf{v})$ where $\mathbf{v}$ is a map from argument name to value, where
for each argument $x$, $\mathbf{v}(x)$ is of type $\mathbf{T}(x)$.
For a machine $\mathbf{M}$, we denote its set of instructions as
$\mathcal{I}_\mathbf{M}$.

\casp machine descriptions define the semantics of an instruction
as a partial function (partial due to the possibility of runtime
errors) from a machine state to a new machine
state accompanied by a \textit{branching state}:
\begin{align*}
	\llbracket (op, \mathbf{v}) \rrbracket_\mathbf{M} : \Sigma_\mathbf{M} \rightharpoonup
	\Sigma_\mathbf{M} \times \left(\mathbb{N} \cup \{\cdot, \mathtt{ext}\}\right)
\end{align*}
The branching state indicates the destination of a branch instruction.
Operations in \casp may only branch forward, so the branching state indicates
that execution should proceed with either
the next instruction ($\cdot$), an external assembler label ($\mathtt{ext}$),
or the instruction after skipping $n$ instructions forward ($\mathbb{N}$) (where
$n$ should be at least 1).

A \textit{program} is a series of instructions.
We define the result of running a program $P$ (a list of instructions) on a
machine state $(\rho, \sigma) \in \Sigma_\mathbf{M}$ using a partial
function $run_\mathbf{M} : \mathcal{P}_\mathbf{M} \times \Sigma_\mathbf{M}
\rightharpoonup \Sigma_\mathbf{M}
\times bool$ where $\mathcal{P}_\mathbf{M}$ is the set of programs for machine
$\mathbf{M}$ and the result includes a boolean representing the
\texttt{branchto} value, indicating whether the program jumped to the special
external assembler label.

\begin{align*}
run_\mathbf{M}(P, (\rho, \sigma)) =
\begin{cases}
(\rho, \sigma, false) & P = nil \\
(\rho', \sigma', true) &
\llbracket head(P) \rrbracket_\mathbf{M}(\rho, \sigma) =
  (\rho', \sigma', \mathtt{ext}) \\
run(drop(P, 1), (\rho', \sigma')) &
\llbracket head(P) \rrbracket_\mathbf{M}(\rho, \sigma) =
  (\rho', \sigma', \cdot) \\
run(drop(P, n), (\rho', \sigma')) &
\llbracket head(P) \rrbracket_\mathbf{M}(\rho, \sigma) =
  (\rho', \sigma', n) \wedge n \leq len(P) \\
\bot &
\llbracket head(P) \rrbracket_\mathbf{M}(\rho, \sigma) = \bot \\
\bot &
\llbracket head(P) \rrbracket_\mathbf{M}(\rho, \sigma) =
  (\rho', \sigma', n) \wedge n > len(P) \\
\end{cases}
\end{align*}
The function $head(\ell)$ is the first element of a list $\ell$,
$drop(\ell, n)$ is the tail of the list $\ell$ after removing $n$ elements,
and $len(\ell)$ is the length of the list $\ell$.

\subsubsection*{Other elements.}
\casp files may define type aliases, values, functions, and procedures
and use them in the
definitions of operations.
(Functions are pure; procedures may alter machine state.)
Both the MIPS and ARM descriptions define \verb|word| as a type of 32 bits (line 3)
and \verb|register| as the type of 32-bit registers (line 4).
The value \verb|wordsize|, discussed in \autoref{sec:ale}, is also specified as 
$32$ on both 32-bit MIPS and 32-bit ARMv7 (line 2)
and $64$ for x86\_64 (not shown).
Function \verb|valid_gpreg| checks whether the given operand is a general purpose register (line 21-23 in \autoref{fig:casp-all}A and line 22 in \autoref{fig:casp-all}B).
This prevents generating code that tries to use special registers or
control registers that happen to be the same size as general-purpose
registers and thus have the correct type to pass as operands.
\verb.LW. in MIPS uses a function \verb.sign_extend. (line 37 in \autoref{fig:casp-all}A, definition not shown),
which implements sign extension on bitvectors.
\verb|LDR_imm| in ARM uses a function \verb|zero_extend| (line 36 in \autoref{fig:casp-all}B),
which implements (unsigned) zero-extension.
Common functions and procedures, such as those used for bitvector manipulation, can be defined
in separate files and reused by inclusion in different machine descriptions.

It is also possible to define machine invariants.
These are predicates
that must hold true before and after block execution.
The \textbf{\texttt{invariant}} statement in line 13 in \autoref{fig:casp-all}A declares that the
value in register \verb|r0| is always zero.


\subsubsection*{Execution model}
\casp is not Turing-complete; machine models are finite-state and evaluation of
\casp functions, procedures, and operations always terminates.
We deliberately restrict \casp's ability to model control flow:
\casp operations may only branch forward, so assembly snippets modeled in \casp
are loop-free.
The semantics of \casp do not include clocks and timers, concurrency, hazards,
or weak memory models.
The minimalistic design of \casp allows symbolic execution to generate complete
logical descriptions of our models and helps to simplify synthesis.
We show in \autoref{sec:validation} that \casp remains sufficiently
expressive to support assembly synthesis for many OS components.

\smallskip

Because \casp models ISAs at the assembler level, \casp models need not be wire-
or bit-level accurate.
Our goal is to generate code to be assembled and linked into an (existing) OS,
using that OS's compiler and toolchain.
Thus, \casp descriptions need not capture phenomena hidden by the
assembler (e.g., branch delay slots on MIPS).
Moreover, machine descriptions need contain only those parts necessary to
synthesize targeted \MD OS components; a surprising finding of this work is
how little of an ISA needs to be specified to synthesize significant parts of
an operating system.
Although we anticipate a world where vendors provide descriptions of ISA
semantics, we currently write \casp machine descriptions manually.

\section{Lowering to \MD specifications with the \lowertool\label{sec:lowering}}

Synthesizing an OS component for a particular machine architecture requires
a \emph{\MD specification} (\circled{E} in \autoref{fig:overview}), i.e., a pre- and
postcondition expressed in terms of a particular machine
architecture's state and semantics.
Meanwhile, the \ale specification for an OS component is written in terms of
abstract functions and abstract machine state, describing
an abstract
machine.
\emph{Lowering}\footnote{
In abstract interpretation, a concretization function typically maps
an abstract state to a \emph{set} of concrete states.
Our \lowering chooses a single concrete specification, so we avoid the
term ``concretization''. However, like a concretization function,
lowering files map something abstract (an \ale specification) to something
concrete (a \MD specification).
}
is the process of
bridging this gap, that is,
specializing a \MI \ale specification to a \MD specification.

As explained in \autoref{sec:ale}, this requires a definition of each
abstract \ale element suitable for the selected machine.
One possible source of these definitions is the machine description
itself.
For OS-independent concepts, it is reasonable to create a single
abstract \MI model and instantiate it in each machine description.
For example, in practice every machine of interest has a stack pointer
register; thus we include in each machine description a definition of
\verb|stackpointer| as an alias for the proper register.
This is accompanied by a definition for the expected alignment of the
stack pointer on entry into C code and a boolean indicating whether
stacks conventionally grow up or down.
We also have OS-independent models for calling conventions,
position-independent code, interrupt enable/disable state, and certain
pieces of trap handling state, such as the address at which a trap
occurred.
OS-independent models for cache and TLB flushes are the subject of ongoing
research.
Any OS using our machine descriptions is free to use these models or
ignore them.
For example, our interrupt state model consists of one definition:
\verb|interrupts_are_on|.
This is sufficient for the OSes in our use cases, but an OS
with traditional interrupt priorities needs a more
complex model; for example it might associate an interrupt mask with
each priority level and provide a definition that returns the
hardware's current interrupt mask.
A slightly more complex model might provide a definition that
returns the current priority level by interpreting the hardware
interrupt mask.
That interpretation is itself OS-dependent, and since OS-dependent
definitions do not properly belong in the machine description itself,
they must appear elsewhere.
We place them in \emph{lowering files}
(\circled{C} in \autoref{fig:overview}).

%
%

Lowering files are \casp-language files that contain OS-specific and
machine-specific definitions suitable for instantiating \ale
abstractions.
To improve sharing, lowering files are organized in terms of
\emph{modules}; each file can contain arbitrarily many modules
and each module is a collection of arbitrary \casp declarations and,
optionally, a frame clause.
Both lowering file modules and \ale specifications can \verb|import|
zero or more lowering modules by name.

The \lowertool (\circled{D} in \autoref{fig:overview}) reads the
lowering files, extracts requested modules from them, and lowers \ale
specifications by
replacing \MI types, definitions, and specifications with their \MD
counterparts.
The definitions that go into the lowering files must be written as
part of producing a new OS port.
Most of them derive from OS design decisions about how the OS
interacts with \MD abstractions.

Other lowering definitions arise from required
machine-\emph{specific} logic: for example, MIPS and RISC-V both
have a ``global pointer'' register, which is used (in some cases)
to optimize access to global data.
When applicable, it needs to be initialized to a fixed (but
linker-chosen) value in key places.
This requires a suitable set of abstractions to use in specifications.
The lowering files for MIPS and RISC-V supply the needed
definitions; for other machines the definitions are empty.
Note that while the set of possible machine-specific concepts
requiring hooks in \MI specifications or code is infinite, and any new
machine that appears can produce novel ones that require more hooks to
be added (or even require OS \MI code to be altered to accommodate the
new ideas), it should be stressed that
this is not a new problem, in the sense that it occurs regardless
of what porting technology the OS uses.

\begin{figure}
\noindent\begin{minipage}[t]{.49\textwidth}
\begin{lstlisting}[language=casp, title=Listing A. MIPS lowering file\label{fig:stack-lowering-mips}, frame=r]
lowering disp_defs {
  let pc_reg: register = r5
  let disp_reg: register = r6
  let disp_area_reg: register = r4
  let DISP_MAX: int = 268
  def get_crit_ptr base: word->word=
    base b+ 0x00000058 }
lowering may_use_flags { }
lowering disp_scratch {modify: r2}
\end{lstlisting}
\end{minipage}\hfill
\begin{minipage}[t]{.49\textwidth}
\begin{lstlisting}[language=casp, title=Listing B. ARMv7 lowering file\label{fig:stack-lowering-arm}]
lowering disp_defs {
  let pc_reg: register = r14
  let disp_reg: register = r2
  let disp_area_reg: register = r1
  let DISP_MAX: int = 270
  def get_crit_ptr base: word->word=
    base b+ 0x00000058 }
lowering may_use_flags {modify:N Z C V}
lowering disp_scratch { }
\end{lstlisting}
\end{minipage}
\caption{Lowering files for \texttt{disp\_check}, for use with the \ale
specification in \autoref{fig:stack-alewife}.}
\label{fig:stack-lowering}
\end{figure}

\autoref{fig:stack-lowering} shows the MIPS and ARMv7 lowering files used to specialize our
example \ale specification in \autoref{fig:stack-alewife}.
Each defines the same three modules: \verb|disp_defs|, \verb|may_use_flags|,
and \verb|disp_scratch|.

The \verb|disp_defs| module supplies values for registers, a constant, and
a function for lookup in the \verb|dispatcher| structure.
Lines 2--4 supply concrete registers for
\verb|pc_reg|, \verb|disp_reg|,
and \verb|disp_area_reg|.
Line 5 declares the concrete value for \verb|DISP_MAX|, which
gives the dispatcher structure size; it is instantiated with $268$ on MIPS 
and $270$ for ARMv7.
Lines 6--7 define the abstract function \verb|get_crit_ptr|,
which computes the address of a specific field in the structure.
The offset is the same for both MIPS and ARMv7, because both
machines have 32-bit words and pointers, but is
different for 64-bit machines.


The \verb|may_use_flags| and \verb|disp_scratch| lowering definitions
relax frame conditions.
Different machines can require different numbers of scratch
registers to accomplish the same task.
Furthermore, on machines that have a flags word, the
specification must include explicit permission to modify the flags.
The \textbf{\texttt{modify}} \emph{frame condition}, as seen in 
\autoref{fig:stack-lowering}, indicates additional machine state 
(processor flags and scratch registers) that
any specification using this lowering module may modify.
The MIPS implementation of \verb|disp_scratch| makes register
\verb|r2| modifiable,
so it can be used as a scratch register (line 9 in \autoref{fig:stack-lowering}A),
while the \verb|may_use_flags| module is empty because there are no
processor flags on MIPS.
ARM needs no scratch register, but \verb|may_use_flags| 
makes four control registers \verb|N|, \verb|Z|, \verb|C| and \verb|V|
(the basic processor flags) modifiable (line 8 in
\autoref{fig:stack-lowering}B).


We use the lowering definitions for two further
purposes: first, to define \MD context structures (such as trap
frames) and predicates about them, and second, to allocate registers
in procedures.

Currently we write all the lowering definitions by hand, including
the identity and number of scratch registers.
However,
we expect other tools in \aquarium can provide some of these
definitions automatically.
In particular, register allocation (both placing values in registers
as they flow from block to block, and choosing scratch registers) is
a well-studied problem.
Extracting offsets and sizes of structures for use in assembly code is
also easily done; almost every OS already has a tool for this\footnote{
For example: \url{https://anonhg.NetBSD.org/src/file/tip/usr.bin/genassym/}
}, and
adjusting the output format to make it available to \aquarium is
trivial.
We have left integrating as-needed use of scratch registers
into the synthesis mechanism as future work.

In Section \ref{sec:validation}, we distinguish lowering
definitions that we expect to be automatically generated (those
defining context structures and those doing register allocation,
including scratch registers) from the other, essential, lowering
definitions that must be written by hand.
The context structure lowerings in particular are large (since they
include multiple assertions about each register involved, and there
are a lot of registers), but 
only the essential lowering definitions represent porting effort.
In fact, we already have a tool in the broader \aquarium ecosystem
that generates context structures and the necessary predicates about
them \cite{dholland-thesis}.

Regardless of whether they are automatically generated or manually written,
for a given OS and architecture, some lowering definitions are used
repeatedly in different procedures.
For example, the module \verb|may_use_flags| (line 8 in \autoref{fig:stack-lowering}B) 
is used in every block that might involve conditional execution, 
because on ARM and other machines with a flags word, 
it must be possible to set the flags so conditional
executions can test them.

The declarations in a lowering module are specifically permitted to
contain memory regions: some specifications require memory
regions on some machines and not others.
For example, the blocks that load the ``global pointer'' register on
MIPS and RISC-V require a memory region (with an assembler label) that
defines the address to be loaded; this is not present on other
machines. 
We do not, however, permit adding new machine instructions in lowering
files.

\begin{figure}
\noindent\begin{minipage}[t]{.49\textwidth}
\begin{lstlisting}[language=casp, title=Listing A. MIPS specification\label{fig:stack-caspspec-mips}, frame=r]
letstate DispMem:
  32 bit 268 len 32 ref memory
frame: modify: r2
let crit_pc: 32 bit =
  [DispMem, 0] b+ 0x00000058
let crit: bool =
    *r5 b< fetch(crit_pc, 32)
pre: ( *r6 == [DispMem, 0] ) &&
    ( *r0 == 0x00000000 )
post: (if crit then *r4 = 0x00000001
    else *r4 = 0x00000000) &&
    ( *r0 == 0x00000000 )
\end{lstlisting}
\end{minipage}\hfill
\begin{minipage}[t]{.49\textwidth}
\begin{lstlisting}[language=casp, title=Listing B. ARMv7 specification\label{fig:stack-caspspec-arm}]
letstate DispMem:
  32 bit 270 len 32 ref memory
frame: modify: N Z C V
let crit_pc: 32 bit =
  [DispMem, 0] b+ 0x00000058
let crit: bool =
    *r14 b< fetch(crit_pc, 32)
pre: *r2 == [DispMem, 0]
post: if crit then *r1 = 0x00000001
    else *r1 = 0x00000000
\end{lstlisting}
\end{minipage}
\caption{Generated \MD specifications for \texttt{disp\_check}.}
\label{fig:stack-caspspec}
\end{figure}

\subsection*{Machine-Dependent Specifications}
With the \casp machine descriptions (\circled{B} in \autoref{fig:overview}) 
and \lowering files (\circled{C} in \autoref{fig:overview}), 
the \lowertool (\circled{D} in \autoref{fig:overview}) generates
\MD specifications (\circled{E} in \autoref{fig:overview}) for \verb|disp_check| in MIPS and ARMv7, 
shown in \autoref{fig:stack-caspspec}.
In \autoref{fig:stack-caspspec}A, 
the \MD specification also includes the MIPS invariant in the pre- and
postcondition, ensuring that \verb|r0| is always zero (line 9 and 12 in \autoref{fig:stack-caspspec}A).
We include more details about the \lowering procedures in Appendix \ref{sec:ale_exec}.



\section{Synthesis with the \casp Synthesis Engine}\label{sec:impl}


The \casp \synth (\circled{F} in \autoref{fig:overview}) operates in
two modes.
In synthesis mode, it takes as input
a \casp machine description (\circled{B} in \autoref{fig:overview}) and
a \MD specification generated by the \lowertool
(\circled{E} in \autoref{fig:overview})
and synthesizes and verifies assembly programs (\circled{G} in \autoref{fig:overview}).
In verify-only mode, it takes as input
a sequence of operations, a \MD specification, and a \casp machine
description and it verifies that the operations satisfy the specification.
We use this latter mode extensively to debug both specifications and
machine descriptions.

\autoref{fig:stack-caspprog} shows the MIPS and ARMv7 sequences synthesized
for the \verb|disp_check| example.
\autoref{fig:stack-caspprog}A uses \verb|LW| to
load the contents of the memory location \verb|[DispMem,88]|
into scratch register \verb|r2|
and then uses \verb|SLTU| to compare and set \verb|r4| based
on the result of the comparison.
By contrast, \autoref{fig:stack-caspprog}B
uses \verb|LDR_imm| to load the value stored in \verb|[DispMem,88]|
into \verb|r0|, then uses \verb|CMP_reg| to
compare it with \verb|r14|.
Finally, it uses \verb|MOV_imm| conditionally twice to set \verb|r0| to
either \verb|0| or \verb|1|.
Implicit frame conditions prevent the synthesized code from modifying other
general-purpose registers or other fields of the control registers.
It is possible that there exist multiple sequences that satisfy the pre- and postconditions;
our \synth takes the first specification-satisfying sequence of
operations it finds (working from smaller to larger programs) as the
synthesized result, and it may produce different results in multiple runs.
If the specification is incomplete, the synthesized results can be functionally different.
This is common in assembly, because, in general, it does not matter what gets left behind in scratch registers.
After verification, our \synth extracts an operation sequence to a
syntactically valid assembly program, Figures \ref{fig:stack-caspprog}C and D.

\begin{figure}
\noindent\begin{minipage}[t]{.49\textwidth}
\begin{lstlisting}[language=casp, title=Listing A. MIPS operation sequence\label{fig:stack-caspprog-mips}]
  (LW r2 r6 0x0058)
  (SLTU r4 r5 r2)
\end{lstlisting}
\end{minipage}\hfill
\begin{minipage}[t]{.49\textwidth}
\begin{lstlisting}[language=casp, title=Listing B. ARMv7 operation sequence\label{fig:stack-caspprog-arm}, frame=l]
  (LDR_imm r0 r2 0x058 0b1110)
  (CMP_reg r14 r0 0b1110)
  (MOV_imm r0 0x0 0x01 0b0011)
  (MOV_imm r0 0x0 0x00 0b0010)
\end{lstlisting}
\end{minipage}
\noindent\begin{minipage}[t]{.49\textwidth}
\begin{lstlisting}[language=alewife, title=Listing C. MIPS assembly program\label{fig:stack-caspasm-mips}]
  lw $2, 88($6)
  sltu $4, $5, $2
\end{lstlisting}
\end{minipage}\hfill
\begin{minipage}[t]{.49\textwidth}
\begin{lstlisting}[language=alewife, title=Listing D. ARMv7 assembly program\label{fig:stack-caspasm-arm}, frame=l]
  ldr r0, [r2, #88]
  cmp lr, r0
  movlo r0, #0x00000001
  movhs r0, #0x00000000
\end{lstlisting}
\end{minipage}
\caption{Synthesized operation sequences and the verified assembly programs for \texttt{disp\_check}.}
\label{fig:stack-caspprog}
\end{figure}

Our \synth uses symbolic execution to compile a program to a
representation in
first-order logic, allowing the use
of a satisfiability-modulo-theories (SMT) solver to perform the synthesis
itself.
It uses several 
solvers: Boolector \cite{boolector}, Z3 \cite{Z3}, and Yices \cite{yices}.
Our representation uses the quan\-ti\-fier-free logic of finite bitvectors
\texttt{QF\_BV} for efficient SMT solving.
We developed several novel optimization techniques to improve the
scalability of program synthesis, especially synthesis for assembly language.
In the rest of this section,
we describe the main components of our \synth and several
optimizations critical for scaling (\autoref{subsec:opts}).

\subsection{Syntax-Guided Synthesis}

Our \synth implements syntax-guided synthesis for assembly programs.
We model assembly programs as sequences of instructions, each of
which has a name and zero or more operands (arguments).
As in standard syntax-guided synthesis~\cite{6679385, sketching-structures},
we use a \textit{symbolic program} composed of symbolic instructions to represent
sets of possible programs.
Symbolic programs are parameterized by \textit{control variables} that represent
the choice of operation (i.e., the opcode) and arguments.

Given a symbolic program and a specification,
the goal of synthesis is to select a correct program by choosing an appropriate
value for each control variable.
We first use symbolic execution (\autoref{subsec:symbolicexec}) to
generate correctness conditions,
then use Counterexample-Guided Inductive Synthesis (CEGIS) to find a
satisfying assignment of the control variables.


The CEGIS loop iterates \emph{guess} and \emph{verify} phases until it either
obtains a correct program or fails.
In the guess phase, the solver tries to find an assignment to the control
variables, producing a concrete program candidate.
In the verify phase, the solver tries to find an initial state, called
a \emph{counterexample}, that satisfies the
precondition but causes the candidate to falsify the postcondition.
If it succeeds in finding such a counterexample, we attempt another guess; otherwise, the candidate is a correct
program.

\subsection{Symbolic Execution}
\label{subsec:symbolicexec}

Symbolic execution for program synthesis explores every execution path to produce a
summary of a program's semantics.
Since evaluation in \casp always
terminates, we do not need any special techniques to approximate loop semantics.

We implemented a symbolic execution engine to efficiently execute assembly
programs described in \casp.
Like both Sketch~\cite{solarLezama08} and Rosette~\cite{rosette}, we represent
symbolic values with a DAG and take advantage of this representation to evaluate
operations concretely whenever possible.
We incorporate simplification techniques from both Sketch and Rosette for
efficiency.
From Sketch, we incorporate backwards value-set analysis, enabling us to
aggressively simplify pre- and postconditions.
From Rosette, we incorporate symbolic reflection and structural merging,
enabling us to represent higher-level structures (such as \casp values and
machine states) separately from SMT constraints, while curtailing the path
explosions associated with symbolic pointers.

Although we built our \synth as a standalone system,
we believe that we could have extended Rosette, which would
have allowed us to take advantage of its symbolic execution engine.
For instance, it should be possible to incorporate the simplification
techniques from Sketch into Rosette
either by extending Rosette's internal symbolic value simplifier or by using
Rosette's symbolic reflection interface.
Rosette's default type-directed merge alone does not automatically handle the
path explosions associated with symbolic pointers, since the merging behavior we
desire depends on the pointer's memory region.
However, it should be possible to implement
our desired merging behavior using Rosette's symbolic reflection interface.
Ultimately we chose to implement our own engine, because we found it easier to debug. In retrospect, our optimizations would have been more generally useful had we implemented them as Rosette extensions.

\paragraph{Symbolic pointers}
\casp's semantics of bitvectors and pointers requires special treatment for
efficient symbolic execution.
Symbolic pointers generate two primary problems for symbolic execution: path
explosion on reading memory and path explosion due to irregular
representation.
Careful rewriting is often necessary to prevent combinatorial blowup in
symbolic execution; we developed a set of rewriting rules that work well
for \casp machine descriptions.

\paragraph{Accessing memory at a symbolic offset}
Suppose $p$ is a symbolic pointer $(\ttmemid, N)$ where $\ttmemid$ is a memory region and $N$
is a symbolic offset.
To represent the result of reading memory at $p$, we elected to use an
if-then-else (ITE) of the values in $\ttmemid$ at the possible values of $N$, similarly
to \textsc{Angr}~\cite{angr}.
Likewise, when writing to memory at a symbolic offset, we update the memory at
each possible offset to an appropriately-guarded ITE of the old and new values.

We considered using uninterpreted functions (UIFs) to represent memory, similar
to Serval \cite{serval}.
However, in early experimentation, we found that UIFs produced
significant slowdown in candidate guessing compared to the ITE-based approach.
Although UIFs productively accelerate verification for Serval, our \synth spends
more than 90\% of its run time in candidate guessing.
Furthermore, most of our use cases have fairly small memory regions.
In combination with our pointer merging approach, this means that
the extra merging generated by the ITE-based approach does not significantly
affect the performance of symbolic execution.

\paragraph{Path explosion on reading memory}
Accessing memory with a symbolic pointer can lead to path explosion.
Taming this path explosion requires structural merging for pointer values.
However, the merge rule must be selected carefully, as we illustrate with the
following example.
To merge $m$ concrete pointers $q_i$ for $i \in \{1, \dots, m\}$, a direct
approach would be to merge the regions and offsets, producing a pointer $q$ with
a symbolic region and a symbolic offset.
But reading memory at $q$ may result in path explosion: since we could end up
reading $m$ offsets from each of $m$ distinct regions, the resulting ITE could
contain up to $m^2$ distinct branches.
Moreover, all but $m$ of these branches are infeasible.

To avoid this inefficiency, our \synth merges only the offsets of pointers that
point into the same region.
For example, we rewrite the symbolic value \texttt{(ite g (R1, x) (R1, y))}
(representing the symbolic pointer $(R1, x)$ if \texttt{g} is true and symbolic
pointer $(R1, y)$ otherwise)
as \texttt{(R1, (ite g x y))}, because the symbolic pointers reference the same
memory region.
However, we will not rewrite the symbolic value \texttt{(ite g (R1, x) (R2,
y))}.
If $q$ is produced by this merge strategy, reading from $q$ can produce at
most $m$ branches, one for each of the original concrete $q_i$.

\paragraph{Path explosion due to irregular representation}

Representing symbolic values of type $\bvaltyp{\ttC}$ requires care to avoid
what \citet{sympro} refer to as \textit{irregular representation}.
A symbolic $\bvaltyp{\ttC}$ value may be a bitvector, a pointer, or a complex
ITE that depends on guards or path conditions introduced by state merging.
When this structure has a ``regular'' (canonical) representation, structural
merging can efficiently generate compact results.
If the representations are not forced to be regular, they will often diverge
structurally and execution results will grow combinatorially, producing path
explosion.

For example, consider the result of executing a symbolic \texttt{ADD}
instruction such as \texttt{ADD R1 ? R2} (where \texttt{?} is a symbolic
register) on an initial machine state $M$.
Register \texttt{R2} in the final machine state contains a symbolic value with
up to one branch per possible register.
The symbolic values in the initial machine state can themselves be ITEs of
bitvectors and pointers, so the result of executing this \texttt{ADD} is often a
nested ITE structure.
Without careful canonicalization, as the number of symbolic instructions grows,
the ITEs become deep, leading to path explosion when the resulting value is
later used.

Avoiding such blowups is critical for performance.
We selected structural merging rules to ensure that symbolic $\bvaltyp{\ttC}$
values have a regular representation.
Our \synth rewrites symbolic values that contain both bitvector and pointer
branches so that bitvector values are in a separate branch from pointer values,
in addition to using the pointer merging approach described above.
It is necessary to apply this canonicalization recursively to flatten nested
ITEs.
Besides mitigating path explosion, this also simplifies the path conditions that
result from accessing memory at an address that may not be a pointer.

\subsection{Synthesis Optimizations}
\label{subsec:opts}

Our \synth employs several synthesis optimizations:
read/write constraints, state gating, dependency constraints, and
rule-based decomposition.
Read/write constraints, dependency constraints, and rule-based decomposition
are generally applicable techniques, although our specific rules
are mostly assembly language specific.
Read/write and dependency constraints are particularly powerful
in assembly language synthesis but likely apply to other imperative contexts.
Applying these optimizations produces an average speedup of \EVALNUM{$21\times$}
on our use cases, achieving \EVALNUM{$1380\times$} in one case.
We further evaluate these optimizations in \autoref{sec:optperf}.

\subsubsection{Read/Write Constraints}
\label{subsec:rdwt}

We add constraints to restrict the registers that an operation is allowed to
read, based on the specification and preceding operations.
Specifically, an operation may read only those registers either mentioned by the
specification (including frame conditions) or written by preceding operations.
We call these \textit{available} registers.
For each operation, we add constraints to ensure that any registers it reads are contained in the available register set.

We note that in some machines, operations may read some registers
(e.g.,~control registers) implicitly.
We annotate these registers with the \textbf{\texttt{control}} keyword (as shown
in \autoref{fig:casp-all}) and include them in the set of available registers
for each operation.

Intuitively, registers not in the available register set contain arbitrary
values, since they are neither constrained by the specification nor defined by
preceding operations.
As such, we expect that they should not be read in
candidate programs.

\subsubsection{State Gating}
\label{subsec:gating}
State gating removes registers and operations that are not needed to produce a
correct implementation for a \MD specification.
If a register is not mentioned in a precondition, postcondition, or frame
condition of a specification, it contains an arbitrary value that is not
restricted by the specification, and we consider it irrelevant.
A register not mentioned in the postcondition or in a frame condition may not be
modified by a correct implementation; registers that are additionally not
mentioned in the precondition also do not contain information that needs to be
read.

We cannot delete all unnecessary registers immediately, because assembly
instructions often access registers implicitly; for example, many instructions
in 32-bit ARM have conditional versions that access flag bits in the control register.
We analyze the machine description to find the registers that operations can access
together with relevant registers and add these to the set of registers that
should be retained.
We forbid read/write access to all other registers by removing them from the
machine description.
We additionally remove operations that require an access to an
irrelevant register, either explicitly or when all possible values for
an operand are irrelevant.

\subsubsection{Dependency Constraints}
\label{subsec:dep}

We analyze the specification and machine description to discover dependency
constraints to further prune the search space.
Where a specification ensures that the final value of a
location is uniquely determined by the initial state, we determine which
values in the initial state affect it.
We then require that guessed programs exhibit these dependencies, i.e.,~the
computation of the final value does indeed depend on the relevant parts of the
initial state.

We find locations $\ell$ that are uniquely determined by the initial state by querying
whether there exists an initial state for which the postcondition holds for two
distinct values in $\ell$:
\begin{align*}
\exists \sigma, \sigma', \sigma''\ .\
\text{pre}(\sigma)
\wedge \text{post}(\sigma, \sigma')
\wedge \text{post}(\sigma, \sigma'')
\wedge \sigma'[\ell] \neq \sigma''[\ell]
\end{align*}
If this query is UNSAT, then for any initial state, the final value in the
location is uniquely determined.

For such a location $\ell$, we then perform a series of SMT queries to identify
on which universally quantified variables in the precondition the final value
of $\ell$ depends.
Specifically, for each variable $x$ in the precondition, we ask whether there
exists an initial state $\sigma$ and distinct values $v_1$ and $v_2$ such that
$\sigma[x \mapsto v_1]$ and $\sigma[x \mapsto v_2]$
produce two different correct values for location $\ell$:
\begin{align*}
\exists \sigma, \sigma', \sigma'', v_1, v_2\ .\ 
&\text{pre}(\sigma[x \mapsto v_1])
\wedge \text{pre}(\sigma[x \mapsto v_2]) \\
\wedge\ &\text{post}(\sigma[x \mapsto v_1], \sigma')
\wedge \text{post}(\sigma[x \mapsto v_2], \sigma'') \\
\wedge\ &\sigma'[\ell] \neq \sigma''[\ell]
\end{align*}
If this query is SAT, then $\ell$ depends on $x$.

We add constraints to ensure that guessed programs exhibit these dependencies;
we analyze the symbolic program to compute an assertion on control
variables that ensures that variable $x$ is used in the computation of location
$\ell$.\footnote{The assertion is a necessary but not sufficient condition that
$x$ is used in the computation of $\ell$, i.e., our program analysis
overapproximates when $x$ might be used.}
Intuitively, these assertions are effective in accelerating guessing, because
they enable the solver to quickly reject programs that
do not exhibit the appropriate dependencies
and therefore cannot possibly be correct.

\subsubsection{Rule-Based Decomposition}\label{subsec:dec}

\begin{algorithm}
\begin{algorithmic}[1]
\Procedure{Synthesize}{$s$}
  \Comment{Given starting specification $s$.}
  \State $S \gets \{ \text{Hole}($s$, 1) \}$
  \Comment{Begin with hole for one instruction.}
  \While{$S \neq \emptyset \wedge \text{not timed out}$}
    \State $T \gets \textsc{TakeBest}(S)$
    \State $h \gets \textsc{FindFirstHole}(S)$
    \If{$h = \bot$}
      \Comment{No holes remain; $S$ is a complete program.}
      \State \textbf{return} $S$
    \EndIf
    \State $\text{Hole}(g, n) \gets h$
    \Comment{Otherwise, $h$ is the current hole with spec $g$ and size $n$.}
    \State $P \gets \textsc{Sygus}(g, n)$
    \If{$P \neq \bot$}
      \Comment{Syntax-guided synthesis succeeded at $n$ instructions.}
      \State $S \gets S \cup \{ T[h \mapsto \text{Prog}(P)] \}$
      \Comment{Replace current hole with found program $P$.}
      \State \textbf{continue}
    \EndIf
    \For{$(g_1, \dots, g_m) \gets \textsc{Decompose}(g)$}
      \Comment{For each subgoal decomposition, create}
      \State $N \gets \text{Seq}(\text{Hole}(g_1, 1), \dots,
                                 \text{Hole}(g_m, 1))$
      \Comment{a sequence of holes, one per subgoal.}
      \State $S \gets S \cup \{ T[h \mapsto N] \}$
      \Comment{Replace current hole with sequence.}
    \EndFor
    \State $S \gets S \cup \{ T[h \mapsto \text{Hole}(\sigma, n+1)] \}$
    \Comment{Increase current hole's search size.}
  \EndWhile
\EndProcedure
\end{algorithmic}
\caption{Rule-based decomposition procedure.}
\label{alg:decomposition}
\end{algorithm}

Our \synth augments syntax-guided synthesis with a heuristic rule-based
decomposition algorithm to achieve performance and scalability.
Rule-based decomposition uses a set of \MI \textit{decomposition rules} that
decompose a specification into specifications for smaller subprograms
(``subgoals'').
Our algorithm combines decomposition and syntax-guided synthesis in an iterative
scheme, alternately using rules to decompose subgoals and using pure syntax-guided
synthesis to discharge subgoals.
The procedure is described in \autoref{alg:decomposition}.

Rule-based decomposition allows us to improve performance over
syntax-guided synthesis.
Syntax-guided synthesis scales poorly with increasing program length;
decomposing the specification into subgoals for smaller programs reduces the
size of each syntax-guided synthesis invocation, significantly accelerating
synthesis overall.
Meanwhile, using syntax-guided synthesis as a final code generation stage
allows our rules to remain \MI.

\paragraph{Search Algorithm}
During synthesis, our \synth performs a search over possible trees of rule
applications to the input specification instance, maintaining a set of trees
encountered ($S$ in \autoref{alg:decomposition}).
Each tree $T$ corresponds to one possible subdivision of the specification into
subgoals; each subgoal has an instruction count $n$, which is an upper bound on
the number of instructions that will be used to satisfy the subgoal.
In each iteration, the \synth selects a tree from this set and a subgoal from
the tree, and invokes syntax-guided synthesis up to $n$ instructions to try to
discharge the subgoal (line 10).
The \synth selects the next tree to attempt according to a cost heuristic
(\textsc{TakeBest}); our heuristic scores a tree by the lengths of its remaining
subgoals, weighting each subgoal exponentially in its length.

If syntax-guided synthesis fails on a subgoal, our \synth attempts to decompose
the subgoal by applying each rule (\textsc{Decompose}, line 15).
Each new tree obtained is added to the set (line 17).
The \synth also retains the original tree, increasing the subgoal instruction
count to $n+1$ for later exploration (line 19).
If syntax-guided synthesis succeeds, the \synth stores the
partially discharged tree and continues (lines 11--13).
If the \synth discharges a tree's last remaining subgoal, it combines the
results to produce a program that is correct with respect to the original goal
(line 7).

\paragraph{Rule Selection}
Our implementation of rule-based decomposition includes five rules. All of our
rules are specific instances of sequential composition; they generate subgoals
such that $\text{post}(g_i) = \text{pre}(g_{i + 1})$. As such, subprograms that
satisfy the generated subgoals may be composed by sequence to produce a correct
program for the input goal.
\begin{itemize}
  \item \textsc{LoadMem}. Given goal $g$, if a value $v$ exists in memory that
        is not in a register, generate two subgoals $(g_1, g_2)$. Subgoal $g_1$ has
        $\text{pre}(g_1) = \text{pre}(g)$ and a postcondition that requires $v$ in a
        scratch register. Subgoal $g_2$ has $\text{post}(g_2) = \text{post}(g)$.
  \item \textsc{LoadLbl}. Given goal $g$, if a label $\ell$ exists whose value
        (an address)
        is not in a register, generate two subgoals $(g_1, g_2)$. Subgoal $g_1$ has
        $\text{pre}(g_1) = \text{pre}(g)$ and a postcondition that requires
        $\ell$ in a
        scratch register. Subgoal $g_2$ has $\text{post}(g_2) = \text{post}(g)$.
  \item \textsc{SetReg}. Given goal $g$, if $\text{post}(g)$ contains a register
        $r$ whose value is uniquely determined and is wrong, then generate two
        subgoals $(g_1, g_2)$. Subgoal $g_1$ has $\text{pre}(g_1) =
        \text{pre}(g)$ and a postcondition demanding the correct value
        in register $r$. Subgoal $g_2$ has $\text{post}(g_2) = \text{post}(g)$.
  \item \textsc{SetMem} is akin to \textsc{SetReg}, but for memory locations.
  \item \textsc{FindPtr}. When a subgoal $g$ was generated by \textsc{LoadMem} or
        \textsc{SetMem}, generate subgoals $(g_1, g_2)$. Subgoal $g_1$ has
        $\text{pre}(g_1) = \text{pre}(g)$ and a postcondition demanding that a
        scratch register contain a pointer to the relevant address. Subgoal
        $g_2$ has $\text{post}(g_2) = \text{post}(g)$.
\end{itemize}

We base our rule selection on observations of common properties of OS assembly code and assembly languages.
First, OS assembly code typically performs simple calculations and otherwise
serves only to move or set state.
Often, these moves and sets are performed by independent program fragments.
Second, many assembly languages support arithmetic over only a subset of their
registers; a significant fraction of assembly programs
is dedicated to moving values for calculations then storing the
results.
We use rules that generate (1)~independent subgoals for
programs to satisfy independent pieces of state and (2)~subgoals that move data
to and from general-purpose registers.
Due to our choice of rules, our algorithm does not speed up synthesis for
programs that mainly perform arithmetic.

Our rules are nondeterministic: more than one rule may be applicable to a
specification, and one rule may apply several ways.
For instance, \textsc{SetReg} may apply to several registers in a specification.
Our rules are not, in general, invertible~\cite{LiangM2009}. Applying an invertible rule to a derivable goal will produce only derivable subgoals. Applying non-invertible rules may require backtracking.
Our rules are also incomplete: it is possible that no rule applies to a given
specification instance, in which case our \synth regresses to normal syntax-guided
synthesis.
However, our rules are sound: if synthesis succeeds on the generated subgoals, then the resulting program is correct against
the input specification instance.

\paragraph{Separation Logic}
Internally, we use a separation-logic-like representation of the specification,
inspired by \textsc{SuSLik}'s~\cite{suslik} use of separation logic to represent
synthesis goals for imperative heap-manipulating programs.
This allows our rules to structurally decompose specifications by simple
pattern-matching.

\section{Validation}
\label{sec:validation}

We validate the expressiveness of \casp by successfully modeling four real machine
architectures---32-bit MIPS, 32-bit ARM, 32-bit RISC-V and x86\_64---and
using them to synthesize runnable code.

From two preexisting OSes---%
Barrelfish \cite{barrelfish} and
OS/161, a BSD-like teaching OS \cite{os161}---%
we took eight complete machine-dependent
procedures (written either entirely in assembly language
or in a mixture of C and assembly language)
and split them into semantically meaningful steps.
We implement each step with a block of code, which is either
a synthesis use case or out of scope for synthesis.
We evaluate \CASENUM{50} blocks in all:
\CASENUM{35} are synthesis use cases,
and \CASENUM{15} are out of scope.
%
Of the latter, \CASENUM{10} are single call or return instructions, \CASENUM{two} are
single system call instructions, and the other \CASENUM{three} are
machine-\emph{specific} exception operations.
Section \ref{sec:usecases} discusses the use cases in more detail.

We validate the expressiveness of \ale by successfully writing
\MI specifications for each of the blocks not explicitly out of scope.


We validate our implementation as follows:
\begin{itemize}
\item
We use the \lowertool to lower each \MI specification for each
architecture, generating a total of \CASENUM{140} \MD
specifications for synthesis.
\item
We use our \synth to verify hand-written assembly code for all of these \MD
specifications.
\item
We use our \synth to synthesize and verify assembly code for all \CASENUM{140}
\MD specifications.
\end{itemize}

In the remainder of this section,
we explain the targeted machine-dependent procedures
and the four real machine architectures modeled in Sections \ref{sec:usecases} and \ref{sec:machdesp}, respectively.
Sections \ref{sec:lowerinfo} and \ref{sec:proginfo} report the information about the corresponding files,
while Sections \ref{sec:performance} and \ref{sec:optperf} discuss verification and synthesis performance.


\subsection{Use cases}\label{sec:usecases}

We selected two Barrelfish and six OS/161 complete machine-dependent
procedures, most of which were originally written entirely in assembly.
Each procedure comes from a specific port of the original OS and
performs some specific,
recognizable \MI task in the appropriate \MD manner for its target
machine.
We selected the procedures
without significantly refactoring the original OS kernel.
The procedures include interrupt and exception handling, system
call handling, user-level startup code, context switches, and other
small OS components.
We chose the procedures to be representative of tasks that
machine-dependent kernel code and user-level OS code must perform in
assembly language:
changing processor state and manipulating registers in ways not
expressible in C.
There are two chief categories missing: one is kernel startup code,
which often contains large amounts of machine-\emph{specific} code,
and the other is memory management code, such as cache control and TLB
management.
We successfully synthesized some memory management code blocks, but do
not include them in the paper: it is difficult to find an example in
existing code bases that is both manageable and explainable without
extensive refactoring.
We believe the chosen procedures are largely representative of the
synthesizable assembly programs in our scope.

Note that refactoring existing OSes to have clean \MI interfaces for \MD
code is not a goal for our system or our validation.
For example, the native Barrelfish trap handling code for ARM and x86\_64 is
quite different, because each port was handwritten separately.
Based on advice from the Barrelfish team, our use cases are based on
the ARM version.
This means that while we generate x86\_64 code, that code cannot be
easily incorporated into the
existing x86\_64 Barrelfish port, which needs a different
specification.
However, in all cases the code for the original architecture (ARM for
Barrelfish, MIPS for OS/161) is suitable for incorporation into the
existing OS implementation.
(We include handwritten code for out-of-scope blocks.)
Reorganizing the OSes with a uniform structure across all ports,
and factoring out code that could be \MI but is gratuitously different
in each port,
so that specifications are not needlessly \MD and synthesis results
can be seamlessly plugged in for all
machines, is a large project in its own right and the subject of
ongoing research.

Our goal of synthesizing assembly procedures to implement portions
of \MD OS functionality is difficult due to the challenge of scaling
assembly code synthesis, while
balancing such scalability with the expressiveness and complexity
of the languages that describe
OS functionality and processor semantics.
We adopted a decomposition model, where 
each procedure is explicitly (and manually) divided into a sequence of 
\MI steps each performing a specific recognizable task,
because most of these procedures are too large to synthesize all at once.
(One could imagine letting our \synth run for longer periods of time
to synthesize complete procedures; unfortunately this inherently
requires exponentially longer periods of time.)
A given block might be empty for a given machine, but no block is empty
for all machines.

As we synthesize only loop-free blocks,
we use program labels and external branches for all control flow other than
forward branches within a block.
%

As mentioned earlier, some blocks are out of scope for synthesis
or best handled in other ways,
so we generate blocks in different ways. 
Some of these other generators, such as a context-switch compiler and
a tool for automatically composing blocks, are described in other
work~\cite{dholland-thesis}.
In the following section, we describe each of the blocks amenable to
synthesis.

\subsection*{Barrelfish Use Cases}

\subsubsection*{\textbf{swi-handler (SWI)}}

This is the machine-level system call trap from the ARMv7 implementation.
``SWI'' stands for ``software interrupt'', which is ARM terminology for a
system call trap.
From our machine-independent specifications, we generate comparable
system call handlers for MIPS, RISC-V, and x86\_64.
There are 13 synthesis blocks,
of which two are empty on ARM but were necessary for 
one or more of the other machines.

\begin{enumerate}[topsep=0pt, parsep=0pt]
\item spill:
save some registers to make room to work.

\item check-kern:
check whether the trap came from kernel or user mode,
and branch
to the corresponding code.

\item load-got:
get the global offset table base address on machines that need it
in a register. 
(The global offset table is used for indirect addressing in
position-independent code.)

\item get-disp:
load the address of the current dispatcher, the Barrelfish process
structure equivalent.

\item check-disabled:
check whether the dispatcher was in ``disabled'' mode,
and if so, branch to the subsequent set-disabled block.

\item findpc: load the exception program counter into a general
purpose register on machines where this is needed, which does not
include ARM.

\item check-low: check the exception program counter against the lower
bound of a critical region,
and branch to the set-enabled
block if below it.

\item check-high: check the upper bound and branch to the set-disabled
block if below it.

\item set-enabled:
load the address of the ``enabled'' register save area in the
dispatcher.

\item set-disabled:
load the address of the ``disabled'' register save area.



\item reshuffle:
after saving registers, move values around as needed to avoid
clobbering them in later blocks.
Empty on ARM.


\item initstack:
load the stack pointer to point to the kernel stack.

\item moveargs:
set up the arguments for the C system call handler.



\end{enumerate}

\subsubsection*{\textbf{cpu-start (CS)}}
This is the entry point for application-core kernels\footnote{
A Barrelfish system uses a single core running a special
kernel to initialize the system. The remaining cores then boot with
a kernel for cores running applications, including the system Monitor
process.}
from the ARMv7 port.
It consists of two parts. The first part is considered as a synthesis block:
\begin{enumerate}[topsep=0pt, parsep=0pt]
\item store-got:
on machines that need the global offset table in a register,
fetch its base address from the argument structure, and store it where
the trap handler expects to find it later.
\end{enumerate}
The second part is the same as the initstack block from the swi-handler example (SWI-12),
so we reuse the same \ale specification for this part and 
do not count it as an extra synthesis block for the evaluation.

%

\subsection*{OS/161 Use Cases}
\label{sec:usecase:os161}

\subsubsection*{\textbf{setjmp (SJ)}}
This is the C standard library function \texttt{setjmp}.
It stores callee-saved registers and the stack pointer and returns 0.
It is composed of two blocks:
\begin{enumerate}[topsep=0pt, parsep=0pt]
\item saveregs: save registers.
\item retval: set the return value.
\end{enumerate}

\subsubsection*{\textbf{longjmp (LJ)}}
This is the C standard library function \texttt{longjmp}.
It reloads the registers saved by \texttt{setjmp} and and arranges
to return a caller-supplied value from the \texttt{setjmp} call.
It is compsed of two blocks:
\begin{enumerate}[topsep=0pt, parsep=0pt]
\item loadregs: load registers.
\item retval: set the return value, adjusting it if needed.
\end{enumerate}

\subsubsection*{\textbf{crt0 (CRT)}}
This is the startup code (\texttt{crt0}) for user-level programs.
It contains seven synthesis blocks:
\begin{enumerate}[topsep=0pt, parsep=0pt]
\item initstack: ensure that the stack pointer is aligned correctly
according to the function call ABI.
As OS/161 is a teaching operating system,
it does not assume that the initial stack pointer provided by
the kernel is correctly aligned.
\item savevals: save the values of \texttt{argv} and
\texttt{environ}, passed in registers from the kernel, into the
private global variables \texttt{libc} uses to hold them.
\item initregs: perform machine-dependent register initializations,
such as the global pointer register used by some RISC machines.
\item mainargplace: place the arguments for calling \texttt{main}.
\item exitsave: save the return value from \texttt{main} in a
callee-save register.
\item exitargplace: place the return value from \texttt{main} as the
argument to \texttt{exit}.
\item loop: unconditional branch back to repeat the call to \texttt{exit} in
case it returns.
(While \texttt{exit} normally does not return, \texttt{exit} is also student
code; the loop is a precaution to avoid bizarre behavior
when it does not work as intended.)
\end{enumerate}


\subsubsection*{\textbf{syscallstub (SYS)}}
This is a user-level system call stub.
It has three synthesis blocks:
\begin{enumerate}[topsep=0pt, parsep=0pt]
\item loadnum: load the system call number into the appropriate
register.
Empty on machines where the call number is placed in the system call
instruction.
\item jump:
unconditional jump from per-system-call code to shared code.
\item seterrno:
set the C-level return value and \texttt{errno} from the kernel-level return value.
\end{enumerate}



\subsubsection*{\textbf{cpu-irqoff (IRQ)}}
This procedure turns interrupts off for the
current processor.
It is just one block:
\begin{enumerate}[topsep=0pt, parsep=0pt]
	\item irqoff: turn off the interrupts.
\end{enumerate}
(Turning interrupts on is effectively the same and omitted for
brevity.
A related function for idling the processor consists mostly of
interrupt state manipulation, along with a special instruction for
idling that we do not attempt to synthesize, and is also omitted.)

\subsubsection*{\textbf{thread-switch (TS)}}
This is a complete kernel-level
thread switch; it saves the state of the current thread 
and restores the state of the next thread to run.
It consists of six blocks:
\begin{enumerate}[topsep=0pt, parsep=0pt]
\item entry: make space on the stack to save registers.
\item saveregs: write the old thread's registers to the stack.
\item savestack: save the old thread's stack to its thread structure.
\item loadstack: load the new thread's stack from its thread structure.
\item loadregs: read the new thread's registers from the stack.
\item cleanup: clean up the stack for return.
\end{enumerate}

\subsection{Machine Descriptions}
\label{sec:machdesp}

\subsubsection*{ARM}
The 32-bit ARM model has three-operand instructions, 16 registers, and a status register.
We do not include Thumb instructions, the 16-bit variant of the ARM ISA.
Most ARM instructions are all conditionally executed based on the
flags in the status register.
We model the status register as multiple smaller register fields, as
direct access to the entire status
register is rare in
ARM code.
In ARM, many immediate values are represented as an 8-bit field rotated right
by an even 4-bit shift amount between 0 and 30.
We model this directly and thus emit only valid immediates.
This is necessary, because the ARM assembler will not emit extra
instructions to encode impossible immediates.
We do not include multiple data transfer instructions such as LDM/STM, 
since they can always be replaced by multiple single load/store and arithmetic instructions.

\subsubsection*{MIPS}
Our MIPS model has three-operand instructions and 32 general-purpose registers.
Most kernel-mode phenomena are handled by control registers, 
which are composed of fixed-size fields;
we model each field separately in \casp
and concatenate the fields together when accessed as a register.
In MIPS,
general-purpose register 0 is always zero, and writes to it are
discarded.
We handle this by treating register 0 as special in every instruction
and adding a machine invariant to the \casp machine description (see
\autoref{fig:casp-all}A).
We handle the branch delay slots associated with jumps by using the 
assembler mode that hides branch delay slots.

\subsubsection*{RISC-V}
Our 32-bit RISC-V model has three-operand instructions, 32 general-purpose registers,
and a supervisor-mode status register similar to that of MIPS. 
We again model the status register by separating it into fixed-size
fields, concatenating them together when accessing it as a whole.
As in MIPS, general-purpose register 0 is always zero, and we
encode this as a machine invariant.

\subsubsection*{x86\_64}
We modeled x86\_64 with AT\&T syntax~\cite{x86atnt}. Our x86\_64 model has
two-operand instructions, 16 registers, and a flags register
interrogated by certain branch instructions.  We model the flags
register as separate 1-bit registers.

\begin{table*}[!htbp]
\resizebox{0.8\textwidth}{!}{%
   \begin{tabular}{c|cccccc}
     \hline
             &       &            & General   & Special & Flags & Control \\
Architecture & Lines & Operations & Registers & Registers & Registers & Registers \\
\hhline{=======}
ARM     & 933 & 34 & 15 &   & 4 & 9 (141 total bits) \\
MIPS    & 495 & 37 & 32 & 2 &   & 9 (156 total bits) \\
RISC-V  & 618 & 37 & 32 &   &   & 31 (193 total bits) \\
x86\_64 & 567 & 56 & 16 &   & 4 & 3 (53 total bits) \\ \hline
   \end{tabular}
}
\caption{
Size of the \casp machine descriptions for each architecture.
Control registers are modeled with separate registers for individual
fields, so both the counts and the total addressable space vary
widely depending on the layout.
}
\label{tab:descs}
\end{table*}

\subsubsection*{Completeness}
We implemented 34 operations in ARM, 37 in MIPS, 37 in RISC-V
and 56 in x86\_64.
All models include assembly operations for arithmetic, bitwise logic and
shifts, comparison, moves, memory accesses, conditional branching, and supervisor operations.
This is a small fraction of the total instructions available on each
machine (especially x86\_64) but covers all the basic operations.
Each machine model also includes all the general-purpose registers,
the basic processor flags on machines that have them, and a selection
of control registers as needed for our use cases.
Table \ref{tab:descs} shows line, operation, and register counts for
each machine model.

\subsubsection*{Shortened Machine Descriptions}


Because including the full complement of general-purpose registers
makes even the simplest synthesis problems extremely slow in the
absence of state gating, for each
machine we prepared a single alternate description with most of the
general-purpose registers commented out, retaining only 6--8 of the
most commonly used ones.
We use this shorter description for all the use cases it can support, that is,
those that do not require the registers it comments out.
The blocks that use the full descriptions are:
\begin{itemize}
\item SJ-1, LJ-1, TS-2 and TS-5, whose context operations 
inherently require the full register set.
\item CRT-5 and CRT-6 on MIPS and RISC-V.
\item All the SWI blocks on MIPS except SWI-2 and SWI-12.
\end{itemize}

This trimming reflects what one might readily do by hand in the
absence of state gating.
(It is, indeed, what we did by hand before we had state gating.)
Note that this change is strictly adverse to our evaluation, in that
it makes our baseline considerably more performant.

\subsection{\ale Specifications and Lowering}
\label{sec:lowerinfo}

We were able to express all \CASENUM{35} \MI specifications in \ale.
We supplied lowerings for all $\CASENUM{35} \times 4 = 140$ combinations of \MI specifications and
machine descriptions.

\autoref{tab:lengths1} shows the number of lines of code for each \MI
specification.%
\footnote{All lines-of-code counts exclude blank lines, comments, and standalone close-braces.}
\autoref{tab:lengths2} presents the \textit{total} length of the lowering files for each
OS and architecture.
As discussed in \autoref{sec:lowering}, we divide these into
two categories:
``Manual'' lowering includes OS-and-machine-specific definitions that
need to be hand-written.
``Automatable'' lowering includes definitions that can reasonably be
generated: register allocation, data structure offsets, and
specification fragments for context (register-save) structures.
We cannot ascribe a specific length to the lowerings associated with each \MI specification,
since there is significant sharing and reuse.
Centralizing \MD definitions in a small set of files heavily reduces manual effort;
we consider it an important element of
the expressiveness of \ale and \casp.

Our use cases include \LINENUM{$434$} lines of \ale and
\LINENUM{$67$} lines of manual lowering, for a total of
\LINENUM{$501$} lines of specification.
This in turn allows us to synthesize \LINENUM{$383$} lines of code.
Because the incremental cost of a new port is just a new set of
lowering files, averaged over the four existing ports the ratio of
(manual) lowering files to write to lines of code synthesized is
\LINENUM{$0.175$}.

This is perhaps somewhat optimistic, because the generation of
automatable lowerings requires some specification of its own, some of
which may be machine-dependent and need to be provided for each port.
A full discussion of those considerations is beyond the scope of this
paper, but an absolute upper bound may be given by including the
\LINENUM{$588$} lines of automatable lowerings in the porting effort,
which makes the ratio of lines of lowering files to lines synthesized
\LINENUM{$1.71$}.

Including the \MI specifications the overall effort for four ports is
\LINENUM{$1.31$} lines of specification per line synthesized.
This number is perhaps not as appealing, but we consider the
incremental porting cost the primary concern;
see \autoref{sec:effort} for more context.

\subsection{Hand-written assembly programs}
\label{sec:proginfo}

To help with validation, we obtained 
hand-written assembly for all use cases on all machines 
(we either wrote them manually ourselves or took them from
existing implementations)%
\footnote{
OS/161 ships with MIPS code; we wrote versions for ARM, RISC-V, and x86\_64.
Barrelfish ships with ARM and x86\_64 versions; we wrote MIPS and RISC-V manually,
and restructured the x86\_64 version as necessary to match the structure of the ARM implementation.
}

\autoref{tab:lengths1} reports the size of the assembly program
synthesized for each \MI specification under all four architectures.
Typically, synthesized assembly programs differ significantly in
implementation from our hand-written versions; sometimes, the synthesized
programs are shorter.
For cases where synthesis timed out after 1800 seconds,
the reported number in
parentheses is the number of
assembly instructions in our hand-written implementation.

\begin{table*}[!htbp]
\resizebox{\textwidth}{!}{
\begin{tabular}{c|c|cccc|cccc|cccc}
\hline
\multirow{2}{*}{Spec} &
\multirow{2}{*}{\begin{tabular}[|c|]{@{}c@{}}\ale\\ (lines)\end{tabular}} &
\multicolumn{4}{c|}{Verification Time (ms)} &
\multicolumn{4}{|c|}{Synthesis Time (s)} & 
\multicolumn{4}{|c}{Assembly (lines)} \\
& & ARM & MIPS & RISC-V & x86-64 & ARM & MIPS & RISC-V & x86-64 & ARM & MIPS & RISC-V & x86-64 \\ \hhline{==============}
SWI-1 & 5 & 60 & 130 & 56 & 43 & 5.1 & 0.65 & 45 & 1.0 & 5 & 2 & 7 & 3 \\
SWI-2 & 5 & 41 & 120 & 43 & 34 & 23 & 23 & 9.9 & 0.70 & 3 & 4 & 3 & 2 \\
SWI-3 & 7 & 41 & 53 & 41 & 32 & 0.15 & 0.36 & 0.036 & 0.031 & 1 & 2 & 0 & 0 \\
SWI-4 & 20 & 63 & 150 & 63 & 55 & 15 & 4.6 & 3.8 & 4.1 & 4 & 4 & 4 & 4 \\
SWI-5 & 13 & 45 & 130 & 44 & 37 & 38 & 2.4 & 1.5 & 1.7 & 3 & 2 & 2 & 2 \\
SWI-6 & 6 & 40 & 120 & 43 & 33 & 0.047 & 0.24 & 0.37 & 0.043 & 0 & 1 & 1 & 0 \\
SWI-7 & 12 & 58 & 130 & 47 & 42 & 5.4 & 1.0 & 0.81 & 2.0 & 3 & 2 & 2 & 3 \\
SWI-8 & 12 & 48 & 140 & 48 & 42 & 15 & 1.0 & 0.80 & 2.1 & 3 & 2 & 2 & 3 \\
SWI-9 & 12 & 42 & 120 & 42 & 33 & 1.6 & 1.3 & 1.1 & 0.67 & 3 & 3 & 3 & 3 \\
SWI-10 & 10 & 41 & 120 & 42 & 34 & 1.1 & 0.62 & 0.54 & 0.45 & 2 & 2 & 2 & 2 \\
SWI-11 & 5 & 40 & 120 & 42 & 32 & 0.047 & 0.19 & 0.051 & 0.043 & 0 & 1 & 0 & 0 \\
SWI-12 & 13 & 43 & 55 & 42 & 33 & 1.7 & 0.48 & 1.7 & 0.45 & 2 & 2 & 3 & 2 \\
SWI-13 & 21 & 41 & 120 & 42 & 33 & 0.077 & 1.9 & 0.082 & 0.073 & 0 & 3 & 0 & 0 \\
CS-1 & 15 & 52 & 68 & 49 & 38 & 1.2 & 1.6 & 0.050 & 0.037 & 2 & 3 & 0 & 0 \\
SJ-1 & 10 & 160 & 240 & 290 & 120 & 43 & 33 & 59 & 9.5 & 11 & 11 & 14 & 8 \\
SJ-2 & 5 & 41 & 45 & 41 & 32 & 0.28 & 0.080 & 0.083 & 0.064 & 1 & 1 & 1 & 1 \\
LJ-1 & 10 & 140 & 220 & 250 & 110 & 54 & 32 & 58 & 9.6 & 11 & 11 & 14 & 8 \\
LJ-2 & 13 & 41 & 49 & 44 & 34 & 9.4 & 1.2 & 2.8 & 14 & 2 & 2 & 2 & 3 \\
CRT-1 & 24 & 46 & 65 & 49 & 55 & 8.4 & 19 & 0.69 & 0.44 & 2 & 3 & 1 & 1 \\
CRT-2 & 14 & 48 & 57 & 52 & 40 & 4.2 & 2.6 & 2.4 & 0.63 & 4 & 2 & 4 & 2 \\
CRT-3 & 15 & 40 & 50 & 45 & 33 & 0.062 & 0.20 & 0.20 & 0.074 & 0 & 1 & 1 & 0 \\
CRT-4 & 30 & 43 & 65 & 47 & 35 & 0.066 & 0.089 & 0.077 & 0.079 & 0 & 0 & 0 & 0 \\
CRT-5 & 7 & 41 & 130 & 120 & 33 & 0.22 & 0.23 & 0.19 & 0.20 & 1 & 1 & 1 & 1 \\
CRT-6 & 13 & 42 & 150 & 130 & 35 & 0.45 & 0.44 & 0.25 & 0.24 & 1 & 1 & 1 & 1 \\
CRT-7 & 4 & 40 & 45 & 41 & 32 & 0.042 & 0.053 & 0.057 & 0.060 & 1 & 1 & 1 & 1 \\
SYS-1 & 9 & 39 & 45 & 41 & 32 & 0.035 & 0.079 & 0.082 & 0.064 & 0 & 1 & 1 & 1 \\
SYS-2 & 4 & 40 & 45 & 41 & 32 & 0.043 & 0.054 & 0.057 & 0.060 & 1 & 1 & 1 & 1 \\
SYS-3 & 26 & 46 & 51 & 47 & 34 & 370 & 160 & 170 & 12 & 6 & 6 & 7 & 3 \\
IRQ-1 & 4 & 40 & 46 & 42 & 32 & 0.057 & 1.6 & 0.44 & 0.047 & 1 & 4 & 1 & 1 \\
TS-1 & 14 & 41 & 47 & 41 & 33 & 0.17 & 0.11 & 0.11 & 0.16 & 1 & 1 & 1 & 1 \\
TS-2 & 19 & 120 & 190 & 220 & 110 & 24 & 24 & 47 & 9.1 & 9 & 10 & 13 & 8 \\
TS-3 & 12 & 45 & 51 & 47 & 38 & 0.17 & 0.12 & 0.12 & 0.092 & 1 & 1 & 1 & 1 \\
TS-4 & 12 & 46 & 51 & 47 & 38 & 0.17 & 0.12 & 0.12 & 0.094 & 1 & 1 & 1 & 1 \\
TS-5 & 19 & 120 & 210 & 240 & 110 & 30 & 25 & 46 & 9.4 & 9 & 10 & 13 & 8 \\
TS-6 & 14 & 40 & 46 & 41 & 33 & 0.30 & 0.11 & 0.11 & 0.16 & 1 & 1 & 1 & 1 \\ \hline
\end{tabular}
}

\caption{Performance of our \synth on our \MI specifications. For each, we report:
(1)~the size of the \MI specification,
(2)~verification time for a hand-written implementation (in
milliseconds, averaged across 10 trials), and 
(3)~synthesis time (in seconds,
averaged across 5 trials) with all optimizations enabled.
We also report
(4)~the size of the synthesized assembly program.
}
\label{tab:lengths1}
\end{table*}

\begin{table*}[!htbp]
\resizebox{0.8\textwidth}{!}{%
  \begin{tabular}{c|cccc|cccc}
    \hline
    & \multicolumn{4}{|c|}{Barrelfish} & \multicolumn{4}{|c}{OS/161}    \\ \hhline{=========}
    Architecture             & ARM  & MIPS & RISC-V & x86\_64 & ARM & MIPS & RISC-V & x86\_64 \\ \hline
    Manual Lowering (LOC)    	   & 11 & 10 & 6 & 7 & 8 & 8 & 8 & 9 \\ \hline
    Automatable Lowering (LOC) & 39 & 33 & 39 & 38 & 106 & 110 & 137 & 86 \\ \hline
  \end{tabular}
}
\caption{Length of the lowering files, divided into two categories, for each OS
and architecture.}
\label{tab:lengths2}
\end{table*}

\subsection{Performance}
\label{sec:performance}

We measured the performance of our \synth for verification and
synthesis.
Our measurement platform is an Intel Core i7-7700K clocked at 4.2 GHz with 64 GB
RAM; our tests use only one of the four cores.
We ran five trials for synthesis and ten trials for verification, 
and each trial varied the solver RNG seed.
We also randomized the variable names used for solver communication by
prepending a random five-character alphabetical string, which we find produces
significant variance in solver performance.

\subsubsection*{Verification}
\autoref{tab:lengths1} reports the verification performance of our \synth on all use
cases.
We verified all machine-independent \ale specifications against 
the hand-written assembly programs, which are taken from existing implementations,
to support the validity of our specifications.
In most cases, the \synth verifies both our hand-written programs and
the synthesized implementations in milliseconds.
We are able to verify the whole of thread-switch (TS), our longest hand-written
program at 26 instructions, in less than \EVALNUM{one second}.
During synthesis, our \synth uses verification as part of the CEGIS loop; because guessing
is much slower than verification, this occupies only a small fraction of the total synthesis time,
often $< 1\%$.

\begin{figure}[!b]
  \begin{subfigure}[t]{0.5\textwidth}
  \centering
  \includegraphics[width=\linewidth]{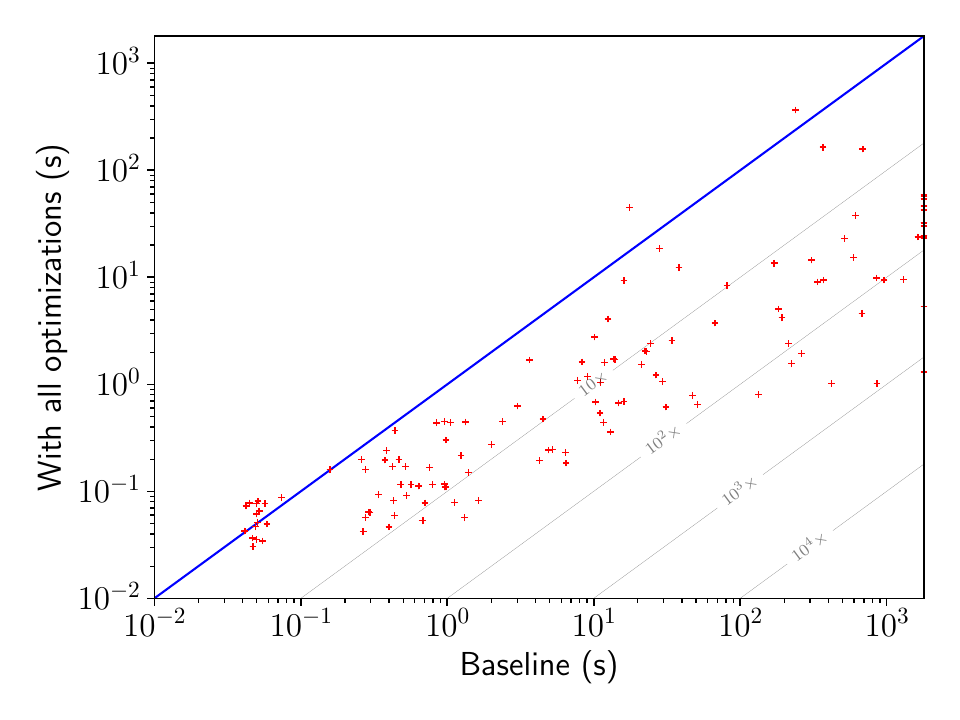}
  \caption{All optimizations vs.~baseline.}
  \label{fig:compopt}
  \end{subfigure}
  
  \begin{subfigure}[b]{0.5\textwidth}
  \centering
  \includegraphics[width=\linewidth]{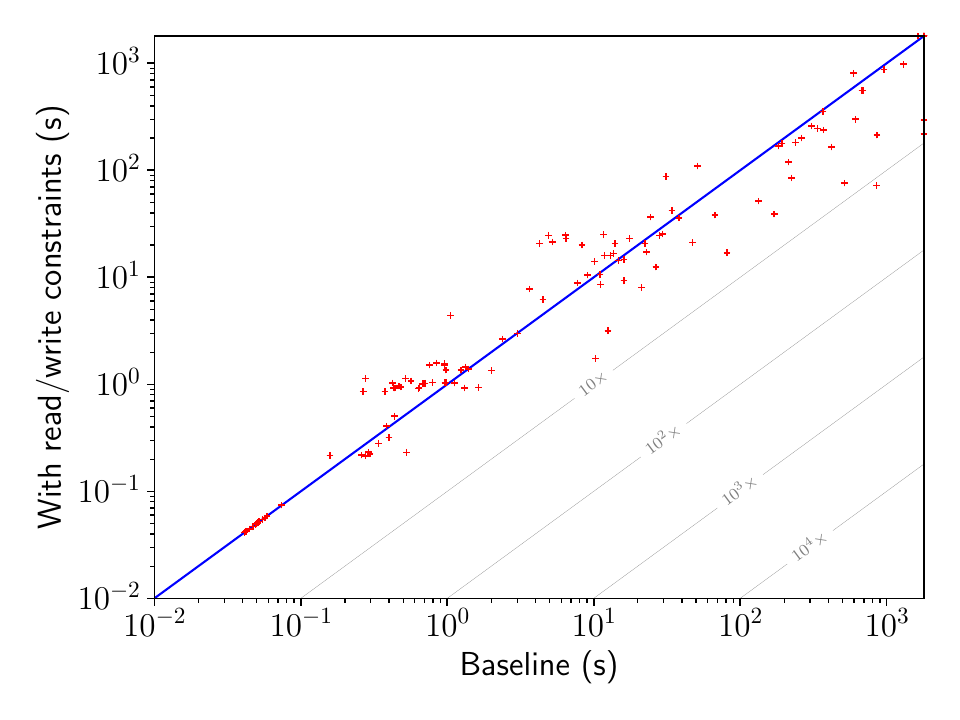}
  \caption{With and without read/write constraints.}
  \label{fig:comprdwt}
  \end{subfigure}
~
  \begin{subfigure}[b]{0.5\textwidth}
  \centering
  \includegraphics[width=\linewidth]{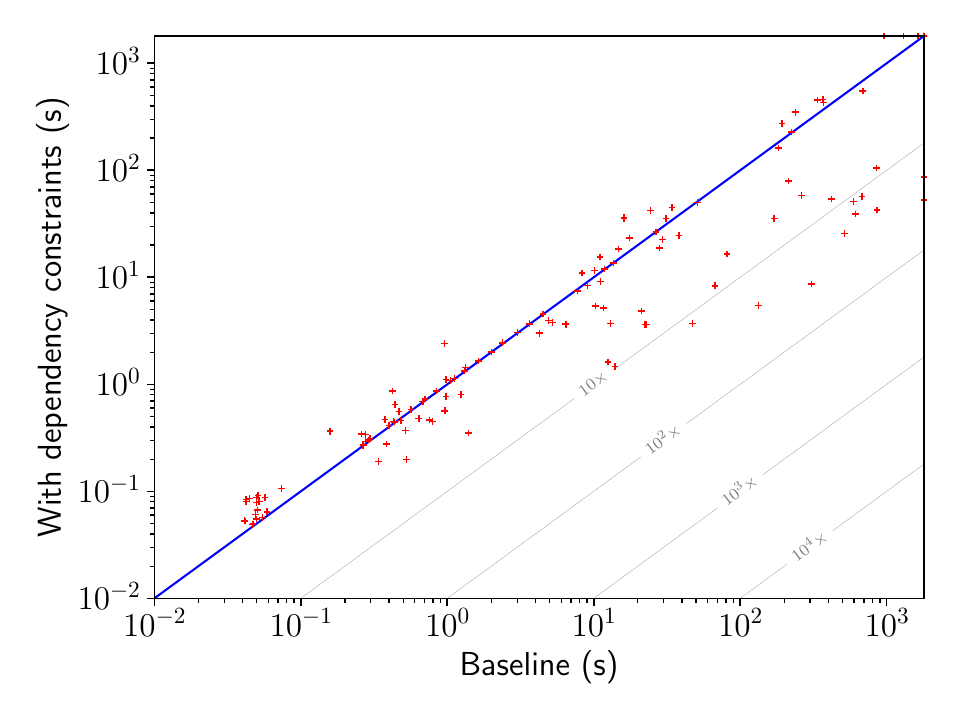}
  \caption{With and without dependency constraints.}
  \label{fig:compdep}
  \end{subfigure}
\caption{Effect of optimizations on synthesis runtimes.
\autoref{fig:compopt} shows all optimizations; the other plots show a
single optimization enabled, all against our baseline.
Each data point represents the runtime under both conditions for one \MI specification
and architecture, averaged over five trials (timeouts are counted as 1800
seconds).
The blue diagonal line represents equal time under both conditions, so that points
below/right of the diagonal line demonstrate better performance
with the optimization enabled.
Gray contours provide guidelines for visually estimating speedup factor.
The upper and right boundaries of the plot represent an 1800 second timeout.}
\end{figure}

\begin{figure}[ht]\ContinuedFloat
	
  \begin{subfigure}[t]{0.5\textwidth}
	\centering
	\includegraphics[width=\linewidth]{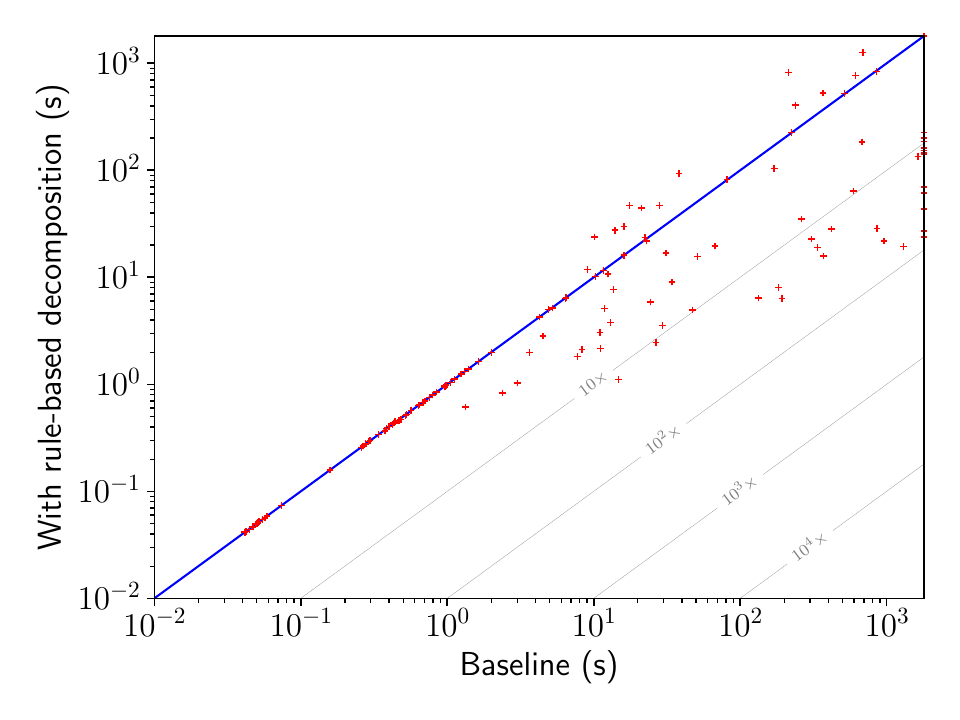}
	\caption{With and without rule-based decomposition.}
	\label{fig:compded}
  \end{subfigure}
~
  \begin{subfigure}[t]{0.5\textwidth} 
  \centering
  \includegraphics[width=\linewidth]{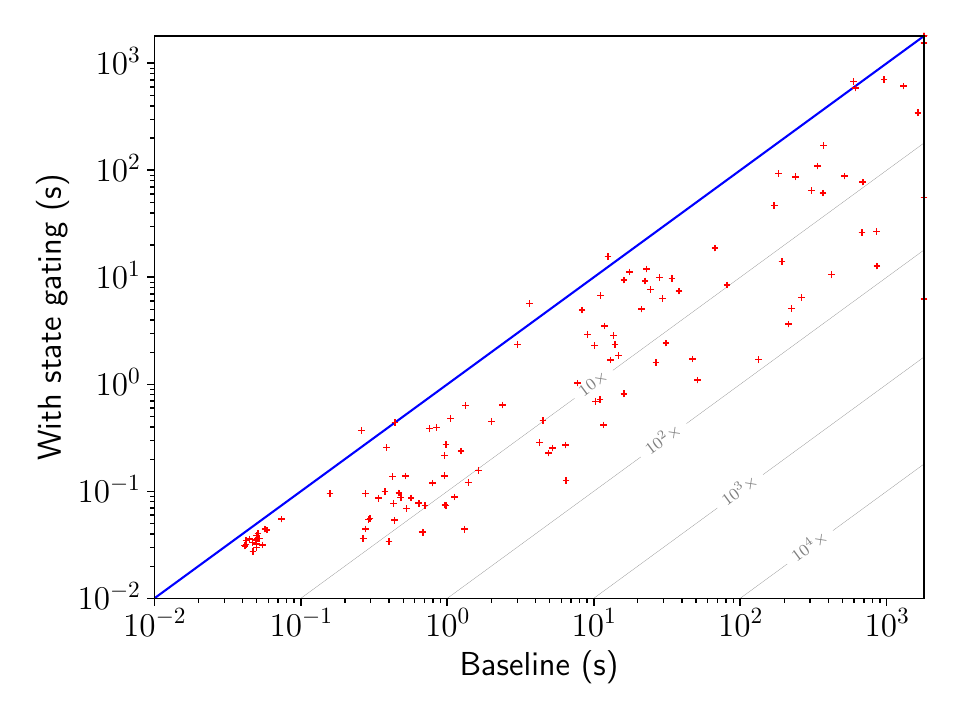}
  \caption{With and without state gating.}
  \label{fig:compgate}
  \end{subfigure}

\caption{Effect of optimizations on synthesis runtimes  (cont.).}
\label{fig:comp}
\end{figure}

\begin{figure}
\centering
\includegraphics[width=0.75\linewidth]{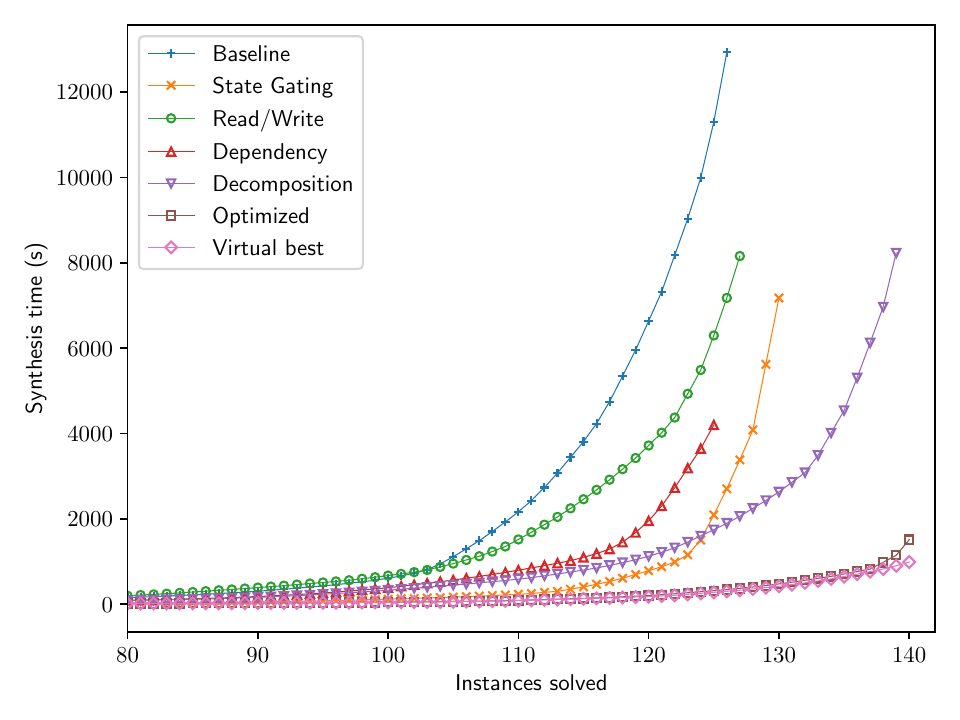}
\caption{Cactus plot comparing each optimization against baseline.
Each plot shows for all \MI specifications the average runtime across each of
five trials that do not time out.
Machine-independent
specifications where all trials timed out are omitted.
The \textit{Optimized} curve shows the performance of our system with all
optimizations enabled together.
The \textit{Virtual best} curve simulates the performance of a system that takes
the minimum time observed for each \MI specification across all trials.}
\label{fig:cactus}
\end{figure}

\subsubsection*{Synthesis}\label{sssec:synthesis}
Synthesis in general is difficult, and 
synthesis of assembly is particularly hard.
The size of the space of possible programs is combinatorial in the number of
registers, memory locations, and immediate values, and
exponential in the number of instructions \cite{plos19}.
This especially affects the performance of syntax-guided synthesis.
As a first-order approximation, in our machine descriptions, the overall sizes
of the search spaces for programs of length $n$
are \EVALNUM{$2^{40.3n}$} in ARM, \EVALNUM{$2^{36.6n}$} in MIPS,
\EVALNUM{$2^{36.3n}$} in RISC-V, and
\EVALNUM{$2^{74.7n}$} in x86\_64.

\autoref{tab:lengths1} summarizes the synthesis performance of our \synth on all
\MI specifications with all optimizations enabled.
We ran the experiments with a half-hour timeout; no cases timed out.
Our synthesis times vary widely with architecture and \MI specification.
Examining these use cases more closely, we see that SWI-1 is a register spill,
requiring a series of stores followed by a subtraction that sets the stack
pointer appropriately: each instruction is relatively independent and no value
in the final state requires multiple arithmetic operations to calculate.
In contrast, SWI-5 loads a value from memory, then uses it to perform a
conditional branch; this requires that the solver reason about conditional
execution and arithmetic to produce a correct program.
This leads to entirely different synthesis times under each of the four
architectures, especially for ARM and RISC-V.
Note that our \synth discharges all two-instruction \MI specifications in less than
ten seconds.


\subsection{Optimizations}
\label{sec:optperf}

Our optimizations are critical for performance and scaling, especially as the
size of the program to synthesize increases.
To measure the effect of each optimization, we compare a baseline of our \synth with no
optimizations enabled against the \synth with each single optimization enabled.
\autoref{fig:comp} and \autoref{fig:cactus} show the results.

Many of our optimizations are heuristic: while accelerating synthesis on the majority
of the use cases, they may sometimes cause synthesis to slow down.
In particular, read/write constraints and dependency analysis work by providing
additional constraints to the SMT solver.
Experimentally, we find that these help in most cases, especially as synthesis
time increases.


Read/write constraints exhibit a significant tradeoff.
They slow synthesis for small programs, but
for programs that take our baseline more than 100 seconds to
synthesize, read/write constraints produce a geometric mean speedup of
\EVALNUM{$1.6\times$}.
We observe a maximum speedup of \EVALNUM{$12\times$}, which occurs on a use case
that takes \EVALNUM{851 seconds} to run on average in the baseline condition.
In the worst case, read/write constraints produce a slowdown of
\EVALNUM{$5.0\times$}, which occurs on a use case that takes \EVALNUM{4.9
seconds} in the baseline condition.

Dependency analysis is often beneficial.
It provides additional constraints when values in the postcondition are
uniquely determined by the initial state.
Dependency analysis makes our \synth up to \EVALNUM{$35\times$} faster, which
occurs on a use case that takes \EVALNUM{307 seconds} to run on average in the baseline
condition.
Dependency analysis produces a geometric mean speedup of \EVALNUM{$1.4\times$}.
Dependency analysis induces an overhead typically less than one second, which is
significant only for small programs; the effect is visible as a small upward
nonlinearity at smaller times in \autoref{fig:compdep}.
In the worst case, dependency analysis makes our \synth~\EVALNUM{$2.5\times$}
slower, which occurs on a use case that takes \EVALNUM{0.96 seconds} in the
baseline condition.

Rule-based decomposition accelerates synthesis when its rules
apply and the use case requires a long program; it has sufficiently low
overhead that it does not impact performance when its rules do not apply.
Rule-based decomposition produces a geometric mean speedup of
\EVALNUM{$2.0\times$}.
In the best case, rule-based decomposition synthesizes programs that are not
accessible to the baseline, such as context switch code; counting timeouts as
1800 seconds, this is a speedup of \EVALNUM{$75\times$}.
Rule-based decomposition can sometimes slow down synthesis by spending time
exploring goal decompositions whose subgoals do not simplify the synthesis task;
for example, when a subgoal demands that an unnecessary value be loaded.
In the worst case, we observe a slowdown of \EVALNUM{$3.8\times$}.

State gating is effective, because it reduces the search space directly, limiting
the options available to the SMT solver, without ruling out possibly correct
programs.
State gating makes our \synth up to \EVALNUM{$290\times$} faster, with a geometric mean
speedup of \EVALNUM{$4.5\times$}.
In the worst case, we observe a slowdown of \EVALNUM{$1.6\times$} on a use case
requiring an average to \EVALNUM{3.6 seconds} to synthesize in the baseline condition.

Used together, our optimizations cut the average total time required to
synthesize all of our use cases by \EVALNUM{$25\times$} (from \EVALNUM{38130}
seconds to \EVALNUM{1511} seconds), with a geometric mean speedup of
\EVALNUM{$9.2\times$}.
One use case is accelerated \EVALNUM{$1380\times$}: while timing out in the baseline
condition, our optimizations allow us to synthesize it in just over one second.

Currently, our \synth is single-threaded; we included the \textit{virtual best}
curve in \autoref{fig:cactus} as a demonstration of the performance approachable
by a \textit{parallelized} implementation, which would run several different
\synth configurations and random seeds for a single synthesis problem, taking
the first result returned.

\subsection{Our Experience Writing Specifications}
While we have not conducted an extensive user study to determine how easily
architects and kernel designers can produce specifications, we have one
related user study and our own experience from which to draw conclusions.
In collaboration with colleagues in Human Computer Interaction (HCI),
a subset of our team developed an interactive tool that let users
give hints to the \synth~\cite{assuage}.
Although the study did not ask users to produce specifications, they
had to understand enough about the specifications and the architecture
to make suggestions about synthesis strategies, inclusion/exclusion of
instructions, and hypotheses about the structure of the code that
should be produced.
All users, from novice to expert were able to give hints that improved
the time to synthesize kernel components.

Our own experience writing specifications is more directly relevant to
the work presented here.
Recall that synthesis in \aquarium is designed to decouple expertise
requirements: computer architects write machine specifications and
kernel developers write kernel specifications.
The team responsible for writing all the specifications presented
in this paper included
one experienced kernel developer and two others with minimal kernel
development experience.
None of the team were computer architects.
Nonetheless, we found that it took approximately one day to write and
debug a \casp file for a new architecture, even when the architecture
was significantly different from others we had written (e.g., producing
the \casp file for MIPS after we had produced \casp files for only
x86-64 and ARM).

We found that writing the Alewife specifications took us about the same
amount of time it takes to manually produce a port of the machine dependent
parts of a system. However, once written, every
additional port can use these specifications.
In other words, using Aquarium is a time saver
as soon as you start on a second port.
Even our team member who had never written kernel code before found it
reasonably straight forward to produce specifications.
Further, we discovered that we could use existing ports to debug both
our \ale and \casp specifications.
If a \casp specification is incorrect, and we try to verify the code in
an existing port, the verification fails\footnote{This could also be due
to a bug in the existing port, but we did not encounter any such situations.}.
Similarly, if we write an incorrect kernel specification, verification of
the existing code also fails.
Overall, we found that Aquarium let us achieve our goal of transforming the
$N$ ports for $M$ operating systems problem into something that took time
proportional to $N + M$ rather than $N * M$, as is the case for traditional
porting.

\section{Discussion}
\label{sec:disc}


\subsection{\aquarium Ecosystem}
\label{sec:disc-aquarium}

The \aquarium code generation model is to decompose procedures into
blocks and generate the blocks one at a time, by synthesis or by other
techniques.
These blocks must then be recombined
into functions and files, connected to
other parts of the system via include files, and ultimately compiled
and linked using \verb.make. or a similar tool.
\aquarium handles this with separate tools that construct complete procedures
from assortments of code blocks \cite{dholland-thesis}.
We have omitted discussion of this \aquariumX of tools and languages from this paper,
which focuses exclusively on our program synthesis tools.

\subsection{Effort/Benefit Tradeoff}
\label{sec:effort}

As noted in \autoref{sec:lowerinfo}, the line counts of specification
and lowering relative to output code are perhaps larger than one might
like.
When the ratio of specification to output code exceeds $1$, it is
tempting to think that just writing the code directly might be
preferable.
Nonetheless, several factors suggest that synthesis is still promising.

First, verifying handwritten code still requires specifications (and
lowerings), so these need to be written regardless, and synthesis is
always strictly less work.

Second, it is important to interpret these numbers in the correct
context.
Our specification-to-code ratio is reasonably competitive with
other synthesis and verification approaches,
especially that of other work in assembly-language synthesis and verification.
We ask how much specification a user would need to write to specify
assembly for a new architecture.
JitSynth~\cite{jitsynth}, another assembly synthesis approach, does not report a
spec-to-code ratio, but does at least require the user to write the state
equivalence between the source and target register machines.
Vale~\cite{bond17} reports $2014$ lines of spec, $3213$ lines of implementation, and
$13558$ lines of proof for $6943$ generated lines of assembly (excluding lines of
specification, implementation, and proof associated with definitions of the ARM
and Intel semantics).
This corresponds to a spec-to-code ratio of about $0.75$, not counting lines of
proof; we are unsure how much additional code would be required to extend Vale
to another architecture.
The SeL4 proof files in the \texttt{spec/machine} directories contain $2600$
lines, while the assembly files in the \texttt{arch} directory contains $3380$
lines; trimming comments and newlines, these counts are about $2000$ and about
$1900$ lines respectively.
This corresponds to a spec-to-code ratio of about $1.05$, which we expect to
correspond to the ratio of specification needed to extend SeL4 to a new
architecture.
We think it is reasonable to claim that the
specification effort for a new port is at least comparable to that of
similar work.

Third, the cost of producing OS specifications is amortized across multiple
ports: the same specifications can be used for multiple architectures.
We could dilute this ratio to an arbitrarily low number by including more
architectures in our evaluation.
Moreover, having \MI OS specifications available provides additional benefits,
such as enabling verification and analyses to reduce bugs and vulnerabilities.
This benefit is also portable to new architectures, which is not true of
existing assembly verification approaches.
Given that low-level code is often extremely difficult to debug, this is
potentially a significant improvement.

Fourth, the synthesis approach reduces the level of mastery of
architectural details required of the porter.
A porter need only follow the machine manual and not necessarily learn it well
enough to write assembly code from scratch.
As an example, the authors of this paper had no prior experience writing RISC-V
assembly.
We found the control register access instructions complex; especially when first
starting with the architecture, we found it convenient to let our \synth
figure out how to use them.
The required level of knowledge of the ported OS is reduced as well.
Currently, porting an OS to a specific architecture using \aquarium requires the
porter to provide a lowering for each \ale specification.
Only the lowering file, which relates a particular machine description to the
OS specification, requires knowledge of both the machine and the OS, and the
lowering file is relatively small, offering considerable savings of time and effort.
Many of the blocks in our use cases require few or no
block-specific lowering definitions.
For many of the rest, the lowering definitions required are
straightforward and easily explained; the porter can provide them
without necessarily having to fully understand the blocks that use
them.
However, this does not apply to all the ported code; some code,
such as the swi-handler (SWI) example, still requires expert attention.
It is expected that with more ports, a
considerable fraction of the port code will be derivable from
specifications and comparatively simple lowering definitions,
especially as the rest of the \aquarium ecosystem comes on
line.

Note also that in some cases the need for porters to think about
particular pieces of code disappears entirely.
For example, the only lowering definitions specific to the OS/161
thread-switch (TS) example  are ``Data-Structure'' definitions
that can be
automatically generated.
Consequently with our approach, the thread switch code no longer
requires any attention from the porter: it simply happens, just as
all the kernel's \MI code simply happens when recompiled.

There are nonetheless cases where the cost/benefit ratio is an issue,
especially for blocks that demand additional functionality
or expressiveness.
One example of the latter is memory barrier instructions, discussed
further in \autoref{sec:limit}.




\subsection{Machine-Independence}
\label{sec:disc-mi}

It is tempting to assume that \ale specifications cannot ``really'' be
\MI and that all the important parts of each specification must
actually be in the lowering files.
This is, perhaps surprisingly, not true.
Operating systems already have a \MI model of the abstractions in
and functions of the \MD code, which
underlies the interface to that code.
Our \ale specifications elucidate and formalize this model.
They can also, it turns out, frequently extend it: the kernel's
interface to \MD code is necessarily function-level, but our
specifications are block-level.
Consequently any \MI structure in the steps that need to be taken,
which is frequently substantial, can be written directly into the \ale
specifications.
For example, in the thread-switch (TS) example, there are six
steps, and the nature of each and the order in which they appear is
necessarily the same on every machine, because at a suitable level of
abstraction, they must each do the same thing.

Furthermore, even for operations that cannot be further decomposed,
such as turning on interrupts, the specification for the operation is
the \MI specification for what the kernel expects to happen: some
piece of machine state, corresponding to the interrupt state, is
toggled on.
While the machine-level state and instructions that correspond to \MI
functionality often vary from machine to machine, higher-level OS code can and
will use this block without regard for the implementation details.
We can write a \MI specification in terms of an abstract
version of this state, then use the \lowertool to lower the
specification so it references the correct
control register field(s) for each target machine.
Then the \synth can synthesize correct code given the machine description.

%

An OS architect writing \ale specifications must understand the abstract model of
the underlying machine represented by the \MI-to-\MD interface.
In the case of Barrelfish, where the kernel is built as
position-independent code to facilitate the ``multikernel''
architecture,
this must take into account differences in the handling of
position-dependent code on different machines.
If porting to a new machine changes the behaviors observable from the \MI code,
then the \MI specifications may need to be strengthened.
This is a common problem in OS porting.
Every OS embeds a similar understanding in the structure of its \MD interfaces,
and altering it can require altering those interfaces and the OS's \MI code.

\subsection{Formal ISA Descriptions}

We expect machine descriptions to ultimately be supplied by architecture
vendors.
Architecture vendors already internally maintain extensively tested executable
formalizations of their architectures; for instance, ARM internally employs the
ARM Specification Language to describe processor semantics, and uses it to test
both Verilog and physical hardware.
Fully formalizing the semantics of an ISA is a challenging problem; despite
this, recent work in formalizing ARM~\cite{reid-fmcad16} and various other
architectures~\cite{armstrong-popl19} supports the assertion that
vendor-supplied ISA descriptions suffice to produce accurate formal descriptions
of ISA semantics.
Moreover, the RISC-V foundation has officially standardized an executable formal
specification of the RISC-V ISA semantics in Sail~\cite{riscv-spec-review}.
We expect that in the future, ISA semantics formalized by the architecture
vendor will become increasingly available and are here to stay.

We used a custom language for machine descriptions, \casp, for two reasons.
First, at the time of development, formalized ISA semantics were
less available than they are today.
Second, our synthesis proof-of-concept requires only a subset of the ISA
semantics, and as observed by Bourgeat et al.~\cite{bourgeat-riscv-spec}, many
formal methods projects use simplified ISA semantics.
\casp itself, as discussed, is not a complete ISA description language; it is
one point on a spectrum of languages with varying levels of expressiveness.


\subsubsection*{ISA instruction selection}
One critical issue is the level of completeness of the ISA description. A more
complete description could enable synthesis of more OS components; on the
other hand, including more instructions negatively impacts the synthesis
performance and the ease of writing machine descriptions.
Fortunately for the latter, expressiveness does not require the ISA description
to include most of the available instructions.
For example, the x86 architecture contains many instructions that are not
necessary for full expressiveness, such as the specialized vector operations in
the 64-bit version.
Excluding these reduced the work required for us to produce ISA descriptions,
although we expect that manufacturer-supplied ISA descriptions would be
relatively more complete.

Using only a sufficiently expressive ISA description additionally helps with
synthesis performance.
This gives a reasonably sized search space to allow the \synth to
create an initial runnable program.
Once we have a runnable program, we can apply superoptimization~\cite{stokeAsplos} or standard compiler
techniques~\cite{10.1145/349214.349233} to obtain a more efficient program that uses a larger machine model.

Our expressiveness decisions were driven by our use case requirements.
Consequently, as discussed in \autoref{sec:machdesp}, we constructed our machine
models to include at least the following general-purpose components:
\begin{itemize}
\item Basic arithmetic and logical instructions on register-sized values: addition, subtraction, signed
and unsigned comparisons, bitwise operations, shifts.
\item Conditional and unconditional branches. When architectures distinguish
between short and long jumps, short jumps suffice for intra-block branching.
\item Loads and stores for word-sized and byte-sized values, ensuring that we
can express any manipulation of an included memory region.
\item A mechanism to load the address of an assembler label.
\end{itemize}
We also include the full set of general-purpose registers for each machine and
any special-purpose registers used by ordinary code, such as the flags word.

Beyond these, we include a selection of pertinent control registers
and the instructions that manipulate them,
as this is a common OS operation.
In general, each block we synthesize needs access to conceptually similar pieces
of machine state in each architecture.

\subsection{Limitations}\label{sec:limit}

\subsubsection*{What we do not synthesize}

To strike the balance between expressiveness and complexity of the languages, 
there are important pieces of OS code that we do not synthesize.
We do not synthesize call and return instructions (nor do
we synthesize system call instructions).
Procedure calls are different from ordinary assembly instructions.
There are two ways one could imagine synthesis interacting with
procedure calls.
The first way is to let the specification explicitly indicate what procedure to call.
This is the approach we take; we generate the call and return
instructions with other \aquarium tools~\cite{dholland-thesis}.
The second way is to make the specification more expressive about procedure calls and
use synthesis to select a particular call from a set of choices.
This is a higher-level synthesis task, which is orthogonal to our goal of 
synthesizing assembly language.


Furthermore, there are small portions of OS code that are not merely
\MD, but are machine-specific, 
which cannot be expressed in a machine-independent way.
Well-known examples include the code to set up x86 segment tables and
SPARC register window trap handling.
A less well-known example that
affects our use cases is that ARM exception handlers must manipulate
exception-related processor modes that do not exist on other
architectures.

\subsubsection*{Scalability of assembly synthesis}

Furthermore, scalability poses a serious challenge.
Synthesis of assembly is particularly hard 
since the characteristics of assembly languages
cause a combinatorial explosion in the number of possible
instructions that must be considered \cite{plos19}.
This detrimentally affects the performance of syntax-guided synthesis.

This difficulty is a general feature of assembly languages.
Even though
our synthesis is parameterized on a \casp machine description, an approach
specialized for a single machine or a low-level intermediate language
must still address the fundamental problem of search space size.
Although our optimizations significantly reduce the
search space, it is still quite easy to write specifications for
which synthesis
always times out.

Currently, using pure inductive synthesis, we can generate up to five
(occasionally only four) instructions in reasonable time; with our
rule-based decomposition synthesis
technique, we can synthesize programs up to twelve instructions for certain
well-structured specifications.
Therefore, we expect to be able to handle most small pieces of assembly code;
besides the
assembly blocks demonstrated in our use cases, \ale and \casp can probably also
handle
frame push and pop operations, stack handling,
and simple bit manipulations without further extension.
Although this is sufficient for many \MD OS
components, some require significantly longer implementations.  For example,
the MIPS general exception handler in NetBSD 9 is about 100 instructions.
For these our methodology of decomposition into steps becomes critical.

Ultimately, while blocks of four or five instructions are small, they
are still useful, since
other tools in our ecosystem allow composing them into larger chunks.
For example, as discussed in Sections \ref{sec:usecase:os161} and
\ref{sec:disc-mi},
the thread-switch (TS) example is
composed of six steps.
Two of these we compile; the other four we synthesize.
Those four are all one instruction each on most or all
architectures.
Yet combined, these six blocks are a complete kernel-level thread
switch, similar to that found in any operating system kernel.
The setjmp (SJ) and longjmp (LJ) examples are similarly
decomposed; combining the pieces gives the complete implementations.
Other examples are a subject of ongoing research; this includes the
crt0 (CRT) use cases.

\subsubsection*{Concurrency and time}

\casp and \ale do not model concurrency or time.
Reasoning about time typically involves temporal logics
that are not SMT solvable.
This is likely a fatal obstacle for working with real-time operating
systems (especially hard real-time), but for conventional systems, \MD
code handles time only when interacting with timer hardware.
We have not addressed this issue in current work; in future
work we plan to investigate whether the temporal aspects of timer
control can be adequately abstracted away such that the low-level
specifications (and thus code synthesis) can be expressed in \ale.

We chose not to burden \casp with support for reasoning about concurrent
executions.
Our use cases do not require concurrency support.
This is representative of the majority of the \MD part of an OS:
most of it executes below the level where concurrent things happen.
(For example, the thread switch code in our use cases executes on a
single CPU, on CPU-private data, and with interrupts already disabled by
higher-level \MI code; it itself need not reason about other CPUs,
other threads running time-sliced on the same CPU, or about being
interrupted.)

However, this means we cannot model instructions related to concurrent
executions.
The \MI code in a typical OS is written in terms of a small set of
concurrency primitives.
Producing these is part of generating a new port, but one we decided
not to tackle in this work.
Currently, they are provided by the porter as prewritten assembly
blocks that can be composed with other blocks via other \aquarium
tools.
For a typical OS this means implementations of spinlocks, a small set
of atomic operations, and a small set of memory barriers.
Note that all of these are also small blocks of assembly code and
likely amenable to synthesis using other tools,
though the specifications are nontrivial \cite{memsynth} and
(particularly for memory barriers, which are always either zero or one
instruction) likely to significantly outweigh the
output code.

\subsubsection*{Performance}
Finally, we do not reason about the performance of synthesized code.
Although the \synth chooses the shortest sequence of instructions that
meets a specification, this may not correlate to performance.
We assume that optimization can be performed after synthesis and block
composition.
Furthermore, low-level OS code typically does not involve expensive computation.
We discuss the related problem of synthesizing the fastest possible assembly
program, or superoptimization~\cite{Superoptimizer}, in \autoref{sec:rw}.

\subsection{Other Approaches Considered}
We explored several different approaches to improve scalability of synthesis.
We briefly summarize our explorations.

\subsubsection*{SMT encodings}
Before building our \synth, we experimented with an assembly synthesizer
written in Rosette~\cite{rosette} and
explored language designs and theory combinations.  We found that combinations
of integer and bitvector theories led to poor performance.  We also
experimented with using the theory of arrays for updates to registers and
memory, but found it markedly slower and abandoned it in favor of a
direct encoding.

\subsubsection*{Accumulation algorithm}
Since our implementation of syntax-guided synthesis proceeds in stages, we experimented with an \emph{accumulation} algorithm.
In general, in stage $n$, we attempt to synthesize an $n$-operation program.
Starting at stage $0$ (which succeeds only if the specification is vacuous),
our \synth proceeds to stage $n+1$ when synthesis at stage $n$ fails,
iterating until synthesis succeeds or times out.
The accumulation method attempts to reuse work done at stage $n$ in
later stages.
Specifically, in a later stage $n+k$ we use the last guessed program from stage $n$ as
the initial $n$ operations, and attempt to synthesize only the last $k$ operations.
Intuitively, the $n$-operation program might be ``almost correct,'' since it avoids many generated counterexamples.
Accumulation accelerates synthesis when the final candidate
guessed during synthesis for length $n$ is a prefix of a correct program.
Trying to synthesize $k$ operations is typically much faster than synthesizing $n+k$ operations, so accumulation could be worthwhile and relatively cheap.
If none of the accumulated program prefixes can be extended to a correct program, stage $n+k$ proceeds to attempt to synthesize all $n+k$ operations.
When we evaluated the effectiveness of accumulation
we found that it can help, but infrequently. When it does helps, it can result in
significant improvements, making
syntax-guided synthesis up to $7\times$ faster on a
couple of our use cases. However, in the worst case, accumulation caused synthesis to take
twice as long.

\subsubsection*{Instruction dependency graph}
We experimented with adding constraints to require specific dependencies between instructions, for example, requiring the locations read by the third instruction to intersect with the locations written by the first instruction.
In essence, this enforces a given dependence graph over instructions in a block.
Although this can significantly reduce synthesis times, 
we could not find suitable heuristics for indentifying appropriate dependencies, since implementations of the same block on different machines often requires different dependency graphs.


\subsubsection*{Multiple counterexamples per loop}
We experimented with collecting multiple counterexamples per CEGIS loop
iteration, instead of the standard single counterexample per loop.
We analyzed the effect on the total number of CEGIS iterations and the synthesis time.
Although additional counterexamples consistently reduced the number of
iterations, we found it difficult to determine an effect on overall synthesis
time.
We also tried retaining counterexamples across stages as part of our
accumulation approach.
This also led to fewer iterations but
sometimes longer synthesis times.
Ultimately, we found that additional counterexamples increased both the variance
and average of the total synthesis time.

\subsubsection*{Candidate buckets}
Due to the high time cost of generating program candidates with multiple counterexamples,
we also experimented with grouping candidates across CEGIS iterations into
several candidate \emph{buckets} and generating counterexamples that worked for
all candidates in a bucket.
This method led to less synthesis time for one CEGIS iteration
but more iterations in total for some use cases.
The total synthesis time tended to increase overall.

\subsubsection*{Interactive assembly synthesis}
Assuage~\cite{assuage}, based on our \synth, presents a parallel
interactive assembly synthesizer that engages the user as an active
collaborator. Users, who are familiar with the high-level concepts of
assembly languages, provide multiple types of guidance during
synthesis, while the synthesizer provides users with different
representations of progress feedback. Assuage enables synthesis to
scale beyond current limits.

\subsection{Correctness versus Security}
It is reasonable to ask whether the \aquarium approach with program synthesis
improves system security.

\subsubsection*{Guaranteed correctness can improve security}
The \MD parts of an operating system rarely implement policy;
they translate policy decisions into hardware implementations.
In this sense, a correct implementation does enforce certain
security properties, and a verified implementation does provide
some extra security assurance.
However, in general one must also verify that the \MI code invokes the
\MD operations at the correct points, which is a much larger issue we
do not attempt to address.
Also note that many of the possible errors in low-level code do not
have subtle consequences: the system hangs or crashes, or violently
corrupts memory.
Identifying the \emph{source} of such problems can be extremely
expensive, and verification can save much debugging time; however,
identifying their \emph{existence} is typically less challenging, so
they are relatively unlikely to survive long enough to become
vulnerabilities in the field.

For example, machine dependent code must properly update page tables.
Failure to perform such operations correctly, such as setting the
wrong page permissions or using the wrong address-space ID, can create
security vulnerabilities.
However, verifying the page table updates will not protect against the
\MI virtual memory system failing to revoke access permissions at the
proper times.
Correspondingly, setting the wrong page permissions will in general
result in denying accesses that should succeed, which will cause
programs (or the system) to crash, which in turn is relatively
unlikely to be missed during development.

\REVISION{
Furthermore, incorporating security related restrictions into synthesis is 
an interesting direction for future investigation.
Integrating them with the specification
or even adding as extra constraints can, in general, allow rejecting
vulnerable assembly programs during synthesis, which would provide
further security guarantees.
}


%

\subsubsection*{Synthesis is not a security panacea}
There is a large class of security bugs that arise when microarchitectural
implementation details leak through to the instruction level.
Spectre and Meltdown~\cite{SpectreMeltdown}, and Rowhammer~\cite{8708249}
are, perhaps, the most well-known of such attacks.
Synthesis and verification do nothing to address such issues.
\casp specifications describe the instruction set architecture;
nothing in the
Aquarium ecosystem is aware of microarchitectural implementation
details, so there is no way to reason about any unwanted effects that
might arise.
However, given rules for mitigating such vulnerabilities, we expect to
be able to synthesize code that adheres to those rules.
Further, as always, verification against a specification is only as
good as the correctness of the specification.

\subsection{Future Work}

This work demonstrates the feasibility of assembly program synthesis for \MD OS components.
However, there is plenty of opportunity for improvement and many additional directions to explore.
In addition to further improving the scalability of assembly synthesis, 
we see several opportunities to further reduce the effort necessary to port OSes with \aquarium.

%

Ports have many similarities; it is standard practice to begin a port by copying an
existing implementation for a similar machine~\cite{cpminer}.
Since an existing OS implementation often exists, it may be possible to
synthesize OS specifications of \MD components from existing implementations or
to accelerate synthesis by reusing existing implementations.
A key challenge is how to appropriately abstract away from the specifics of the
machine.
We anticipate that accomplishing such synthesis would need a human in the loop.
Similarly, we anticipate synthesizing parts of a lowering file from a given \ale
specification and a \casp machine description iteratively with a human in the loop.
However, this interaction sacrifices automation 
and requires further maintenance of the specifications and descriptions.

\smallskip

While developing machine descriptions, we found that a useful way to
test the correctness of the descriptions was to create \MD
specifications that should be unrealizable and use synthesis to try to
find an assembly program satisfying the specification.
For example,
ARM has certain flag register fields that must not be changed by
arithmetic instructions.
Given a specification that forced arithmetic
operations and required illegal field changes, successful synthesis
implies an error in the formalization of the machine.
This approach
helped us find errors in our \casp descriptions.
More generally, the use of program synthesis to find errors in semantics is worthy of further exploration.

%
\REVISION{
There is a truism about machine descriptions, in the context of
compilers, to the effect that they always eventually become
Turing-complete as they get extended to handle unexpected new cases.
It would be interesting to develop a scheme for evaluating how
declarative (as opposed to ``code-like'') a description language is.
In general, more declarative descriptions are easier to work with but
less expressive and are prone to needing extensions after their
initial development, whereas more code-like descriptions are more
expensive to use but more expressive.
}

\section{Related Work}\label{sec:rw}

Our work lies at the intersection of the programming languages and operating
systems.

\subsubsection*{Program Synthesis}
Modern program synthesis began in 2006 with the introduction of
\textsc{Sketch}~\cite{solarLezama06},
in which a programmer provides a partial program and
specifications for ``holes'' in the program, and \textsc{Sketch} fills in the holes.
Following this work, counter-example guided inductive synthesis
(CEGIS)~\cite{solarLezama08} generalizes
sketching for infinite state programs.
However, for infinite state programs, the verifier is
typically an unsound, bounded model checker.
Although we use CEGIS, we restrict ourselves to
finite state programs.

Another category of program synthesis, programming by example (PBE),
takes input-output examples to
demonstrate desired program behaviors~\cite{10.1007/978-3-319-21690-4_23, PGL-010}.
In the assembly programming context, examples that include concrete
machine state are much more challenging to provide than are
formal specifications.

While the DSLs used in early CEGIS systems, e.g.,
\cite{solarLezama05,solarLezama06}, handle many of the same
concepts as does \casp (e.g., precise bit manipulation), the
problem of synthesizing assembly code is significantly harder than that
of synthesizing high-level languages.
OS machine code manipulates only untyped memory and worse, all state is global.
While programs in high-level languages contain these features, they are typically used
parsimoniously. 
In contrast, untyped memory and global state are
inherent to, and unavoidable in, assembly code.
Consequently, assembly code synthesis has enormous search spaces that are exponential
in program length and combinatorial in
the number of instructions (at least the dozens),
registers (also dozens),
memory locations (possibly 
the hundreds or thousands if one considers kernel structures, such as
page tables),
and immediate values. 

\subsubsection*{Superoptimization and Assembly Synthesis}
To the best of our knowledge, the first encoding of assembly language into
SMT was the Stoke algorithm for stochastic superoptimization~\cite{stokeAsplos}.
The superoptimization problem for machine
code~\cite{Superoptimizer,greenthumb,bansalAiken06} is: given a piece of
machine code and an execution context, find the lowest cost semantically
equivalent piece of code.  Stoke formulates correctness and performance
improvement into a cost function, then uses a Markov Chain Monte Carlo
sampler to find an implementation that, with high probability, outperforms the
original (already optimized by LLVM~\cite{LLVM}).  
Superoptimizers work on straight-line, intraloop, intraprocedural
code.

Superoptimization avoids some challenges presented by the general
synthesis problem: first, a superoptimizer can fail to optimize a snippet, since the original snippet can be used instead.
In contrast, if \aquarium fails to synthesize a
snippet, it is a showstopper.  Second, superoptimization is bootstrapped by a
correct snippet of code to be optimized; in contrast, we have only a
declarative, machine independent specification and must produce a working
implementation.
This rules out, for example, subdividing the problem by computing
intermediate machine states to use as subgoals.

Srinivasan and Reps also consider the problem of synthesizing assembly
programs from semantic specifications~\cite{mcsynth1,mcsynth2,mcsynth3},
developing several synthesis optimizations similar to ours.
Like us, they use a divide-and-conquer scheme that attempts to find independent
sub-specifications that can be solved separately.
Whereas their algorithm divides specifications greedily, our rule-based
decomposition (\autoref{subsec:dec}) divides lazily, when we find that a
specification cannot be solved by a one-instruction program.
We use an ordering heuristic to select the split that should be considered next.
Their conquer phase exhaustively enumerates instruction opcode and register
arguments, using CEGIS to solve for immediate arguments.
Using enumeration means that their conquer phase does not scale beyond two- or
three-instruction programs, whereas our fully solver-based CEGIS can synthesize
certain sequences up to nine instructions long.
Their footprint-based pruning approach can be compared to our read/write
constraints.
Their bits-lost pruner is similar to our dependency constraints
(\autoref{subsec:dep}) in that it prunes program prefixes that discard necessary
dependencies, but it employs finer-grained input tracking -- at the level of
bits rather than locations -- and coarser availability tracking -- at the level
of whole program state rather than locations.
Since their source code is not available, we cannot run a
direct performance comparison.

Van Geffen et al.~\cite{jitsynth} synthesize components of JIT compilers that
produce sequences of assembly instructions which correspond to instructions in
kernel-level DSLs such as eBPF.
They use a metasketch-based approach that optimizes synthesis using syntactic
constraints on the output assembly programs.
Their pre-load sketches load immediate values used by the rest of the program;
our rule-based decomposition can do this in some cases, in particular when the
program requires a value found at a label or in memory.
Their read-write sketches constrain which instructions can be used depending on
whether they access registers, memory, or both.
This optimization can be compared to our read/write constraints and dependency
constraints, which use assertions to restrict the search space according to a
sound summary of instruction semantics.
Our approach is more precise in that it can constrain instructions based on
accesses to specific registers and memory locations as opposed to accesses to
any register or any memory location.

\subsubsection*{Optimization in Synthesis}
Our dependency analysis represents one way to use abstraction to constrain
search.
Other works have found this general approach effective.
In some sense CEGIS is such an approach; the abstraction it considers is the
program behavior on a specific set of inputs.
The Storyboard Programming~\cite{storyboards} framework starts from sketches and
synthesizes imperative programs that manipulate data structures using
specifications that constrain the desired abstract behavior.

Cosette~\cite{cosette,cosette2} searches for input tables that distinguish
between two SQL queries by causing them to produce different outputs.
To accelerate search, Cosette constrains input tables to contain only those tuples
that can possibly affect the query output.
Cosette finds these constraints by using the abstract semantics of SQL to obtain
\textit{provenance predicates} representing facts about input tuples that could
pass through a query.
Provenance predicates resemble our dependency analysis inasmuch as they are
constructed by following dependencies of output on input; Cosette uses the
abstract semantics of SQL to make provenance predicate discovery tractable.
Our approach differs in that we reduce a \textit{program} search space by
introducing constraints on program abstract behavior.

Much like our \casp DSL is an executable encoding of instruction semantics,
Vale~\cite{bond17} encodes the semantics of each assembly instruction as an
snippet of Dafny~\cite{dafny}. While Vale enables
verification of assembly snippets using Dafny's pre- and postconditions,
we use \ale, a purpose-built DSL for specifications. 
Vale does not enable synthesis.
An interesting avenue of future work would be to
extend our synthesis work to the domain of cryptographic functions
with Vale's use cases.

\subsubsection*{Synthesis for Operating Systems}
In the late 1980's and early 1990's, several groups experimented with OS
customization~\cite{spin95,vino96,exo95},
performance optimization \cite{liedtke95},
and specializing code to improve performance \cite{scout96,pu88}.
The specialization approach produced two interesting OS synthesis
projects: The Synthesis Kernel~\cite{pu88} and Scout~\cite{scout96},
both of which were based on the observation that OSes must handle
all cases, although the vast majority of the time they are executing
specific cases, where many of the decisions throughout a code path
are determined a priori by system call parameter values. Creating
versions of these code paths specialized to a given parameter
improves performance. The Synthesis Kernel was custom-built to support
this kind of synthesis.
Scout took a similar approach, synthesizing fast-path implementations
for network functionality.

The Synthesis Kernel and Scout both differ from \aquarium in two ways.
First, they create customized code paths from existing
general components, which differs from \aquarium's synthesis
using high level, machine-independent specifications.
Second, they perform runtime synthesis, relying on
knowing specific parameter values.
These approaches are orthogonal to ours;
it is possible to combine such runtime approaches with
modern program synthesis techniques.

More recently, Termite~\cite{termite} and Termite-2~\cite{termite2}
synthesize device drivers.
These use a different approach to synthesis (reactive synthesis rather
than CEGIS) and synthesize to a C subset rather than assembly.
More importantly, though, they concentrate on figuring out \emph{what}
to do, that is, what sequence of state changes to make, whereas our
work concentrates primarily on figuring out \emph{how} to make state
changes, that is, the exact assembly instructions.
One could imagine a system where something akin to \casp figured out
how to issue the device state changes directed by Termite-2, or where
something akin to Termite-2 was used to figure out the specifications
for the assembly code blocks \casp generates.
Thus this work is complementary to ours.



\section{Conclusions}\label{sec:conclusion}

We present the \aquarium approach to program synthesis,
part of a collection of languages and tools
to automatically construct machine-dependent OS code.
This paper focuses on \MD assembly code synthesis.
We developed two domain specific languages, \ale and \casp,
to specify OS functionality and describe machine instruction set architectures, respectively.
The \lowertool compiles \MI specifications into \MD specifications, 
while the \casp \synth synthesizes the implementations in assembly language for specific machines.

With \ale, we have made a first approach to \MI specification of \MD components.
We can readily specify some parts of the OS (e.g., interrupt and cache handling)
and not so readily specify others (e.g., boot code, drivers, concurrency
primitives).
We also demonstrated a collection of optimizations for improving synthesis performance.
These do not fully address the scalability challenges
but do improve program synthesis for an unusual domain.

We validated our tools with \CASENUM{35} synthesis use cases taken from two preexisting operating systems,
deployed on four real machine architectures.
For most of our use cases, synthesis finishes in a reasonable time, typically a
few minutes, although some use cases are beyond our ability to synthesize.

\begin{acks}
    This article is based on work supported by the U.S. Air Force and DARPA under contract FA8750-16-C-0045.
    The views, opinions, or findings expressed are those of the authors and
    should not be interpreted as representing the official views or
    policies of the U.S. Department of Defense or the U.S. Government.
    
    We acknowledge the support of the Natural Sciences and Engineering Research Council of Canada (NSERC).
    Nous remercions le Conseil de recherches en sciences naturelles et en génie du Canada (CRSNG) de son soutien.

    We are grateful for the support of the Barrelfish team from ETH
    Zurich, and, especially, to Reto Achermann for answering our
    many questions about Barrelfish and its ecosystem.

\end{acks}

\ifanon
\bibliography{paper,selfcite-anon}
\else
\bibliography{paper,selfcite}
\fi
\appendix

%
%
%


\section{\casp and \ale Language Definition}

%


\newcommand{\mynote}[1]{\text{{\begin{tiny}\textbf{MK: }#1\end{tiny}}}}

\newcommand{\mypara}[1]{\textbf{#1.}}

\newcommand{\ie}{\textit{i.e.,}}
\newcommand{\eg}{\textit{e.g.,}}
\newcommand{\etc}{\textit{etc.}}

\renewcommand{\ensm}[1]{\ensuremath{#1}}

\newcommand{\eps}{\ensm{\epsilon}}

\newcommand{\holds}{\ensm{\vdash}}
\newcommand{\sholds}{\ensm{\models}}

\newcommand{\tsp}{\thinspace}

\newcommand{\bitnot}{\ensm{\mathtt{bnot}}\xspace}
\newcommand{\bitadd}{\ensm{\mathtt{b+}}\xspace}
\newcommand{\bitsub}{\ensm{\mathtt{b-}}\xspace}
\newcommand{\bitmul}{\ensm{\mathtt{b*}}\xspace}
\newcommand{\bitdiv}{\ensm{\mathtt{b/}}\xspace}
\newcommand{\bitlt}{\ensm{\mathtt{\mathop{b\mathord{<}}}}\xspace}
\newcommand{\bitle}{\ensm{\mathtt{\mathop{b\mathord{<=}}}}\xspace}
\newcommand{\bitgt}{\ensm{\mathtt{\mathop{b\mathord{>}}}}\xspace}
\newcommand{\bitge}{\ensm{\mathtt{\mathop{b\mathord{>=}}}}\xspace}
\newcommand{\bitsle}{\ensm{\mathtt{\mathop{bs\mathord{<=}}}}\xspace}
\newcommand{\bitsge}{\ensm{\mathtt{\mathop{bs\mathord{>=}}}}\xspace}
\newcommand{\bitslt}{\ensm{\mathtt{\mathop{bs\mathord{<}}}}\xspace}
\newcommand{\bitsgt}{\ensm{\mathtt{\mathop{bs\mathord{>}}}}\xspace}
\newcommand{\bitand}{\ensm{\mathtt{band}}\xspace}
\newcommand{\bitor}{\ensm{\mathtt{bor}}\xspace}
\newcommand{\bitxor}{\ensm{\mathtt{bxor}}\xspace}
\newcommand{\intdiv}{\ensm{\mathtt{div}}\xspace}
\newcommand{\setdiff}{\ensm{\mathtt{s-}}\xspace}
\newcommand{\xor}{\mathbin{\oplus}}

\newcommand{\defeq}{\ensm{=}\xspace}
\newcommand{\ttinclude}{\ensm{\mathtt{include}}\xspace}
\newcommand{\ttskip}{\ensm{\mathtt{skip}}\xspace}
\newcommand{\ttsemi}{\ensm{\mathit{;}}\xspace}
\newcommand{\ttcol}{\ensm{\mathit{:}}\xspace}
\newcommand{\ttcomma}{\ensm{\mathit{,}\ }\xspace}
\newcommand{\ttdot}{\ensm{\mathit{.}}\xspace}
\newcommand{\tttrue}{\ensm{\mathtt{true}}\xspace}
\newcommand{\ttfalse}{\ensm{\mathtt{false}}\xspace}
\newcommand{\tterr}{\ensm{\mathit{err}}\xspace}
\newcommand{\ttassert}{\ensm{\mathtt{assert}}\xspace}
\newcommand{\ttassertfalse}{\ensm{\mathtt{fail}}\xspace}
\newcommand{\ttwrong}{\ensm{\mathit{wrong}}\xspace}
\newcommand{\ttinvalid}{\ensm{\mathit{invalid}}\xspace}
\newcommand{\ttvalid}{\ensm{\mathit{valid}}\xspace}
\newcommand{\itbinop}{\ensm{\mathit{binop}}\xspace}
\newcommand{\itsetop}{\ensm{\mathit{setop}}\xspace}
\newcommand{\itboolop}{\ensm{\mathit{boolop}}\xspace}
\newcommand{\itbinrel}{\ensm{\mathit{binrel}}\xspace}
\newcommand{\itbintop}{\ensm{\mathit{bintop}}\xspace}
\newcommand{\itbvecop}{\ensm{\mathit{bvecop}}\xspace}
\newcommand{\itunop}{\ensm{\mathit{unop}}\xspace}
\newcommand{\itatomic}{\ensm{\mathit{atomic}}\xspace}
\newcommand{\ttvec}{\ensm{\mathit{vec}}\xspace}
\renewcommand{\ttid}{\ensm{\mathit{id}}\xspace}
\newcommand{\ttxmoduleid}{\ensm{\mathit{{x}_{module}}}\xspace}
\newcommand{\ttxblockid}{\ensm{\mathit{{x}_{block}}}\xspace}
\newcommand{\ttfuncid}{\ensm{\mathit{x_{func}}}\xspace}
\newcommand{\ttprocid}{\ensm{\mathit{x_{proc}}}\xspace}
\newcommand{\ttxop}{\ensm{\mathit{x_{op}}}\xspace}
\newcommand{\ttregid}{\ensm{\mathit{x_{reg}}}\xspace}
\newcommand{\ttregidi}{\ensm{\mathit{x_{{\mathit{reg}}_i}}}\xspace}
\renewcommand{\ttmemid}{\ensm{\mathit{x_{mem}}}\xspace}
\renewcommand{\ttmemidi}{\ensm{\mathit{x_{{\mathit{mem}}_i}}}\xspace}
\newcommand{\ttlabelid}{\ensm{\mathit{x_{label}}}\xspace}
\newcommand{\tttypeid}{\ensm{\mathit{x_{\tau}}}\xspace}
\newcommand{\ttextid}{\ensm{\mathit{x_{ext}}}\xspace}
\newcommand{\ttidents}{\ensm{\mathit{identifiers}}\xspace}
\newcommand{\ttunit}{\ensm{\mathit{()}}\xspace}
\newcommand{\ttargs}{\ensm{\mathit{args}}\xspace}
\newcommand{\ttstr}{\ensm{\mathit{string}}\xspace}
\newcommand{\ttget}{\ensm{\mathtt{get}}\xspace}
\newcommand{\ttassign}{\ensm{\mathtt{assign}}\xspace}
\newcommand{\ttindex}{\ensm{\mathit{index}}\xspace}
\newcommand{\ttusing}{\ensm{\mathit{with}}\xspace}
\newcommand{\ttread}{\ensm{\mathtt{read}}\xspace}
\newcommand{\ttfetch}{\ensm{\mathtt{fetch}}\xspace}
\newcommand{\ttwrite}{\ensm{\mathtt{write}}\xspace}
\newcommand{\ttstore}{\ensm{\mathtt{store}}\xspace}
\newcommand{\ttat}{\ensm{{\mathit{\small @}}}\xspace}
\newcommand{\ttdecl}{\ensm{\mathit{decl}\xspace}}
\newcommand{\ttdecls}{\ensm{\mathit{decls}\xspace}}
\newcommand{\ttlet}{\ensm{\mathtt{let}}\xspace}
\newcommand{\ttletstate}{\ensm{\mathtt{letstate}}\xspace}
\newcommand{\ttdef}{\ensm{\mathtt{def}}\xspace}
\newcommand{\ttproc}{\ensm{\mathtt{proc}}\xspace}
\newcommand{\itlet}{\ensm{\mathit{let}}\xspace}
\newcommand{\itlets}{\ensm{\mathit{lets}}\xspace}
\newcommand{\tttype}{\ensm{\mathtt{type}}\xspace}
\newcommand{\ttstate}{\ensm{\mathtt{state}}\xspace}
\newcommand{\ttcontrol}{\ensm{\mathtt{control}}\xspace}
\newcommand{\ttdontgate}{\ensm{\mathtt{dontgate}}\xspace}
\newcommand{\ttastate}{\ensm{\mathtt{state}}\xspace}
\newcommand{\ttdefop}{\ensm{\mathtt{defop}}\xspace}
\newcommand{\ttend}{\ensm{\mathtt{end}}\xspace}
\newcommand{\ttmach}{\ensm{\mathit{machine}}\xspace}
\newcommand{\ttprog}{\ensm{\mathit{program}}\xspace}
\newcommand{\tttxt}{\ensm{\mathtt{txt}}\xspace}
\newcommand{\tthex}{\ensm{\mathtt{hex}}\xspace}
\newcommand{\ttdec}{\ensm{\mathtt{dec}}\xspace}
\newcommand{\ttbin}{\ensm{\mathtt{bin}}\xspace}
\newcommand{\ttlbl}{\ensm{\mathtt{lbl}}\xspace}
\newcommand{\ttsem}{\ensm{\mathtt{sem}}\xspace}
\newcommand{\ttinit}{\ensm{\mathit{init}}\xspace}
\newcommand{\ttfunc}{\ensm{\mathit{function}}\xspace}
\newcommand{\ttval}{\ensm{\mathit{val}}\xspace}
\newcommand{\ttvalue}{\ensm{\mathit{Value}}\xspace}
\newcommand{\ttcast}{\ensm{\mathtt{coerce}\tsp\mathtt{as}}\xspace}
\newcommand{\tthavoc}{\ensm{\mathit{nondet()}\xspace}}
\newcommand{\ttaddress}{\ensm{\mathit{Addr}\xspace}}
\newcommand{\ttis}{\ensm{\mathtt{is}\xspace}}
\newcommand{\ttcptr}{\ensm{\mathtt{ptr}}\xspace}
\newcommand{\ttcPtr}{\ensm{\mathtt{Ptr}}\xspace}
\newcommand{\tteptr}{\ensm{\mathtt{ptr}}\xspace}
\newcommand{\ttvars}{\ensm{\mathit{vars}}\xspace}
\newcommand{\ttvals}{\ensm{\mathit{values}}\xspace}
\newcommand{\ttglobals}{\ensm{\mathit{globals}}\xspace}
\newcommand{\ttterm}{\ensm{T}\xspace}
\newcommand{\ttor}{\ensm{\mathit{or}}\xspace}
\newcommand{\ttfresh}{\ensm{\mathit{fresh}}\xspace}
\newcommand{\ttmod}{\ensm{\mathtt{mod}}\xspace}
\newcommand{\ttthen}{\ensm{\rhd}\xspace}

\newcommand{\exprfail}{\ttassertfalse}

\newcommand{\ttbaseof}{\ensm{\mathtt{base}}}
\newcommand{\ttoffsetof}{\ensm{\mathtt{offset}}}
\newcommand{\ttasbitv}{\ensm{\mathit{(as\ttbit)}}}

\newcommand{\clholds}{\ensm{\vdash_{pred}}\tsp}
\newcommand{\clog}{\ensm{\phi}}
\newcommand{\calog}{\ensm{\clog_\forall}}
\newcommand{\celog}{\ensm{\clog_\exists}}
\newcommand{\cor}{\ensm{\vee}}
\newcommand{\cand}{\ensm{\wedge}}
\newcommand{\cimplies}{\ensm{\rightarrow}}
\newcommand{\ceq}{\ensm{=}}

\renewcommand{\bvtyp}[1]{\ensm{#1\ \ttbit}}
\newcommand{\cptrtyp}[1]{\ensm{\bvtyp{#1}}}
\renewcommand{\bvaltyp}[1]{\ensm{\bvtyp{#1}}}
\renewcommand{\bvltyp}[1]{\ensm{#1\ \ttloc}}
\newcommand{\bvstyp}[1]{\ensm{#1\ \ttloc\ \ttset}}


\renewcommand{\ttC}{\ensm{\mathit{C}}\xspace}
\newcommand{\ttR}{\ensm{\mathit{R}}\xspace}
\newcommand{\ttInt}{\ensm{\mathit{Int}}\xspace}
\newcommand{\ttVec}{\ensm{\mathit{Vec}}\xspace}

\newcommand{\ttmap}{\ensm{\mathit{map}}\xspace}

\newcommand{\cfuntyp}{\atyp_{\mathit{func}}}
\newcommand{\cbasetyp}{\atyp_{\mathit{base}}}
\newcommand{\cbasetypi}{\atyp_{{\mathit{base}}_i}}
\newcommand{\cbasetypone}{\ensm{\atyp_{{\mathit{base}}_1}}}
\newcommand{\cbasetyptwo}{\ensm{\atyp_{{\mathit{base}}_2}}}
\newcommand{\cregtyp}{\atyp_{\mathit{reg}}}
\newcommand{\cregstyp}{\atyp_{\mathit{regs}}}
\newcommand{\cmemtyp}{\atyp_{\mathit{mem}}}
\newcommand{\clbltyp}{\atyp_{\mathit{label}}}
\newcommand{\cstatetyp}{\atyp_{\mathit{state}}}

\newcommand{\ttmem}{\ensm{\mathtt{mem}}\xspace}
\newcommand{\ttlabel}{\ensm{\mathtt{label}}\xspace}
\renewcommand{\ttloc}{\ensm{\mathtt{reg}}\xspace}

\newcommand{\ttlen}{\ensm{\mathit{len}}\xspace}
\renewcommand{\ttmlen}{\ensm{\mathtt{len}}\xspace}
\renewcommand{\ttbit}{\ensm{\mathtt{bit}}\xspace}
\newcommand{\ttref}{\ensm{\mathtt{ref}}\xspace}
\renewcommand{\ttwptr}{\ensm{\mathtt{ptr}}\xspace}
\renewcommand{\ttwvec}{\ensm{\mathtt{vec}}\xspace}
\newcommand{\ttBit}{\ensm{\mathtt{Bits}}\xspace}
\newcommand{\ttset}{\ensm{\mathtt{set}}\xspace}
\renewcommand{\ttbool}{\ensm{\mathtt{bool}}\xspace}
\renewcommand{\ttint}{\ensm{\mathtt{int}}\xspace}
\newcommand{\ttstring}{\ensm{\mathtt{string}}\xspace}
\newcommand{\ttn}{\ensm{\mathit{n}}\xspace}
\newcommand{\ttk}{\ensm{\mathit{k}}\xspace}
\newcommand{\ttK}{\ensm{\mathit{K}}\xspace}
\newcommand{\tti}{\ensm{\mathit{i}}\xspace}
\newcommand{\ttj}{\ensm{\mathit{j}}\xspace}
\newcommand{\ttb}{\ensm{\mathit{b}}\xspace}
\newcommand{\ttv}{\ensm{\mathit{v}}\xspace}
\newcommand{\tildev}{\ensm{\mathit{\tilde{v}}}\xspace}
\newcommand{\ttr}{\ensm{\mathit{r}}\xspace}
\newcommand{\tte}{\ensm{\mathit{e}}\xspace}
\newcommand{\ttS}{\ensm{\mathit{S}}\xspace}
\renewcommand{\ttN}{\ensm{\mathit{N}}\xspace}
\newcommand{\ttx}{\ensm{\mathit{x}}\xspace}
\newcommand{\tty}{\ensm{\mathit{y}}\xspace}
\newcommand{\ttz}{\ensm{\mathit{z}}\xspace}
\newcommand{\ttl}{\ensm{\mathit{l}}\xspace}
\newcommand{\ttT}{\ensm{\mathit{T}}\xspace}
\newcommand{\ttm}{\ensm{\mathit{m}}\xspace}
\newcommand{\ttM}{\ensm{\mathit{M}}\xspace}
\newcommand{\ttV}{\ensm{\mathit{V}}\xspace}

\newcommand{\ttconst}{\ensm{const}\xspace}

\newcommand{\globalv}[1]{\ensm{#1}\xspace}

\newcommand{\ttchoice}{\ensm{\mathit{\ ?\ }}\xspace}
\newcommand{\ttinbracket}[1]{\ensm{\mathit{[}#1\mathit{]}}\xspace}
\renewcommand{\ttinparen}[1]{\ensm{\mathit{(}#1\mathit{)}}\xspace}
\newcommand{\ttinbrace}[1]{\ensm{\mathit{\{}#1\mathit{\}}}\xspace}
\newcommand{\inhline}[1]{\ensm{\vert #1\vert}\xspace}

\newcommand{\bitconstprefix}[1]{\ensm{\mathtt{0b}#1}}
\newcommand{\bitconst}{\ensm{\bitconstprefix{\ttC}}}
\newcommand{\hexconst}{\ensm{\mathtt{0x}\ttC}}

\newcommand{\sizeof}[1]{\ensm{\ttsizeof(#1)}\xspace}

\newcommand{\setlit}{\ensm{\ttinbrace{\ttx_1, \ldots, \ttx_k}}\xspace}
\newcommand{\intsetlit}{\ensm{\ttinbrace{\ttC, \ldots \ttC}}}
\newcommand{\ttsetemp}{\ensm{\ttinbrace{}}\xspace}
\newcommand{\sizeofset}[1]{\ensm{\|{#1}\|}\xspace}
\newcommand{\setmemberof}[2]{\ensm{#1\tsp{\in}#2}}
\newcommand{\setcomplement}[1]{\overline{\ensm{#1}^c}}

\newcommand{\program}{\ensm{\ttprog}}
\newcommand{\machine}{\ensm{\ttmach}}
\newcommand{\decls}{\ensm{\ttdecls}}
\newcommand{\declim}{\ensm{\ttsemi\ }}
\newcommand{\decl}{\ensm{\ttdecl}}
\newcommand{\defops}{\ensm{\mathit{defops}}\xspace}
\newcommand{\defop}{\ensm{\mathit{defop}}\xspace}
\newcommand{\defeffectfunc}{\ensm{\ttdefsproc}}

\newcommand{\meminit}[1]{\ensm{\ttinit\ \ttindex\ \tti\ \ttusing\ (#1)}} 

\newcommand{\bfe}{\ensm{\mathbf{e}}}
\newcommand{\bferr}{\ensm{\mathtt{crash}}}

\newcommand{\subst}[2]{\ensm{[#1\mathbf{\setminus} #2]}}

\newcommand{\lderef}[1]{\regderef}
\newcommand{\memread}[1]{\ttfetch(#1,\ttC)}
\newcommand{\fetch}[2]{\ttfetch(#1,#2)}
\newcommand{\absfetch}[1]{\ttfetch(#1)}
\newcommand{\regderef}[1]{\:^*#1}
\newcommand{\dholds}{\ensm{\vDash}}
\newcommand{\tholds}{\ensm{\vdash}}
\newcommand{\twfholds}{\ensm{\vdash_{\texttt{wf}}}}
\newcommand{\sqright}{\rightsquigarrow}
\newcommand{\sqrightm}[1]{\rightsquigarrow_{#1}}
\newcommand{\assign}[2]{\ensm{\ttstore(#1) \leftarrow #2}}
\newcommand{\callstmt}[2]{\ensm{\ttcall\ #1\ #2}}

\newcommand{\sseq}[1]{\ensm{\overline{#1}}}
\newcommand{\pseq}[2]{\ensm{\widehat{#1\ttcol #2}}}

\renewcommand{\atyp}{\ensm{\tau}\xspace}
\newcommand{\funtyp}[2]{\ensm{#1 \rightarrow #2}\xspace}
\newcommand{\proctyp}[1]{\ensm{#1 \rightarrow ()}\xspace}

\newcommand{\hastyp}[2]{\ensm{#1\ \mathit{:}\ #2}}
\newcommand{\oftyp}[2]{\ensm{#1\ \mathit{:}\ #2}}


\renewcommand{\memtyp}[3]{\ensm{#1\ \ttbit\ #2\ \ttmlen\ #3\ \texttt{ref}}}
\renewcommand{\labeltyp}[1]{\ensm{#1\ \ttlabel}}
\renewcommand{\nbitmemtyp}[1]{\memtyp{\_}{\_}{#1}}

\newcommand{\cptr}[2]{\ensm{\ttcptr(#1,#2)}}
\newcommand{\eptr}[2]{\ensm{\tteptr(#1,#2)}}
\newcommand{\ptrtyp}{\ensm{\ttaptr}}
\renewcommand{\ptrform}[2]{\ensm{\ttinparen{#1 \ttcomma #2}}}
\newcommand{\lblform}[1]{\ensm{\ttinbrace{#1}}}


\newcommand{\isptr}[1]{\ensm{\ttis\ttcptr(#1)}}
\newcommand{\addrof}[1]{\ensm{\ttbaseof(#1)}}
\newcommand{\offsetof}[1]{\ensm{\ttoffsetof(#1)}}
\newcommand{\bvcast}[1]{\ensm{\ttasbitv(#1)}}

\newcommand{\tyvar}{\ensm{\alpha}}

\newcommand{\bmapsto}[2]{\ensm{\mathit{[}{#1}\rightarrowtail{#2}}\mathit{]}}

\newcommand{\fapp}[2]{\ensm{#1\ (#2)}}
\newcommand{\papp}[2]{\ensm{#1\ (#2)}}

\newcommand{\offset}[2]{\ensm{#1\mathit{@}#2}}

\newcommand{\irule}[2]{\ensm{\inferrule{#1}{#2}}}
\newcommand{\actrans}[1]{\ensm{\mathcal{AC}\llbracket#1\rrbracket}}
\newcommand{\actrule}[1]{\ensm{\typenv, \kenv \holds \actrans{#1}}}
\newcommand{\acrule}[1]{\ensm{\typenv, \kenv \holds \actrans{#1}}}
\newcommand{\acruleCC}[1]{\ensm{\typenvCC, \kenv \holds \actrans{#1}}}
\newcommand{\actyptrans}[1]{\ensm{\mathcal{AC}\llbracket#1\rrbracket}}
\newcommand{\actyprule}[1]{\ensm{\typenv, \kenv \holds \actyptrans{#1}}}
\newcommand{\acsigma}[1]{\ensm{\Sigma(#1)}}
\newcommand{\acphi}[1]{\ensm{\Phi(#1)}}
\newcommand{\aclambda}[1]{\ensm{\venv(#1)}}
\newcommand{\acdelta}[1]{\ensm{\genv(#1)}}
\newcommand{\acgamma}[1]{\ensm{\tenv(#1)}}
\newcommand{\st}{\mathbf{.}}

\newcommand{\sITE}[3]{\ensm{\texttt{if}\tsp#1\tsp\texttt{then}\tsp#2\tsp\texttt{else}\tsp#3}}
\newcommand{\eITE}[3]{\ensm{#1\texttt{ ? }#2\texttt{ : }#3}}

\newcommand{\sLet}[4]{\ensm{\texttt{let}\tsp\oftyp{#1}{#2}= #3\tsp\texttt{in}\tsp#4}}
\newcommand{\eLet}[4]{\ensm{\texttt{let}\tsp\oftyp{#1}{#2}= #3\tsp\texttt{in}\tsp#4}}

\newcommand{\memidx}[1]{\:^*#1}
\newcommand{\sFor}[4]{\ensm{\texttt{for}\tsp#1\in(#2\ldots #3)\ \texttt{do}\tsp#4}}
\newcommand{\smAssign}[2]{\assign{#1}{#2}}
\newcommand{\slAssign}[2]{#1\tsp\ttcol= #2}

\newcommand{\ebranchto}[1]{\ensm{\texttt{branchto}}}
\newcommand{\sBRANCH}[1]{\ensm{\texttt{BRANCH}(#1)}}

\newcommand{\bitidx}[2]{\ensm{#1\ttinbracket{#2}}}
\newcommand{\bitfield}[3]{\ensm{#1\ttinbracket{#2\ttcomma #3}}}

\newcommand{\execmemidx}[2]{#1 (#2)}
\newcommand{\execmenvidx}[2]{#1 (#2)}

\renewcommand{\bnfeq}{\ensm{\Coloneqq}}
\renewcommand{\bnfor}{\ensm{\mid}}


\newcommand{\strconcat}[2]{\ensm{#1\ \texttt{++}\ #2}}
\newcommand{\strof}[1]{\ensm{#1\ttdot\tttxt}}
\newcommand{\hexof}[1]{\ensm{#1\ttdot\tthex}}
\newcommand{\decof}[1]{\ensm{#1\ttdot\ttdec}}
\newcommand{\binof}[1]{\ensm{#1\ttdot\ttbin}}
\newcommand{\lblof}[1]{\ensm{#1\ttdot\ttlbl}}
\newcommand{\strlit}{\texttt{"}$\ldots$\texttt{"}}


\newcommand{\genv}{\Delta} 
\newcommand{\tenv}{\Gamma} 
\newcommand{\venv}{\Lambda}
\newcommand{\env}{\ensm{\tenv}\xspace} 
\newcommand{\wfenv}{\ensm{\genv}}
\newcommand{\typenv}{\ensm{\genv, \tenv}}
\newcommand{\typenvAB}{\ensm{\genv, \tenv'}}
\newcommand{\typenvAC}{\ensm{\genv, \tenv''}}
\newcommand{\typenvBA}{\ensm{\genv', \tenv}}
\newcommand{\typenvBB}{\ensm{\genv', \tenv'}}
\newcommand{\typenvCC}{\ensm{\genv'', \tenv''}}
\newcommand{\subenv}{\ensm{\Sigma}}
\newcommand{\subenvZ}{\ensm{\Sigma_{\mathit{builtin}}}}
\newcommand{\subenvB}{\ensm{\Sigma'}}
\newcommand{\subenvC}{\ensm{\Sigma''}}

\newcommand{\alenv}{\ensm{\Sigma, \Phi}}
\newcommand{\alenvZZ}{\ensm{\Sigma_{\mathit{builtin}}, \Phi_{\mathit{builtin}}}}
\newcommand{\alenvAB}{\ensm{\Sigma, \Phi'}}
\newcommand{\alenvAC}{\ensm{\Sigma, \Phi''}}
\newcommand{\alenvBA}{\ensm{\Sigma', \Phi}}
\newcommand{\alenvBB}{\ensm{\Sigma', \Phi'}}
\newcommand{\alenvCC}{\ensm{\Sigma'', \Phi''}}
\newcommand{\denv}{\ensm{\tenv,\venv}\xspace} 

\newcommand{\subt}{<:}

\newcommand{\nextrulesmall}{\vskip 8pt}
\newcommand{\nextrule}{\vskip 15pt}
\newcommand{\nextclause}{\ensm{}}

\newcommand{\envgoesto}{\ensm{\mapsto}}
\newcommand{\goesto}{\ensm{\rightarrow}}
\newcommand{\dngoesto}{\ensm{\nrightarrow}}
\newcommand{\egoesto}{\ensm{\Downarrow}}
\newcommand{\sgoesto}{\ensm{\Downarrow}}
\newcommand{\opstenv}{\ensm{\Gamma}}
\newcommand{\opsmenv}{\ensm{\sigma}}
\newcommand{\opsrenv}{\ensm{\rho}}
\newcommand{\memgoesto}{\ensm{\rightarrow}}

\newcommand{\opsenv}{\ensm{\opsrenv, \opsmenv}\xspace}
\newcommand{\opsenvB}{\ensm{\opsrenv', \opsmenv'}\xspace}
\newcommand{\opsenvC}{\ensm{\opsrenv'', \opsmenv''}\xspace}
\newcommand{\opsenvafter}[2]{\ensm{\opsrenv_{#2},\opsmenv_{#1}}\xspace}
\newcommand{\oholds}{\ensm{\vdash}\xspace}

\newcommand{\osreduce}[4]{\osreducetenv{\venv}{#1}{#2}{#3}{#4}}
\newcommand{\osreduceafter}[4]{\osreducetenv{\venv'}{#1}{#2}{#3}{#4}}
\newcommand{\osreducetenv}[5]{\ensm{#1\oholds(#3,#2)\sgoesto(#5,#4,\NEW{\cdot})}}
\newcommand{\osreduceBR}[5]{\osreducetenvBR{\venv}{#1}{#2}{#3}{#4}{#5}}
\newcommand{\osreduceafterBR}[5]{\osreducetenvBR{\venv'}{#1}{#2}{#3}{#4}{#5}}
\newcommand{\osreducetenvBR}[6]{\ensm{#1\oholds(#3,#2)\sgoesto(#5,#4,\NEW{#6})}}

\newcommand{\psreduce}[3]{\preducetenv{\venv}{#1}{#2}{#3}}
\newcommand{\psreduceB}[3]{\preducetenv{\venv'}{#1}{#2}{#3}}
\newcommand{\preducetenv}[4]{\ensm{#1\oholds(#3,#2)\goesto(#4)}}

\newcommand{\psreducemulti}[3]{\preducetenvmulti{\venv}{#1}{#2}{#3}}
\newcommand{\psreducemultiB}[3]{\preducetenvmulti{\venv'}{#1}{#2}{#3}}
\newcommand{\preducetenvmulti}[4]{\ensm{#1\oholds(#3,#2)\goesto^*(#4)}}

\newcommand{\esloclookup}[1]{\ensm{\opsrenv(#1) = \ttv}}
\newcommand{\esmemlookup}[3]{\ensm{\opsmenv(#1, #2) = #3}}
\newcommand{\esred}[4]{\ensm{\venv#4\oholds(\opsenv,#1)#3#2}}
\newcommand{\esreduce}[2]{\esred{#1}{#2}{\egoesto}{}{}}
\newcommand{\esredafter}[4]{\ensm{\venv#4\oholds(\opsrenv', \opsmenv',#1)#3#2}}
\newcommand{\esreduceafter}[2]{\esredafter{#1}{#2}{\egoesto}{}{}}
\newcommand{\esredB}[4]{\ensm{\venv'#4\oholds(\opsenv,#1)#3#2}}
\newcommand{\esreduceB}[2]{\esredB{#1}{#2}{\egoesto}{}{}}
\newcommand{\esredafterB}[4]{\ensm{\venv'#4\oholds(\opsrenv', \opsmenv',#1)#3#2}}
\newcommand{\esreduceafterB}[2]{\esredafterB{#1}{#2}{\egoesto}{}{}}
\newcommand{\esreducevenv}[4]{\ensm{#1\oholds(#3,#2)\egoesto#4}}
\newcommand{\esdnreduce}[2]{\esred{#1}{#2}{\dngoesto}{}{}}


\newcommand{\rbox}[1]{\ensm{r}}
\newcommand{\rboxid}{\rbox{\globid}}
\newcommand{\membox}[1]{\ensm{m}}
\newcommand{\memboxid}{\membox{\globid}}

\newcommand{\SEMdeclZA}[1]{\ensm{  \venv_{\mathit{builtin}}, (\{\}, \{\}) \holds #1 \ttthen \venv }}
\newcommand{\SEMdeclAA}[1]{\ensm{  \venv, (\opsrenv, \opsmenv) \holds  #1  \ttthen \venv  }}
\newcommand{\SEMdeclAB}[1]{\ensm{  \venv, (\opsrenv, \opsmenv) \holds  #1  \ttthen \venv'  }}
\newcommand{\SEMdeclAC}[1]{\ensm{  \venv, (\opsrenv, \opsmenv) \holds  #1  \ttthen \venv''  }}
\newcommand{\SEMdeclBC}[1]{\ensm{  \venv', (\opsrenv, \opsmenv) \holds #1  \ttthen \venv''  }}
\newcommand{\SEMdeclAx}[2]{\ensm{  \venv, (\opsrenv, \opsmenv) \holds  #1  \ttthen \venv[#2]  }}
\newcommand{\SEMdeclZ}[1]{\ensm{  \holds, (\opsrenv, \opsmenv)  #1  \ttthen \venv_0  }}
\newcommand{\SEMdeclA}[1]{\ensm{  \holds, (\opsrenv, \opsmenv)  #1  \ttthen \venv  }}
\newcommand{\SEMdeclB}[1]{\ensm{  \holds, (\opsrenv, \opsmenv)  #1  \ttthen \venv'  }}
\newcommand{\SEMZZA}[1]{\ensm{  \venv_{\mathit{builtin}}, (\{\}, \{\}) \holds #1 \ttthen \venv }}
\newcommand{\SEMAAAA}[1]{\ensm{  \venv, (\opsrenv, \opsmenv) \holds  #1  \ttthen \venv, (\opsrenv, \opsmenv)  }}
\newcommand{\SEMAAB}[1]{\ensm{  \venv, (\opsrenv, \opsmenv) \holds  #1  \ttthen \venv' }}
\newcommand{\SEMAA}[1]{\ensm{  \holds  #1  \ttthen \venv, (\opsrenv, \opsmenv) }}
\newcommand{\SEMBA}[1]{\ensm{  \holds  #1  \ttthen \venv', (\opsrenv, \opsmenv) }}

\newcommand{\SEMmachZ}[1]{\ensm{  \holds  #1  \ttthen \venv_0  }}
\newcommand{\SEMmachA}[1]{\ensm{  \holds  #1  \ttthen \venv  }}
\newcommand{\SEMmachB}[1]{\ensm{  \holds  #1  \ttthen \venv'  }}
\newcommand{\SEMmachAB}[1]{\ensm{  \venv \holds  #1  \ttthen \venv'  }}

\newcommand{\ttae}{\ensm{\mathit{e}}}
\newcommand{\ttce}{\ensm{\mathit{c\text{-}e}}}
\newcommand{\italewifespec}{\ensm{\mathit{ale\text{-}spec}}}
\newcommand{\italewifebinding}{\ensm{\mathit{ale\text{-}binding}}}
\newcommand{\italedecls}{\ensm{\mathit{ale\text{-}decls}}}
\newcommand{\italedecl}{\ensm{\mathit{ale\text{-}decl}}}
\newcommand{\itaxioms}{\ensm{\mathit{axioms}}\xspace}
\newcommand{\itderivations}{\ensm{\mathit{derivations}}\xspace}
\newcommand{\itdeclarations}{\ensm{\mathit{declarations}}\xspace}
\newcommand{\itdecl}{\ensm{\mathit{decl}}\xspace}
\newcommand{\itdecls}{\ensm{\mathit{decls}}\xspace}
\newcommand{\itdefinitions}{\ensm{\mathit{definitions}}\xspace}
\newcommand{\itblocks}{\ensm{\mathit{blocks}}\xspace}
\newcommand{\ttaxiom}{\ensm{\mathit{axiom}}\xspace}
\newcommand{\ttBlock}{\ensm{\mathtt{Block}}\xspace}
\newcommand{\itblock}{\ensm{\mathit{block}}\xspace}
\newcommand{\ttblock}{\ensm{\mathtt{block}}\xspace}
\newcommand{\itspec}{\ensm{\mathit{spec}}\xspace}
\newcommand{\itinsts}{\ensm{\mathit{insts}}\xspace}
\newcommand{\itinst}{\ensm{\mathit{inst}}\xspace}
\newcommand{\ttspecdef}{\ensm{\mathit{define\text{-}spec}}\xspace}
\newcommand{\ttblockdef}{\ensm{\mathit{define\text{-}block}}\xspace}
\newcommand{\ttblockbind}{\ensm{\itblock\text{-}\itlet}\xspace}
\newcommand{\ttblockbinds}{\ensm{\itblock\text{-}\itlets}\xspace}
\newcommand{\itregion}{\ensm{\mathit{region}}\xspace}
\newcommand{\itregions}{\ensm{\mathit{regions}}\xspace}
\newcommand{\ttspecdecls}{\ensm{\itspec\text{-}\itdecls}\xspace}
\newcommand{\ttspecdecl}{\ensm{\itspec\text{-}\itdecl}\xspace}
\newcommand{\ttCallable}{\ensm{\mathit{Callable}}\xspace}
\newcommand{\ttcallable}{\ensm{\mathit{callable}}\xspace}
\newcommand{\ttFrame}{\ensm{\mathit{Frame}}\xspace}
\newcommand{\ttframe}{\ensm{\mathit{frame}}\xspace}
\newcommand{\ttframes}{\ensm{\mathit{frames}}\xspace}
\newcommand{\ttframec}{\ensm{\mathit{frame}_\mathit{map}}\xspace}
\newcommand{\ttframea}{\ensm{\mathit{frame}_\mathit{ale}}\xspace}
\newcommand{\ttmayframe}{\ensm{\mathit{may\text{-}frame}}\xspace}
\newcommand{\ttmustframe}{\ensm{\mathit{must\text{-}frame}}\xspace}
\newcommand{\ttpre}{\ensm{\mathit{pre}}\xspace}
\newcommand{\ttpost}{\ensm{\mathit{post}}\xspace}
\newcommand{\itmodules}{\ensm{\mathit{modules}}\xspace}
\newcommand{\itmodule}{\ensm{\mathit{module}}\xspace}
\newcommand{\itmap}{\ensm{\mathit{mapping}}\xspace}
\newcommand{\ttaptr}{\ensm{\mathit{Pointer}}\xspace}
\newcommand{\ttatype}{\ensm{\mathtt{type}}\xspace}
\newcommand{\ttafunc}{\ensm{\mathtt{func}}\xspace}
\newcommand{\ttaregion}{\ensm{\mathtt{region}}\xspace}
\newcommand{\ttaval}{\ensm{\mathtt{value}}\xspace}
\newcommand{\ttamaps}{\ensm{\mathit{Type Maps}}\xspace}
\newcommand{\ttadefs}{\ensm{\mathit{definition}}\xspace}
\newcommand{\ttaderive}{\ensm{\mathit{derivation}}\xspace}
\newcommand{\ttareq}{\ensm{\mathtt{require}}\xspace}
\newcommand{\ttaprov}{\ensm{\mathtt{provide}}\xspace}
\newcommand{\ttimport}{\ensm{\mathtt{import}}\xspace}
\newcommand{\ttalower}{\ensm{\mathtt{lower\text{-}with}}\xspace}
\newcommand{\ttfreads}{\ensm{\mathtt{read}}\xspace}
\newcommand{\ttfwrites}{\ensm{\mathtt{reg\text{-}modify}}\xspace}
\newcommand{\ttfmwrites}{\ensm{\mathtt{mem\text{-}modify}}\xspace}
\newcommand{\ttmodule}{\ensm{\mathtt{module}}\xspace}
\newcommand{\ttmayfreads}{\ensm{\mathtt{may\text{-}access}}\xspace}
\newcommand{\ttmayfwrites}{\ensm{\mathtt{may\text{-}modify}}\xspace}
\newcommand{\ttmustfreads}{\ensm{\mathtt{must\text{-}access}}\xspace}
\newcommand{\ttmustfwrites}{\ensm{\mathtt{must\text{-}modify}}\xspace}
\newcommand{\ttcasp}{\ensm{\mathit{casp}}\xspace}
\newcommand{\ttnil}{\ensm{\mathtt{nil}}\xspace}
\newcommand{\ttaloc}{\ensm{\mathtt{loc}}\xspace}
\newcommand{\ttnilptr}{\ensm{\mathit{Null}}\xspace}
\newcommand{\ttamap}{\ensm{\mathtt{Map}}\xspace}
\newcommand{\ttadt}{\ensm{\mathit{adt}}\xspace}
\newcommand{\ttmatch}{\ensm{\mathtt{match}}\xspace}
\newcommand{\ttwith}{\ensm{\mathtt{with}}\xspace}
\newcommand{\ttctor}{\ensm{\mathit{Ctor}}\xspace}
\newcommand{\ttisctor}{\ensm{\mathit{is}\text{-}\ttctor}\xspace}
\newcommand{\ttacases}{\ensm{\mathit{ale}\text{-}\mathit{match}\text{-}\mathit{cases}}}
\newcommand{\ttmatchor}{\ensm{\mathtt{\mid}}}
\newcommand{\ttany}{\ensm{\mathtt}{\_}}
\newcommand{\ttmaybe}{\ensm{\mathtt{Maybe}}}

\newcommand{\ttforall}{\ensm{\mathtt{forall}}\xspace}
\newcommand{\ttexists}{\ensm{\mathtt{exists}}\xspace}
\newcommand{\ttderef}{\ensm{\mathtt{deref\text{-}loc}}\xspace}

\newcommand{\ttsizeof}{\ensm{\mathtt{sizeof}}\xspace}

\newcommand{\ttmapemp}{\ensm{\ttnil}}

\newcommand{\bigtuple}{\ensm{\Pi}}
\newcommand{\bigsum}{\ensm{\Sigma}}

\newcommand{\aloctyp}{\ensm{\ttaloc}}
\newcommand{\amaptyp}[2]{\ensm{#1\tsp#2\tsp\ttamap}}

\newcommand{\alog}{\ensm{\phi}}
\newcommand{\auqonlylog}{\ensm{\alog_{\forall\:^*}}}
\newcommand{\auqlog}{\ensm{\alog_{\forall\:^*\exists\:^*}}}
\newcommand{\aeqlog}{\ensm{\alog_{\exists\:^*}}}
\newcommand{\aor}{\ensm{\vee}}
\newcommand{\aand}{\ensm{\wedge}}
\newcommand{\aimplies}{\ensm{\rightarrow}}
\newcommand{\aeq}{\ensm{=}}

\newcommand{\adtyp}{\ensm{\atyp_{\ttadt}}}
\newcommand{\casptyp}{\ensm{\atyp_{\ttcasp}}}
\newcommand{\loctyp}{\ensm{\atyp_{\ttloc}}}

\newcommand{\abitctor}{\ensm{\mathtt{Bitv}}\xspace}
\newcommand{\acptrctor}{\ttcPtr}
\newcommand{\abitvectortyp}{\ensm{\mathtt{bitvector}}}
\newcommand{\eMatch}[2]{\ensm{\ttmatch\tsp#1\tsp\ttwith\tsp#2}}
\newcommand{\ector}[1]{\ensm{\ttctor\tsp#1}}
\newcommand{\equant}{}
\newcommand{\eForall}[3]{\ensm{\ttforall\tsp #1\in #2. #3}}
\newcommand{\eExists}[3]{\ensm{\ttexists\tsp #1\in #2. #3}}

\newcommand{\tLet}[3]{\ensm{\ttlet\tsp\oftyp{#1}{#2}=#3}}
\newcommand{\tDef}[3]{\ensm{\ttdef\tsp{#1}\tsp{#2}=#3}}
\newcommand{\tDefs}[3]{\ensm{\ttdefs\tsp{#1}\tsp{#2}=#3}}

\renewcommand{\aprimitivetyp}{\atyp_{\mathit{base}}}
\newcommand{\afunctiontyp}{\atyp_{\mathit{func}}}
\newcommand{\aeffectfunctyp}{\atyp_{\mathit{proc}}}
\newcommand{\aregsettyp}{\atyp_{\mathit{regs}}}
\newcommand{\amemorytyp}{\atyp_{\mathit{mem}}}
\newcommand{\afuntyp}{\funtyp{\sseq{\cbasetypi}}{\aprimitivetyp}}
\newcommand{\abvtyp}{\bvtyp{\!}}
\newcommand{\awvectyp}[1]{\ensm{#1\ \ttwvec}}
\newcommand{\awptrtyp}[1]{\ensm{#1\ \ttwptr}}
\newcommand{\abvstyp}{\bvstyp{\!}}
\newcommand{\abvltyp}{\bvltyp{\!}}
\newcommand{\ttaset}{\abvstyp}
\newcommand{\aoption}{\ttC\ \vert\ \ttx}

\newcommand{\avectyp}{\mathtt{mem}}

\newcommand{\aderef}[1]{\ttderef(#1)}


\newcommand{\atenv}{\tenv}
\newcommand{\avenv}{\venv}
\newcommand{\agenv}{\genv}
\newcommand{\arholds}{\ensm{\vdash_a}\xspace}
\newcommand{\arenv}{\atenv,\agenv,\avenv}

\newcommand{\aenvrewrite}[4]{\ensm{#1\arholds#2\goesto_a #3\tholds #4}}
\newcommand{\arewrite}[2]{\ensm{\aenvrewrite{\arenv}{#1}{\tenv,\genv,\venv}{#2}}}

\newcommand{\kenv}{\ensm{\Sigma}\xspace}
\newcommand{\Kdeclxx}[3]{\ensm{#1 \tholds #2 \ttthen #3}}
\newcommand{\KdeclAA}[1]{\Kdeclxx{\kenv}{#1}{\kenv}}
\newcommand{\KdeclAB}[1]{\Kdeclxx{\kenv}{#1}{\kenv'}}
\newcommand{\KdeclBC}[1]{\Kdeclxx{\kenv'}{#1}{\kenv''}}
\newcommand{\KdeclAC}[1]{\Kdeclxx{\kenv}{#1}{\kenv''}}
\newcommand{\Kmachinex}[2]{\ensm{\tholds #1 \ttthen #2}}
\newcommand{\KmachineZ}[1]{\Kmachinex{#1}{\kenv_0}}
\newcommand{\KmachineA}[1]{\Kmachinex{#1}{\kenv}}
\newcommand{\KmachineB}[1]{\Kmachinex{#1}{\kenv'}}

\subsection{Notation}
In the abstract syntax, type judgments, and semantics judgments we use
italics for metavariables (e.g., $v$) and for words corresponding
to types in the abstract syntax (e.g., $\mathit{declaration}$).
We use \texttt{typewriter} font for words that correspond to language
keywords.
The notation $\overline{\alpha_i}$ means ``a sequence of one or more
$\alpha$, each to be referred to elsewhere as $\alpha_i$''.
If there are no references outside the overbar, the $i$ subscript may
be left off.
Epsilon ($\epsilon$) appearing in syntax represents an empty
production.
The notation \texttt{"}$\ldots$\texttt{"} represents a string literal
with arbitrary contents.

Bitvectors (machine integers) may be any width greater than zero.
Bitvector \emph{constants} are represented as \bitconstprefix{\ttC},
which can be thought of as an explicit sequence of zeros and
ones.
The number of bits in a bitvector constant (that is, the number of
digits) gives its type.
Thus, \bitconstprefix{00} and \bitconstprefix{0000} are different.
In the concrete syntax, bitvector constants whose size is a multiple
of 4 can also be written in the form \hexconst, where \ttC is a hexadecimal constant.
These are desugared in the parser and not shown further in this
document.

The \casp and \ale syntaxes are disjoint.
Some elements are the same in each, but these are specified
separately regardless.
They use the same metavariables as well, which should not be mixed;
any language construct in a judgment should be all \casp or all
\ale.
In a few places mixing is needed, in which case the translation
defined in Section \ref{sec:ale_exec} is applied to allow inserting
\ale fragments into \casp terms.
The \ale rules in Section \ref{sec:ale_check} do use the same
\emph{environments} as the \casp rules.
These should be construed as holding only \casp elements.
Further details can be found in Section \ref{sec:ale_check}.


\subsection{\casp{} Overview}
\label{sec:lang}
This section covers the abstract syntax for \casp.

\casp is a register-transfer-list style language:
it models instructions as non-Turing-complete procedures that update a
machine state.
Its executable semantics are covered in Section \ref{sec:exec}.

We model the machine from the assembly-language programmer's perspective.
In particular, we do not treat memory as a huge block of address space
but instead treat it in small chunks passed in from somewhere else.
We model both control registers and general-purpose
registers as well as other machine state such as whether interrupts are enabled.

Furthermore we must allow assembler labels, which have addresses,
but those addresses are not resolved until after programs are compiled
and linked and must be treated as abstract.

\begin{figure*}
\centering
\begin{minipage}{0.50\textwidth}
\begin{align*}
	& & \textbf{(\casp Types)} \\
	& \atyp \bnfeq & \cbasetyp \bnfor \cmemtyp \bnfor \cfuntyp \\
	& & \\
	& \cbasetyp \bnfeq & \ttunit \bnfor \ttint \bnfor \ttbool \bnfor \ttstring \bnfor
	\tttypeid \bnfor \bvaltyp{\ttC} \\
	& \bnfor & \cregtyp \bnfor \cregstyp \bnfor \clbltyp\\
	& \cregtyp \bnfeq & \bvltyp{\ttC} \\
	& \cregstyp \bnfeq & \bvstyp{\ttC} \\
	& \clbltyp \bnfeq & \labeltyp{\ttC} \\  
	& & \\
	& \cmemtyp \bnfeq & \memtyp{\ttC_1}{\ttC_2}{\ttC_3} \\
	& \cfuntyp \bnfeq & \funtyp{\sseq{\cbasetypi}}{\cbasetyp} \\
	& & \\
	& & \textbf{(\casp{} Operators)} \\
	& \itunop \bnfeq & - \bnfor \bitsub\ \bnfor \neg \bnfor \bitnot \\\ 
	& \itbinop \bnfeq & =\ \bnfor\ \neq\ \bnfor\ + \bnfor - \bnfor * \bnfor / \ \bnfor\ <\ \bnfor\ <=\ \bnfor\ >\ \bnfor\ >= \\
	& \bnfor & \&\& \bnfor || \bnfor \xor \\
	& \bnfor & >> \bnfor {>>}_S \bnfor << \bnfor \bitand \bnfor \bitor \bnfor \bitxor \\
	& \bnfor & \bitadd \bnfor \bitsub \bnfor \bitmul \bnfor \bitdiv \\
	& \bnfor & \bitlt \bnfor \bitle \bnfor \bitgt \bnfor \bitge \\
	& \bnfor & \bitslt \bnfor \bitsle \bnfor \bitsgt \bnfor \bitsge\\
	& \bnfor & \cup \bnfor \cap \bnfor \subseteq \bnfor \setminus
\end{align*}
\end{minipage}
\hfill\vline\hfill
\begin{minipage}{0.48\textwidth}	
\begin{align*}
	& & \textbf{(\casp Values)} \\
	& \tildev \bnfeq & \ttv \bnfor \ttassertfalse \\
	& \ttv \bnfeq & \tttrue \bnfor \ttfalse \bnfor \ttC \bnfor \bitconst
	\bnfor \text{\strlit{}} \\
	& \bnfor & \ttr \bnfor \ptrform{\ttmemid}{\ttC} \\
	& & \\
	& & \textbf{(\casp{} Expressions)} \\
	& \tte \bnfeq & \tildev \bnfor \ttx \\
	& \bnfor & \strof{\tte} \\
	& \bnfor & \fapp{\ttfuncid}{\sseq{\tte}} \\
	& \bnfor & \itunop\ \tte\ \\
	& \bnfor & \tte_1\ \itbinop\ \tte_2\\
	& \bnfor & \bitidx{\tte}{\ttC} \bnfor \bitfield{\tte}{\ttC_1}{\ttC_2} \\
	& \bnfor & \eLet{\ttx}{\cbasetyp}{\tte_1}{\tte_2} \\ 
	& \bnfor &\sITE{\tte_1}{\tte_2}{\tte_3} \\ 
	& \bnfor & \ptrform{\ttmemid}{\tte} \\
	& \bnfor & \regderef{\tte} \bnfor \memread{\tte}\\
	& \bnfor & \NEW{\texttt{\ebranchto{\ttextid}}} \\
	& \bnfor & \setlit \\
	& \bnfor & \sizeofset{\tte} \bnfor \setmemberof{\tte_1}{\tte_2}
\end{align*}
\end{minipage}
\caption{\casp types, values, operators, and expressions.\label{fig:tr_casp_types1}}
\end{figure*}

\mypara{Notation}
We use the following metavariables:

\begin{tabular}{llr}
\ttx, \tty, \ttz & Program variables (binders) \\
\ttr            & Registers (abstract) \\
\ttC            & Integer constants (written in decimal) \\
\bitconst       & Bitvector constants (written in binary) \\
\atyp           & Types \\
\ttv            & Values \\
\tte            & Expressions \\
\ttS            & Statements \\
\tti, \ttj      & Rule-level integers \\
\end{tabular}

(Other constructs are referred to with longer names.)

A number of constructs are lists (with a
null case and a cons case) -- these are written out in longhand so
that typing and semantic judgments can be applied explicitly to each
case, in order to, for example, propagate environment updates correctly.

\mypara{Identifiers and Variables}
Identifiers are divided syntactically into \NNEW{seven} categories:
\begin{itemize}
\item \ttmemid are identifiers bound to memory regions, which are
second-class.
\item \ttfuncid are identifiers bound to functions, which are second-class.
\item \ttprocid are identifiers bound to procedures, which are second-class.
\item \ttxop are identifiers bound to instructions (``operations''), which
are akin to procedures but distinguished from them.
\item \tttypeid are identifiers for type aliases, which are bound to base types 
in declarations.
\item \ttxmoduleid are the names of ``modules'', which are used to select
among many possible groups of lowering elements.
\item Other identifiers \ttx are used for other things, and should be
assumed to \emph{not} range over the above elements.
\end{itemize}
Note that all identifiers live in the same namespace, and rebinding or
shadowing them is not allowed.
All these identifiers can be thought of as variables, in the sense
that they are names that stand for other things.
All of them are immutable once defined, including the ordinary
variables \ttx that contain plain values.

\mypara{Types}
Types are divided syntactically into base types (integers, booleans,
strings, bitvectors, etc.) and others (memory regions and functions).
Functions may handle only base types.
Furthermore, memory regions and functions are
second-class for reasons discussed below and are excluded in various
places in the syntax and the typing rules.

We use index typing to capture the bit width of values.

\mypara{Registers}
Registers are represented in the specification with the metavariable
$r$, which stands for the underlying abstract identity of a register.
Declaring a register, e.g., with 
$\ttlet\ttstate\ \ttx\ \ttcol\ \bvltyp{\ttC}$
as shown in \autoref{fig:tr_casp_types2},
allocates a fresh register $r$ and binds the variable \ttx to it.
A subsequent declaration of the form
$\ttlet\ \tty\ \ttcol\ \bvltyp{\ttC} = \ttx$
creates another variable \tty that refers to the same underlying
register.
One might think of registers as numbered internally.
We use the form $\ttlet\ttstate\ \ttcontrol\ \ttx\ \ttcol\ \bvltyp{\ttC}$ 
to declare specific \textit{control} registers, which are treated
differently by the framing rules.
The additional keyword \texttt{dontgate} inhibits state gating for the
register; this should be used for flags registers and anything similar
that is implicitly used by all ordinary code.

Some registers
have associated with them a
text form, which is declared
separately and is the form an assembler expects to parse.
The \synth uses this to extract an assembly program from \casp's
internal representation.
It is referred to by attaching the suffix \strof{} to the/a register
variable.
As some registers are not
directly addressable by the assembler (e.g., they might be subfields of some
larger addressable unit or non-addressable internal state), not all
registers have a text form.
This is not readily checked statically without additional information,
so invalid \strof{}
references fail at assembly code extraction time.

The type of a register is \bvltyp{\ttC}, which is a register that holds
a \ttC-bit bitvector.
The bitvector value in question can be updated by assigning a new
value; this is a statement ($\slAssign{\tte_1}{\tte_2}$) and can only
happen in places where
statements are allowed.
The construct $\regderef{\tte}$ reads a register.

The reader will note that the semantics rules for machines and
declarations do not provide initial values for registers.
Instead, executions are defined in terms of some initial register
state (and also some memory state), which is required to have the
right registers to match the machine definition.
This allows reasoning about the execution of programs and program
fragments in terms of many or all possible initial states.
These issues are discussed further below.

\mypara{Memory}
A memory region has the type \memtyp{\ttC_1}{\ttC_2}{\ttC_3}, shown in \autoref{fig:tr_casp_types1}.
This refers to a memory region that has $\ttC_2$ cells, each of which
stores a bitvector of width $\ttC_1$.
This memory region is addressed with pointers of width $\ttC_3$.
Note that we assume byte-addressed machines, and for the purposes of
both this specification and our implementation, we assume bytes are 8
bits wide.
(This restriction could be relaxed if we wanted to model various
historic machines.)
Thus a memory region of type \memtyp{32}{4}{32} has 4 32-bit values in
it, which can be addressed at byte offsets 0, 4, 8, and 12.
Memory regions and registers are mutable state.

Memory regions are named with identifiers.
These names, and memory regions themselves, are not first class;
variables are not allowed to range over them.
Also note that memory regions are a property of programs (and thus are
declared in specifications) and not a property of the machine as a
whole.

\mypara{Pointers}
A pointer literal has the form \ptrform{\ttmemid}{\ttC}, in which
\ttmemid is the name of a memory region and \ttC is the offset, shown in \autoref{fig:tr_casp_types1}.
Because memory regions are second-class, \ttmemid must be one of the
available declared memory regions.
Pointer literals exist in the abstract syntax, but are not allowed in
the concrete syntax except in specifications.
The only way to get a pointer value is to look up a label (discussed
immediately below) or have it provided in a register as part of the
initial machine state.

A pointer literal is treated as a bitvector of the width appropriate for pointers to the memory region, so one
can appear in a register or in memory.
However, we enforce a restriction (not captured in the semantic
judgments so far) that no value in the initial machine state, whether in a
register or in memory, is a pointer unless required to be so by the
precondition part of the specification.
All other values are restricted to be plain bitvector values.

Addition and subtraction are allowed on pointers and they change the offset, 
but other bitvector operations (e.g., multiply) are disallowed
and fail.
Similarly, attempting to fetch from or store to a plain bitvector that
is not a pointer fails.
Note however that we do not statically distinguish pointers and plain
bitvectors.
(We could have used flow-sensitive typing to reason about when
registers and memory cells contain pointers and when they do not; but
this adds substantial complexity and for our problem domain does not
provide significant value.)
Instead, we step to failure at runtime.
This can be seen in the semantic judgments.

Fetching from a pointer takes the form \memread{\tte}.
Storing to a pointer takes the form \smAssign{\tte_1, C}{\tte_2}.
The extra constant \ttC specifies the width of the cell pointed to.
(This is not an offset.)
Because we do not check pointers statically, we do not know the
memory region being pointed to and cannot look up its cell size; thus
we need the width explicitly for typing.
It is checked at runtime.

\begin{figure*}
\centering
\begin{minipage}{0.45\textwidth}
\begin{align*}
	& & \textbf{(\casp{} Statements)} \\
	& \ttS \bnfeq & \ttS\ttsemi\ \ttS \\
	& \bnfor & \papp{\ttprocid}{\sseq{\tte}} \\
	& \bnfor & \sLet{\ttx}{\cbasetyp}{\tte}{\ttS} \\
	& \bnfor & \sFor{\ttx}{\ttC_1}{\ttC_2}{\ttS} \\
	& \bnfor & \sITE{\tte}{\ttS_1}{\ttS_2} \\
	& \bnfor & \slAssign{\tte_1}{\tte_2} \\
	& \bnfor & \smAssign{\tte_1, C}{\tte_2} \\
	& \bnfor & \NEW{\sBRANCH{\tte}} \\
	& \bnfor & \ttassert(\tte) \\
	& \bnfor & \ttskip \\
	& \bnfor & \bferr
\end{align*}
\end{minipage}
\hfill\vline\hfill
\begin{minipage}{0.54\textwidth}
\begin{align*}
	& & \textbf{(\casp{} Declarations)} \\
	& \decls \bnfeq\ & \eps\ \bnfor \decl \ttsemi\ \decls \\
	& \decl \bnfeq & \tttype\ \tttypeid = \cbasetyp \\
	& \bnfor & \ttlet\ \ttx\ \ttcol\ \cbasetyp = \tte \\
	& \bnfor & \ttlet\ \strof{\ttx} = \text{\tte} \\
	& \bnfor & \ttdef\ \ttfuncid\ \funtyp{\sseq{\ttx_i\ \ttcol\ \cbasetypi}}{\cbasetyp} = \tte \\
	& \bnfor & \ttproc\ \ttprocid\ \proctyp{\sseq{\ttx_i\ \ttcol\ \cbasetypi}} = \ttS \\
	& \bnfor & \ttlet\ttstate\ \ttx\ \ttcol\ \cregtyp \\ 
	& \bnfor & \ttlet\ttstate\ \ttcontrol\ \ttx\ \ttcol\ \cregtyp \\ 
	& \bnfor & \ttlet\ttstate\ \ttcontrol\ \ttdontgate\ \ttx\ \ttcol\ \cregtyp \\ 
	& \bnfor & \ttlet\ttstate\ \ttmemid\ \ttcol\ \cmemtyp \\
	& \bnfor & \ttlet\ttstate\ \ttmemid\ \ttcol\ \cmemtyp\ \ttwith\ \NNEW{\ttx} \\
	& & \\
	& \defops \bnfeq\ & \eps \bnfor \defop \ttsemi\ \defops \\
	& \defop \bnfeq\ & \ttdefop\ \ttxop\  \ttinbrace{\tttxt=\tte\ttcomma\ttsem=\ttS} \\
	& \bnfor & \ttdefop\ \ttxop\ \sseq{\ttx_i\ \ttcol\ \cbasetypi}\ \ttinbrace{\tttxt=\tte\ttcomma\ttsem=\ttS}
\end{align*}
\end{minipage}
\caption{\casp statements and declarations.\label{fig:tr_casp_types2}}
\end{figure*}

\mypara{Labels}
As mentioned above, the term ``label'' means an assembler label or linker
symbol.
These are constant addresses that are not known at assembly time, so
we must model them abstractly.

When one declares a memory region, one may attach a label to it, which
is an additional identifier.
This identifier is created as a variable of type \labeltyp{\ttC} in \autoref{fig:tr_casp_types1}.
The value is a pointer to the first entry in the region, and a single
type subsumption rule allows this value to be accessed and placed in
an ordinary register or variable of suitable bitvector type.
The intended mechanism is that for each machine the preferred
instruction on that machine for loading assembler symbols into a
register can be defined to take an operand of type \labeltyp{\ttC},
and then its value can be assigned to the destination register.
This type restriction on the operand is sufficient to synthesize
programs that use labels correctly.

\mypara{\NEW{Code Positions and Branching}}
\casp code blocks may contain branches, either forward within the
block (branching backwards is forbidden) or to a single external
assembler label outside the block.
This model is sufficient for our block-oriented synthesis approach;
more complex control flow can be handled with multiple blocks.

Consequently, a branch may either go to the single external assembler label
(which terminates execution of the current block) or skip zero or more
instructions.
We model branch targets as an 8-bit skip count.
In \autoref{fig:tr_casp_types2}, the statement \sBRANCH{\bitconst{}} skips \bitconst{} instructions;
\sBRANCH{\texttt{0xff}} jumps to the external assembler label.
This statement may be used to define both conditional and
unconditional branch instructions.
Such instructions should be defined to take an operand of type
\bvtyp{8} to choose the branch destination.
This magic number should then be printed to the output assembly text
using the built-in function \texttt{textlabel}, which replaces it with
a valid assembler label, either the selected external label string or
a scratch label attached to the proper destination instruction.

Specifications do not need to be directly concerned with internal
branches, which occur or not as needed.
However, external branches are part of a block's specification;
typically the purpose of a block with an external branch is to test
some condition and then either branch away or fall through to the next
block.
It is thus necessary to be able both to name the external label to
use and to specify the conditions when it should be reached.
For this purpose a predicate \texttt{branchto} is provided.
It may appear in the postcondition and governs the exit
path from the block: if forced to true, the block branches to the
external assembler label, and if false, the block falls through from
its last instruction.
The concrete syntax is
\texttt{branchto(dest)}
which also sets the assembler label used to \texttt{dest}.
It is not valid to name more than one such assembler label.

Note that the assembler labels used in branching are, though also
assembler labels, a separate mechanism unrelated to the labels
attached to memory regions; they are code labels rather than data
labels and inherently work differently.


\mypara{Register Sets}
Register sets are second-class elements intended to exist only as
literals and only as the result of lowering machine-independent
specifications that cannot directly talk about specific registers.
Currently they do not exist in the implementation.
Register sets are not allowed to be operands to instructions to avoid state
explosions when synthesizing.
For simplicity, this restriction is not shown in the abstract syntax or
typing rules.

\mypara{Functions and Procedures}
Functions, defined with \ttdef in \autoref{fig:tr_casp_types2}, are pure functions whose bodies are
expressions.
They produce values.
They can read registers and memory, and can fail, but cannot update
anything.
Procedures, defined with \ttproc, are 
impure and their bodies are statements.
They do not produce values, but they may update the machine state.
They are otherwise similar, and are intended to be used to abstract
out common chunks of functionality shared among multiple instructions
in machine descriptions.
Functions can also be used for state hiding in specifications.

Functions and procedures are second-class; they may be called only by
their own name and may not be bound to variables or passed around.
Furthermore, they are only allowed to handle base types: higher-order
functions are explicitly not supported.

\mypara{Operations}
Operations (defined with \ttdefop in \autoref{fig:tr_casp_types2}) are
machine instructions or short sequences of machine instructions.
An operation takes zero or more operands and transforms
the machine state as defined by one or more statements.
Operands are currently defined as expressions, but are restricted as
follows:
\begin{itemize}
\item They may be values, but not string values, and not \exprfail.
\item They may be variables of register type.
\item They may be variables of label type.
\end{itemize}
This restriction affects what the synthesizer tries to generate; a
broader set of expressions may be accepted for verification or
concrete execution and simply evaluated in place.

In general, we refer to ``instructions'' and ``operations''
interchangeably.
However, there is an important distinction between them:
operations do not necessarily need to be single machine instructions.
The text output to the assembler is arbitrary and if desired can be
computed on the fly based on the operand values.
On some platforms the assembler defines ``synthetic instructions''
that potentially assemble to multiple machine instructions, or to
different sequences based on the constants or registers used.
This facility takes that a step further by allowing the writer of the
machine description to define their own synthetics.
Operations are the units in which \casp{} reasons about machine
operations and the units in which \casp{} generates programs and code
fragments.


\mypara{Other Constructs}
\bitidx{\tte}{\ttC} and \bitfield{\tte}{\ttC_1}{\ttC_2} in \autoref{fig:tr_casp_types1} extract a
single bit and a slice, respectively, from a bitvector.
The offsets are constants; if variable offsets are needed, the value can be shifted first.
The width of the slice must be constant for static typing.

\begin{figure*}
\centering
\begin{minipage}{0.49\textwidth}
\begin{align*}
	& & \textbf{(\casp{} Machines)} \\
	& \machine \bnfeq & \decls\ttsemi\ \defops \\
	& & \\
	& & \textbf{(\casp{} Programs)} \\
	& \itinst \bnfeq & \ttxop \bnfor \ttxop\ \sseq{\tte}\\
	& \itinsts \bnfeq & \eps \bnfor \itinst \ttsemi\ \itinsts \\
	& \program \bnfeq & \itinsts
\end{align*}
\end{minipage}
\hfill\vline\hfill
\begin{minipage}{0.49\textwidth}
\begin{align*}
	& & \textbf{(\casp{} Lowerings)} \\
	& \itmodules \bnfeq & \eps \bnfor \itmodule \ttsemi\ \itmodules \\
	& \itmodule \bnfeq & \ttmodule\ \ttxmoduleid\ \ttinbrace{\ttdecls \ttsemi\ \ttframe} \\
	& & \\
	& & \textbf{(\casp{} Specifications)} \\
	& \ttframe \bnfeq & \ttfwrites\ \ttcol\ \sseq{\ttregidi} \\
	& \bnfor & \ttfmwrites\ \ttcol\ \sseq{\ptrform{\ttmemidi}{\tte_i}}\\
	& \ttframes \bnfeq & \eps \bnfor \ttframe\ \ttframes \\
	& \ttpre \bnfeq & \tte \\ 
	& \ttpost \bnfeq & \tte \\
	& \itspec \bnfeq & \ttdecls \ttsemi\ \ttframes \ttsemi\ \ttpre \ttsemi\ \ttpost
\end{align*}
\end{minipage}
\caption{\casp machines, lowerings, programs, and specifications.\label{fig:tr_casp_types3}}
\end{figure*}


\mypara{Machines, Lowering, Specifications, and Programs}

A \machine~ is the full description of a machine architecture; it
includes declarations (including types, constants, registers,
functions and procedures) and also instructions, shown in \autoref{fig:tr_casp_types3}.
This is typically a standalone file but may be a set of files referenced via
\ttinclude.

A (single) \lowering is a collection of declarations used to instantiate
elements in \ale translations, shown in \autoref{fig:tr_casp_types3}.
These are placed into a \itmodule, with multiple modules per file, so
that the \lowerings associated with multiple related \ale
specifications can be kept together.
The \ttimport \ttxmoduleid directive enables sharing common elements in
one module \ttxmoduleid across multiple specifications.
The modules used to lower a specification are selected using the
\ttalower declaration in \ale.

A \itspec (specification) is a precondition and postcondition, which are boolean
expressions, along with optional permission to modify additional
registers (the \ttframe), shown in \autoref{fig:tr_casp_types3}.
\casp specifications are produced by compiling \ale specifications.
Note that a module can also contain frame declarations; they are
added to any frame conditions provided in the \ale specification.
A code block is permitted to modify any register that is either
explicitly listed in the frame declarations or mentioned in the
postcondition, while it may read any register mentioned in the
precondition and any control register.
This restriction is currently not adequately captured in the
semantics rules.

A \program\ is a sequence of instruction invocations, shown in \autoref{fig:tr_casp_types3}.

\mypara{Built-in Functions}
Here is a partial list of the built-in functions in \casp{} along with their types.
\begin{itemize}
	\item \texttt{empty} : \funtyp{\ttint}{\bvstyp{\ttC}} produces an empty register set of 
	bit size $\ttC$, where $\ttC$ is the value of the first argument.
	Note that this built-in 
	function is dependently	typed and treated specially during 
	type checking. The first argument must be a constant.
	\item \texttt{hex} : \funtyp{\bvtyp{\ttC}}{\ttstring} prints numbers in hexadecimal.
	\item \texttt{bin} : \funtyp{\bvtyp{\ttC}}{\ttstring} prints numbers in binary.
	\item \texttt{dec} : \funtyp{\bvtyp{\ttC}}{\ttstring} prints numbers in decimal.
	\item \texttt{lbl} : \funtyp{\labeltyp{\ttC}}{\ttstring}
	prints labels (it returns the label identifier as a string).
	This is for data labels attached to memory locations.
	\item \NEW{\texttt{textlabel} : \funtyp{\bvtyp{8}}{\ttstring}}
	prints branch offsets as assembler labels.
	This is for code labels, as described above.
	\item \texttt{format} : \funtyp{\ttstring}{\funtyp{\ttstring\
	$\ldots$}{\ttstring}} formats strings.
	The first argument is a format string; the remainder of the
	arguments are substituted into the format string where a
	dollar sign appears followed by the argument number (1-based).
	(A literal dollar sign can be inserted by using
	\texttt{\$\$}.)
	The number of additional arguments expected is deduced from
	the contents of the format string.
	\item \texttt{bv\_to\_len} :
	\funtyp{\ttint}{\funtyp{\bvtyp{\ttC_2}}{\bvtyp{\ttC_1}}}
	returns a new bitvector of size $\ttC_1$ (where $\ttC_1$ is
	the value of the first argument) with the same value
	as the second argument, 
	up to the ability of the new size to represent that value.
	Note that this built-in 
	function is dependently	typed and treated specially during 
	type checking. The first argument must be a constant.
	\item \texttt{bv\_to\_uint} : \funtyp{\bvtyp{\ttC_1}}{\ttint}
	converts a bitvector to unsigned int.
	\item \texttt{uint\_to\_bv\_l} : \funtyp{\ttint\
	}{\funtyp{\ttint}{\bvtyp{\ttC_1}}} converts an
	unsigned int (second argument) into a bitvector of size $\ttC_1$, where $\ttC_1$ 
	is the value of the first argument. Note that this built-in 
	function is dependently	typed and treated specially during 
	type checking. The first argument must be a constant.
	\item \texttt{isptr} : \funtyp{\bvtyp{\ttC}}{\ttbool} tests
	at runtime if a bitvector-typed value is a pointer.
\end{itemize}

Note that some of these functions have their own typing rules, some of
which are polymorphic in bitvector size.
We have not complicated the typing rules presented by including all of
these as special cases.

\mypara{Concrete Syntax}
We do not describe the concrete syntax in detail here; however,
it does not stray very far from the abstract syntax.
The operator precedence and most of the operator spellings are taken
from C but most of the rest of the concrete syntax is ML-style.
There are also a few things desugared in the parser and not shown in the
abstract syntax.
As already mentioned, bitvector constants whose size is a multiple of 4 can
also be written in the form \hexconst.
Syntax of the form \hexof{e}, \binof{e}, and \decof{e} is
converted to the built-in functions \tthex, \ttbin, \ttdec
respectively.
These print either integers or bitvectors as strings in hexadecimal,
binary, or decimal respectively.
The syntax \lblof{\NNEW{\ttx}} is similarly converted to the built-in
function \ttlbl.
This produces the label (that is, the identifier naming the label) as
a string.
Further the concrete syntax supports include files via an \ttinclude
directive, which is useful for sharing common elements between related
descriptions.

\subsection{\casp{} Static Typing}
\label{sec:check}

\begin{figure*}
\textbf{(Type Well-Formedness)}\\
\begin{tabular}{cccc}
\irule{~}{\wfenv\twfholds\ttunit} 	 &  \irule{~}{\wfenv\twfholds\ttint} &
\irule{~}{\wfenv\twfholds\ttbool}	&  \irule{~}{\wfenv\twfholds\ttstring}
\end{tabular}
\nextrulesmall
\begin{tabular}{cccc}
\irule{\ttC > 0}
	{\wfenv\twfholds\bvaltyp{\ttC}}
&  
\irule {\ttC > 0}
	{\wfenv\twfholds\bvltyp{\ttC}}
&
\irule {\ttC > 0}
	{\wfenv\twfholds\labeltyp{\ttC}}
& 
\irule{\ttC > 0}
	{\wfenv\twfholds\bvstyp{\ttC}}
\end{tabular}
\nextrulesmall
\begin{tabular}{ccc}
\irule
	{\ttC_1 > 0,\ \ttC_2 > 0,\ \ttC_3> 0}
	{\wfenv\twfholds\memtyp{\ttC_1}{\ttC_2}{\ttC_3}}
&
\irule{\genv(\ttx)=\atyp\nextclause\\ 
	\wfenv\twfholds\atyp}
	{\wfenv\twfholds\ttx}
& 
\irule
	{\forall i, \wfenv\twfholds\atyp_i \nextclause\\ 
	\wfenv \twfholds \atyp_r}
	{\wfenv\twfholds\funtyp{\sseq{\atyp_i}}{\atyp_r}}
\end{tabular}
\caption{\casp type well-formedness.\label{fig:tr_casp_type_wf}}
\end{figure*}




This section describes the \casp type system.

\mypara{Environments}
The type system uses two environments: $\genv$ maps type alias names
to their expansions, and $\tenv$ maps variable names to types.
Recall from the syntax that type alias names may only expand to base
types; thus
type alias names can be treated as base types.

\mypara{Well-Formedness}
Since types syntactically include type alias names, we need to check
the validity of those names.
We also insist that the widths of bitvectors be greater
than zero.
The judgment for this has the form $\wfenv \twfholds \atyp$, shown in \autoref{fig:tr_casp_type_wf}.
There is an intended invariant that only well-formed types may be
entered into the variable typing environment $\tenv$, so that types
taken out of it do not need to be checked for well-formedness again.

In a typing environment comprised of $\genv$ mapping user-defined type
names (type aliases) to types and $\tenv$ mapping variables to types,
we say that a type is well formed when all type names exist and refer
to well-formed types, and all indices are of type $\ttint$ and positive.

\begin{figure*}
\textbf{(Expression Typing)}\\
\begin{tabular}{cccc}
\irule{~}
		{\typenv\tholds\hastyp{\ttC}{\ttint}}
&
\irule{~}
		{\typenv\tholds\hastyp{\text{\strlit{}}}{\ttstring}}
&
\irule{~}
		{\typenv\tholds\hastyp{\tttrue}{\ttbool}}
&
\irule{~}
		{\typenv\tholds\hastyp{\ttfalse}{\ttbool}}
\end{tabular}
\nextrulesmall
\begin{tabular}{cccc}
\irule
	{\ttC \in {\{0,1\}}^\ttk}
	{\typenv\tholds\hastyp{\bitconst}{\bvtyp{\ttk}}}
&
\irule
	{\wfenv\twfholds\atyp}
	{\typenv \tholds\hastyp{\ttassertfalse}{\atyp}}
&
\irule
	{\tenv(\ttx) = \atyp\nextclause \\ 
	\wfenv\twfholds\atyp}
	{\typenv\tholds\hastyp{\ttx}{\atyp}}
&
\irule
{ \typenv \tholds \tte\ \ttcol\ \cregtyp}
{\typenv\tholds\hastyp{\strof{\tte}}{\ttstring}}
\end{tabular}
\nextrulesmall
\begin{tabular}{cc}
\irule
	{\typenv\tholds\hastyp{\ttfuncid}{\left( \funtyp{\sseq{\cbasetypi}}{\cbasetyp} \right)}\nextclause \\
		\forall \tti,\ \typenv\tholds\hastyp{\tte_i}{\cbasetypi}}
	{\typenv\tholds\hastyp{\fapp{\ttfuncid}{\sseq{\tte_i}}}{\cbasetyp}}
&
\irule
	{\typenv\tholds\hastyp{\tte}{\cbasetypone} \nextclause \\
		\dholds \hastyp{\itunop}{\funtyp{\cbasetypone}{\cbasetyptwo}} }
	{\typenv\tholds\hastyp{\itunop\ \tte}{\cbasetyptwo}}
\end{tabular}

\nextrulesmall
\irule
	{\typenv\tholds\hastyp{\tte_1}{\cbasetypone}\nextclause \\ 
		\typenv\tholds\hastyp{\tte_2}{\cbasetypone} \nextclause \\
		\dholds \hastyp{\itbinop}{\funtyp{\funtyp{\cbasetypone}{\cbasetypone}}{\cbasetyptwo}} }
	{\typenv\tholds\hastyp{\tte_1\ \itbinop\ \tte_2}{\cbasetyptwo}}

\nextrulesmall
\begin{tabular}{cc}
\irule
	{\typenv\tholds\hastyp{\tte}{\bvtyp{\ttC_2}}\nextclause \\ 
		0 \le \ttC_1 < \ttC_2 }
	{\typenv\tholds\hastyp{\bitidx{\tte}{\ttC_1}}{\bvtyp{1}}}
&
\irule
	{\typenv\tholds\hastyp{\tte}{\bvtyp{\ttC_3}}\nextclause \\ 
		0 \le \ttC_1 < \ttC_2 \le \ttC_3 \nextclause \\
		\ttk = \ttC_2-\ttC_1}
	{\typenv\tholds\hastyp{\bitfield{\tte}{\ttC_1}{\ttC_2}}{\bvtyp{\ttk}}}
\end{tabular}

\nextrulesmall
\irule
	{\typenv\tholds\hastyp{\tte_1}{\cbasetyp}\nextclause \\ 
		\ttx\notin\typenv\nextclause \\ 
		\genv, \tenv[\ttx\envgoesto\cbasetyp]\tholds\hastyp{\tte_2}{\atyp_2} }
	{\typenv\tholds\hastyp{\eLet{\ttx}{\cbasetyp}{\tte_1}{\tte_2}}{\atyp_2}}
	
\nextrulesmall
\irule
	{\typenv\tholds\hastyp{\ttb}{\ttbool}\nextclause \\ 
		\typenv\tholds\hastyp{\tte_{1}}{\atyp} \nextclause \\ 
		\typenv\tholds\hastyp{\tte_{2}}{\atyp}}
	{\typenv\tholds\hastyp{\sITE{\ttb}{\tte_1}{\tte_2}}{\atyp}}

\nextrulesmall
\begin{tabular}{cc}
\irule
{\typenv\tholds\hastyp{\tte}{\ttint}\nextclause \\ 
	\typenv\tholds\hastyp{\ttmemid}{\memtyp{\_}{\_}{\ttC}}}
{\typenv\tholds\hastyp{\ptrform{\ttmemid}{\tte}}{\cptrtyp{\ttC}}}
&
\irule
	{\typenv\tholds\hastyp{\tte}{\bvaltyp{\_}}\nextclause \\ \ttC > 0}
	{\typenv\tholds\hastyp{\fetch{\tte}{\ttC}}{\bvaltyp{\ttC}}}
\end{tabular}

\nextrulesmall
\begin{tabular}{ccc}
\irule
	{\typenv\tholds\hastyp{\tte}{\bvltyp{\ttC}}}
	{\typenv\tholds\hastyp{\regderef{\tte}}{\bvaltyp{\ttC}}}
&
\irule
	{\typenv\tholds\hastyp{\NNEW{\ttx}}{\labeltyp{\ttC}}}
	{\typenv\tholds\hastyp{\NNEW{\ttx}}{\bvaltyp{\ttC}}}
&
\irule{~}
	{\NEW{\typenv\tholds\hastyp{\ebranchto{\ttextid}}{\ttbool}}}
\end{tabular}

\nextrulesmall
\begin{tabular}{ccc}
\irule
	{\forall \tti \in(1\ldots\ttk),\ \typenv\tholds \hastyp{\ttx_i}{\bvltyp{\ttC}}}
	{\typenv\tholds\hastyp{\ttinbrace{\ttx_1, \ldots, \ttx_k}}{\bvstyp{\ttC}}}
&
\irule
	{\typenv\tholds\hastyp{\tte}{\bvstyp{\ttC}}}
	{\typenv\tholds\hastyp{\sizeofset{\tte}}{\ttint}}
&
\irule
	{\typenv\tholds\hastyp{\tte_1}{\bvltyp{\ttC}}\ \ \ 
		\typenv\tholds\hastyp{\tte_2}{\bvstyp{\ttC}}}
	{\typenv\tholds\hastyp{\setmemberof{\tte_1}{\tte_2}}{\ttbool}}
\end{tabular}
\caption{\casp typing rules for expressions.\label{fig:tr_casp_type_expr}}
\end{figure*}

\mypara{Expressions}
Expressions produce values that have types.
Because types appear explicitly in some expressions (e.g., \texttt{let}),
we need both environments, so the form of an
expression typing judgment is
$\typenv\tholds\hastyp{\tte}{\atyp}$, shown in \autoref{fig:tr_casp_type_expr}.
This means that we conclude \tte has type \atyp.
Note that the \strof{} form is restricted to registers; it is
specifically for extracting the assembly text form of a register.
We have not written out a separate rule for each unary and binary
operator.
The types of operators are as shown in \autoref{fig:tr_caps_ops}.
Note that the bitvector operators are polymorphic in bit size.

\begin{figure*}
\newcommand{\funtypp}[3]{\funtyp{\funtyp{#1}{#2}}{#3}}
\newcommand{\bvC}{\bvaltyp{\ttC}}
\newcommand{\bvsC}{\bvstyp{\ttC}}
\newcommand{\allC}{$\forall \ttC,\ $}
\textbf{(Operator Expression Typing)} \\
\begin{tabular}{ccccr}
$-$      &&&		& \funtyp{\ttint}{\ttint} \\
\bitsub	 &&&		& \allC\funtyp{\bvC}{\bvC} \\
$\neg$	 &&&		& \funtyp{\ttbool}{\ttbool} \\
\bitnot	 &&&		& \allC\funtyp{\bvC}{\bvC} \\
~	 &&&		& ~ \\
$=$ & $\neq$	&&	& $\forall \cbasetyp,$ \funtypp{\cbasetyp}{\cbasetyp}{\ttbool} \\
$+$ & $-$ & $*$ & $/$		& \funtypp{\ttint}{\ttint}{\ttint} \\
$<$ & $<=$ & $>$ & $>=$		& \funtypp{\ttint}{\ttint}{\ttbool} \\
$\&\&$ & $||$ & $\xor$	&& \funtypp{\ttbool}{\ttbool}{\ttbool} \\
$>>$ & ${>>}_S$ & $<<$	&& \\
\bitand&\bitor&\bitxor	&& \\
\bitadd&\bitsub&\bitmul&\bitdiv & \allC\funtypp{\bvC}{\bvC}{\bvC} \\
\bitlt&\bitle&\bitgt&\bitge & \\
\bitslt&\bitsle&\bitsgt&\bitsge & \allC\funtypp{\bvC}{\bvC}{\ttbool} \\
$\cup$&$\cap$&$\setminus$	&& \allC\funtypp{\bvsC}{\bvsC}{\bvsC}\\
$\subseteq$			&&&& \allC\funtypp{\bvsC}{\bvsC}{\ttbool}
\end{tabular}
\caption{\casp typing rules for \itunop and \itbinop.}\label{fig:tr_caps_ops}
\end{figure*}

Arguably the right hand argument of the shift operators should be
allowed to be a different width.
There is one rule for pointer literals that covers both the expression
and the value form.
There is no rule (either in the typing or in the semantics) that
allows taking a subrange of a memory region as a new smaller region.
We have not needed this for our use cases, so for simplicity we
do not support it.

\begin{figure*}
\textbf{(Statement Typing)}
\nextrulesmall
\begin{tabular}{cc}
\irule
	{\typenv\tholds{\ttS_1}\nextclause \\ 
		\typenv\tholds{\ttS_2} }
	{\typenv\tholds{\ttS_1;\ \ttS_2}}
&
\irule
	{\typenv\tholds\hastyp{\ttprocid}{\left( \proctyp{\sseq{\cbasetypi}} \right)}\nextclause \\
		\forall \tti,\ \typenv\tholds\hastyp{\tte_i}{\cbasetypi}}
	{\typenv\tholds\papp{\ttprocid}{\sseq{\tte_i}}}
\end{tabular}

\nextrulesmall
\begin{tabular}{cc}
\irule
	{\typenv\tholds\hastyp{\tte}{\cbasetyp}\nextclause \\ 
		\ttx\notin\typenv\nextclause \\  
		\typenv[\ttx\envgoesto\cbasetyp]\tholds{\ttS} }
	{\typenv\tholds{\sLet{\ttx}{\cbasetyp}{\tte}{\ttS}}}
&
\irule
	{\ttx\notin\typenv\nextclause \\
		\typenv[\ttx\envgoesto\ttint]\tholds{\ttS}\nextclause}
	{\typenv\tholds{\sFor{\ttx}{\ttC_1}{\ttC_2}{\ttS}}}
\end{tabular}

\nextrulesmall
\begin{tabular}{cc}
\irule
	{\typenv\tholds\hastyp{\tte}{\ttbool}\nextclause \\ 
		\typenv\tholds{\ttS_{1}}\nextclause \\ 
		\typenv\tholds{\ttS_{2}}}
	{\typenv\tholds{\ \sITE{\tte}{\ttS_1}{\ttS_2}}}
&
\irule
	{\typenv\tholds\hastyp{\tte_1}{\bvltyp{\ttC}}\nextclause \\ 
		\typenv\tholds\hastyp{\tte_2}{\bvaltyp{\ttC}}}
	{\typenv\tholds\slAssign{\tte_1}{\tte_2}}
\end{tabular}

\nextrulesmall
\begin{tabular}{ccc}
\irule
	{\typenv\tholds\hastyp{\tte_1}{\bvaltyp{\ttC_1}}\nextclause \\ 
		\typenv\tholds\hastyp{\tte_2}{\bvaltyp{\ttC_2}}}
	{\typenv\tholds\smAssign{\tte_1, \ttC_2}{\tte_2}}
&
\NEW{
	\irule
	{\typenv\tholds\hastyp{\tte}{\bvtyp{8}}}
	{\typenv\tholds{\sBRANCH{\tte}}}
}
&
\irule
	{\typenv\tholds\hastyp{\tte}{\ttbool}}
	{\typenv\tholds{\ttassert(\tte)}}
\end{tabular}
	
\begin{tabular}{cc}
\irule{~}
		{\typenv\tholds{\ttskip}}
&
\irule{~} 
		{\typenv\tholds{\bferr}}
\end{tabular}
\caption{\casp typing rules for statements.\label{fig:tr_casp_type_stmt}}
\end{figure*}

\mypara{Statements}
Statements do not produce values.
We still need both environments, though, so the form of a typing
judgment for a statement is $\typenv\tholds{\ttS}$, shown in \autoref{fig:tr_casp_type_stmt}.
This means that \ttS is well typed.

\begin{figure*}
\textbf{(Declaration Typing)}

\nextrulesmall
\begin{tabular}{cc}
\irule {~}
		{\genv, \tenv\tholds \eps \ttthen \genv, \tenv}
&
\irule 
	{\genv, \tenv\tholds \decl \ttthen \genv', \tenv' \nextclause \\ 
		\genv', \tenv' \tholds \decls \ttthen \genv'', \tenv''}
	{\genv,\tenv\tholds \decl \ttsemi\ \decls \ttthen \genv'',\tenv''}
\end{tabular}

\nextrulesmall
\begin{tabular}{cc}
\irule
	{\wfenv\twfholds\cbasetyp\nextclause \\ 
		\tttypeid\notin\genv, \env\nextclause \\
		\genv'= \genv[\tttypeid\goesto\cbasetyp]}
	{\genv, \tenv\tholds \tttype\ \tttypeid = \cbasetyp \ttthen \genv',\tenv }
&
\irule
	{\typenv \tholds \ttx\ \ttcol\ \cregtyp\nextclause \\
		\genv,\tenv\tholds\hastyp{\tte}{\ttstring} }
	{\genv,\tenv\holds\ttlet\ \strof{\ttx} = \tte\ttthen\genv, \tenv}
\end{tabular}
\nextrulesmall
\irule
	{\wfenv\twfholds\cbasetyp\nextclause \\ 
		\ttx\notin\genv,\tenv\nextclause \\
		\genv, \env\tholds\hastyp{\tte}{\cbasetyp}\nextclause \\ 
		\tenv'=\tenv[\ttx\envgoesto\cbasetyp]}
	{\genv,\tenv\holds\ttlet\ \ttx\ \ttcol\ \cbasetyp = \tte\ttthen\genv, \tenv'}

\nextrulesmall
\irule
	{\wfenv\twfholds \left(\funtyp{\sseq{\cbasetypi}}{\cbasetyp}  \right)\nextclause \\
		\ttfuncid\notin\genv,\tenv \nextclause \\
		\tenv' = \tenv[\forall \tti, \ttx_i\envgoesto\cbasetypi]\nextclause \\
		\genv, \env'\tholds\hastyp{\tte}{\cbasetyp} \nextclause \\
		\tenv'' = \tenv[\ttfuncid\envgoesto\left( \funtyp{\sseq{\ttx_i\ \ttcol\ \cbasetypi}}{\cbasetyp} \right)]}
	{\genv,\tenv\holds\ttdef\ \ttfuncid\ \funtyp{\sseq{\ttx_i\ \ttcol\ \cbasetypi}}{\cbasetyp} = \tte\ttthen\genv, \tenv''}

\nextrulesmall
\irule
{\wfenv\twfholds \left( \proctyp{\sseq{\cbasetypi}} \right)\nextclause \\
	\ttprocid\notin\genv,\tenv \nextclause \\
	\tenv' = \tenv[\forall \tti, \ttx_i\envgoesto\cbasetypi]\nextclause \\
	\genv, \env'\tholds\ttS \nextclause \\
	\tenv'' = \tenv[\ttprocid\envgoesto \left( \proctyp{\sseq{\ttx_i\ \ttcol\ \cbasetypi}} \right)]}
{\genv,\tenv\holds\ttproc\ \ttprocid\ \proctyp{\sseq{\ttx_i\ \ttcol\ \cbasetypi}} = \ttS\ttthen\genv, \tenv''}

\nextrulesmall
\irule
	{\wfenv\twfholds\bvltyp{\ttC}\nextclause \\ 
		\ttx\notin\genv,\tenv \nextclause \\
		\tenv' = \tenv[\ttx\envgoesto\bvltyp{\ttC}]}
	{\genv, \tenv\holds\ttlet\ttstate\ \ttx\ \ttcol\ \bvltyp{\ttC}\ttthen \genv, \tenv'}
	
\nextrulesmall
\irule
{\wfenv\twfholds\bvltyp{\ttC}\nextclause \\ 
	\ttx\notin\genv,\tenv \nextclause \\
	\tenv' = \tenv[\ttx\envgoesto\bvltyp{\ttC}]}
{\genv, \tenv\holds\ttlet\ttstate\ \ttcontrol\ \ttx\ \ttcol\ \bvltyp{\ttC}\ttthen \genv, \tenv'}

\nextrulesmall
\irule
{\wfenv\twfholds\bvltyp{\ttC}\nextclause \\ 
	\ttx\notin\genv,\tenv \nextclause \\
	\tenv' = \tenv[\ttx\envgoesto\bvltyp{\ttC}]}
{\genv, \tenv\holds\ttlet\ttstate\ \ttcontrol\ \ttdontgate\ \ttx\ \ttcol\ \bvltyp{\ttC}\ttthen \genv, \tenv'}

\nextrulesmall
\irule
	{\wfenv\twfholds\memtyp{\ttC_1}{\ttC_2}{\ttC_3}\nextclause \\ 
		\ttmemid\notin\genv,\tenv \nextclause \\
		\tenv' = \env[\ttmemid\envgoesto\memtyp{\ttC_1}{\ttC_2}{\ttC_3}]}
	{\genv, \tenv\holds\ttlet\ttstate\ \ttmemid\ \ttcol\ \memtyp{\ttC_1}{\ttC_2}{\ttC_3} \ttthen \genv, \tenv'}

\nextrulesmall
\irule
	{\wfenv\twfholds\memtyp{\ttC_1}{\ttC_2}{\ttC_3}\nextclause \\
		\wfenv\twfholds\labeltyp{\ttC_3}\nextclause \\ \\
		\ttmemid\notin\genv,\tenv \nextclause \\
		\NNEW{\ttx} \notin\genv,\tenv \nextclause \\ \\
		\tenv' = \tenv[\ttmemid\envgoesto\memtyp{\ttC_1}{\ttC_2}{\ttC_3}] \\ 
		\tenv'' = \tenv'[\NNEW{\ttx}\envgoesto\labeltyp{\ttC_3}]}
	{\genv, \tenv\holds\ttlet\ttstate\ \ttmemid\ \ttcol\ \memtyp{\ttC_1}{\ttC_2}{\ttC_3}
		\ \ttwith\ \NNEW{\ttx} \ttthen \genv, \tenv''}

\nextrule
\begin{tabular}{cc}
\irule {~}
	{\genv, \tenv\tholds \eps \ttthen \genv, \tenv}
&
\irule 
{\genv, \tenv\tholds \defop \ttthen \genv', \tenv' \nextclause \\ 
	\genv', \tenv' \tholds \defops \ttthen \genv'', \tenv''}
{\genv,\tenv\tholds \defop \ttsemi\ \defops \ttthen \genv'',\tenv''}
\end{tabular}

\nextrulesmall	
\irule
{\ttxop\notin\genv,\tenv \nextclause \\
	\genv, \env\tholds\hastyp{\tte}{\ttstring}\nextclause \\
	\genv, \env\tholds\ttS \nextclause \\
	\tenv' = \tenv[\ttxop \envgoesto \funtyp{()} {()}] }
{\genv,\tenv\holds\ttdefop\ \ttxop\ \ttinbrace{\tttxt=\tte\ttcomma\ttsem=\ttS}\ttthen\genv, \tenv'}

\nextrulesmall
\irule
{\wfenv\twfholds \left( \funtyp{\sseq{\cbasetypi}}{()} \right)\nextclause \\
	\ttxop\notin\genv,\tenv \nextclause \\ 
	\forall \tti, \cbasetypi \neq \ttstring \wedge \cbasetypi \neq \ttunit \wedge \cbasetypi \neq \cregstyp \nextclause \\
	\tenv' = \tenv[\forall \tti, \ttx_i\envgoesto\cbasetypi]\nextclause \\
	\genv, \env'\tholds\hastyp{\tte}{\ttstring}\nextclause \\
	\genv, \env'\tholds\ttS \nextclause \\
	\tenv'' = \tenv[\ttxop \envgoesto \left( \funtyp{\sseq{\ttx_i\ \ttcol\ \cbasetypi}} {()}\right) ] }
{\genv,\tenv\holds\ttdefop\ \ttxop\ \sseq{\ttx_i\ \ttcol\ \cbasetypi} \ttinbrace{\tttxt=\tte\ttcomma\ttsem=\ttS}\ttthen\genv, \tenv''}
\caption{\casp typing rules for declarations.\label{fig:tr_casp_type_decl}}
\end{figure*}


\mypara{Declarations}
Declarations update the environment.
The form of a typing judgment for a declaration is
$\genv, \tenv \holds \decl \ttthen \genv', \tenv'$, and a judgment for
a list of declarations has the same form, shown in \autoref{fig:tr_casp_type_decl}.
This means that the declaration (or list) is well typed and produces
the new environment on the right.

We impose an additional syntactic restriction on declarations found in
a machine description (as opposed to the additional declarations that
may appear in a specification): they may not use the expression forms that
refer to machine state (registers or memory),
because when defining the machine, there is no specific machine
state to refer to; any references would need to be quantified.
(That in turn is not allowed; while many SMT solvers now support
quantified expressions, they generally do not perform well.)

\begin{figure*}
\textbf{(Machine Typing)}
\nextrulesmall
\irule 
	{\genv_\mathit{builtin}, \tenv_\mathit{builtin}\tholds \decls \ttthen \genv, \tenv \nextclause \\
		\genv, \tenv \tholds \defops \ttthen \genv', \tenv'}
	{\tholds \decls \ttsemi\ \defops \ttthen \genv', \tenv'}
	
\nextrulesmall
\textbf{(Specification Typing)}
\nextrulesmall
\irule 
{\tholds \machine \ttthen \genv, \tenv \nextclause \\
	\genv, \tenv \tholds \decls \ttthen \genv', \tenv' }
{\tholds \machine \ttsemi\ \decls \ttthen \genv', \tenv'}

\nextrulesmall
\begin{tabular}{cc}
	\irule
	{ \forall \tti,\ \genv, \tenv \tholds \hastyp{\ttregidi}{\bvltyp{\ttC_i}}}
	{\typenv \tholds \ttfwrites\ \ttcol\ \sseq{\ttregidi} }
	&
	\irule
	{ \forall \tti,\ \genv, \tenv \tholds \hastyp{\tte_i}{\ttint} \nextclause \\
		\forall \tti,\ \genv, \tenv \tholds \hastyp{\ttmemidi}{\cmemtyp}}
	{\typenv \tholds \ttfmwrites\ \ttcol\ \sseq{\ptrform{\ttmemidi}{\tte_i}} }
\end{tabular}

\nextrulesmall
\begin{tabular}{cc}
	\irule{~}
	{\typenv \tholds \eps}
	&
	\irule
	{\typenv \tholds \ttframe \nextclause \\ \typenv \tholds \ttframes}
	{\typenv \tholds \ttframe\ \ttframes}
\end{tabular}

\nextrulesmall
\irule 
{\tholds \machine \ttsemi\ \decls \ttthen \genv, \tenv \nextclause \\
	\genv, \tenv \tholds \ttframe \nextclause \\
	\genv, \tenv \tholds \hastyp{\ttpre}{\ttbool} \nextclause \\
	\genv, \tenv \tholds \hastyp{\ttpost}{\ttbool} }
{\machine \tholds \decls \ttsemi\ \ttframes \ttsemi\ \ttpre \ttsemi\ \ttpost}
	
\nextrulesmall
\textbf{(Program Typing)}
\nextrulesmall
\begin{tabular}{cc}
	\irule
	{\typenv\tholds\hastyp{\ttxop}{\left( \funtyp{()}{()} \right)}}
	{\typenv\tholds \ttxop}
	&
	\irule
	{\typenv\tholds\hastyp{\ttxop}{\left( \funtyp{\sseq{\cbasetypi}}{()} \right)}\nextclause \\
		\forall \tti,\ \typenv\tholds\hastyp{\tte_i}{\cbasetypi}}
	{\typenv\tholds \ttxop\ {\sseq{\tte_i}}}
\end{tabular}
\nextrulesmall
\begin{tabular}{ccc}
	\irule {~} {\genv, \tenv \tholds \eps}
	&
	\irule 
	{ \genv, \tenv \tholds \itinst \nextclause \\
		\genv, \tenv \tholds \itinsts}
	{\genv, \tenv \tholds \itinst \ttsemi\ \itinsts}
	&
	\irule 
	{\tholds \machine \ttthen \genv, \tenv \nextclause \\
		\genv, \tenv \tholds \program}
	{\machine \tholds \program}
\end{tabular}
\caption{\casp typing rules for machine, specification, and program.\label{fig:tr_casp_type_mach}}
\end{figure*}

\mypara{Machines}
A machine is some declarations followed by some defops, so the typing
rule is just sequencing, shown in \autoref{fig:tr_casp_type_mach}, 
but there is a wrinkle: the initial
environment for the machine is not an input.
$\genv_\mathit{builtin}$ is the (fixed) environment describing the built-in
type aliases.
(Currently there are none.)
$\tenv_\mathit{builtin}$ is the environment describing the types of
built-in variables.
This notionally includes the built-in functions.
(But as mentioned earlier some of them actually have their own typing
rules.)
The form of a typing judgment for a machine is
$\tholds \machine \ttthen \genv, \tenv$.
This means that the machine description is well typed and provides the
environment on the right for use of other constructs that depend on
the machine.
(Specs and programs are only valid relative to a given machine.)

\mypara{Specifications}
For specifications we need two helper rules, shown in
\autoref{fig:tr_casp_type_mach}:
one that applies an
additional list of declarations to a machine, which has the same
form as the judgment on a machine; and one that says that a frame
(modifies list) is well typed, which has the form
$\typenv \tholds \ttframes$.
This lets us write the real rule, which has the form
$\machine \tholds \itspec$
and means that the specification is well typed under the machine description.

\mypara{Programs}
A program is a sequence of calls to instructions.
We need judgments of the form $\typenv \tholds \itinst$ for a single
instruction and also $\typenv \tholds \itinsts$ for the sequence, shown in \autoref{fig:tr_casp_type_mach}.
There are two cases for a single instruction because of a minor glitch
in formulation: because the overbar notation means ``one or more'',
there are two cases in the syntax for instructions, one for zero
operands and one for one or more operands; we need typing rules for
both cases.
Meanwhile the type entered into $\tenv$ for a zero-operand instruction
is unit to unit, not {\eps} to unit, to avoid needing an additional
form for types just for this case.
(Notice that a one-operand instruction may not have type unit to unit
because unit is not allowed as an instruction operand, so the type is
not ambiguous.)
These rules let us write a judgment for a program, which has the form
$\machine \tholds \program$ and means that the program is well typed
relative to the machine.

\mypara{Soundness}
Our type system is sound: we include the necessary checks and failure
states in the semantics so that evaluation does not get stuck, even
though some properties are not statically checked.

\subsection{\casp{} Semantics}
\label{sec:exec}
This section defines the semantics of \casp.

\begin{figure*}
\textbf{(Expression Semantics)} \\
\nextrulesmall
\begin{tabular}{ccc}
\irule
	{\venv(\ttx)=\ttv}
	{\esreduce{\ttx}{\ttv}}
&
\irule
	{\esreduce{\tte}{\ttr} \nextclause \\
		\venv(\strof{\ttr})=\ttv}
	{\esreduce{\strof{\tte}}{\ttv}}
&
\irule
	{\esreduce{\tte}{\ttr} \nextclause \\
		\strof{\ttr} \notin \venv}
	{\esreduce{\strof{\tte}}{\exprfail}}
\end{tabular}
\nextrulesmall
\irule
	{\forall \tti, \esreduce{\tte_i}{\ttv_i} \nextclause \\ 
		\venv\ttinparen{\ttfuncid} = \{\sseq{\ttx_i}, {\tte}\}\nextclause \\ 
		\esreducevenv{\venv[\forall \tti, \ttx_i \envgoesto \ttv_i]}{\tte}{\opsenv}{\tildev}}
	{\esreduce{\fapp{\ttfuncid}{\sseq{\tte_i}}}{\tildev}}
	
\nextrulesmall
\begin{tabular}{cc}
\irule
	{\esreduce{\tte}{\ttv_1} \nextclause \\
		\tildev_2 = \itunop\ \ttv_1}
	{\esreduce{\itunop\ \tte}{\tildev_2}}
&
\irule
	{\esreduce{\tte_1}{\ttv_1} \ \ \ \ \ 
		\esreduce{\tte_2}{\ttv_2} \ \ \ \ \ 
		\tildev_3 =\ttv_1\ \itbinop\ \ttv_2}
	{\esreduce{\tte_1\ \itbinop\ \tte_2}{\tildev_3}}
\end{tabular}

\nextrulesmall
\begin{tabular}{cc}
\irule
	{\esreduce{\tte}{\bitconst}\nextclause \\   
		\ttC = \ttb_0\ldots\ttb_{\ttC_i}\ldots\ttb_\ttn}
	{\esreduce{\bitidx{\tte}{\ttC_i}}{\ttb_{\ttC_i}}}
&
\irule
	{\esreduce{\tte}{\bitconst}\nextclause \\ 
		\ttC = \ttb_0\ldots\ttb_{\ttC_i}\ldots\ttb_{\ttC_j}\ldots\ttb_\ttn}
	{\esreduce{\bitfield{\tte}{{\ttC_i}}{{\ttC_j}}}{\ttb_{\ttC_i}\ldots\ttb_{\ttC_j}}}
\end{tabular}

\nextrulesmall
\begin{tabular}{cc}
\irule
	{\esreduce{\tte}{\ptrform{\ttmemid}{\ttC}}}
	{\esreduce{\bitidx{\tte}{\_}}{\exprfail}}
&
\irule
	{\esreduce{\tte}{\ptrform{\ttmemid}{\ttC}}}
	{\esreduce{\bitfield{\tte}{{\_}}{{\_}}}{\exprfail}}
\end{tabular}

\nextrulesmall
\irule
	{\esreduce{\tte_1}{\ttv_1} \nextclause \\
		\esreducevenv{\venv[\ttx \envgoesto \ttv_1]}{\tte_2}{\opsenv}{\tildev_2}}
	{\esreduce{\eLet{\ttx}{\cbasetyp}{\tte_1}{\tte_2}}{\tildev_2}}
	
\nextrulesmall
\begin{tabular}{cc}
\irule
	{\esreduce{\tte}{\tttrue}\nextclause \\ 
		\esreduce{\tte_t}{\tildev_t}}
	{\esreduce{\sITE{\tte}{\tte_t}{\_}}{\tildev_t}}
&
\irule
	{\esreduce{\tte}{\ttfalse}\nextclause \\ 
		\esreduce{\tte_f}{\tildev_f}}
	{\esreduce{\sITE{\tte}{\_}{\tte_f}}{\tildev_f}}
\end{tabular}

\nextrulesmall
\begin{tabular}{cc}
\irule
	{\esreduce{\tte}{\ttr}\nextclause \\ 
		\esloclookup{\ttr}}
	{\esreduce{\regderef{\tte}}{\ttv}}
&
\irule
	{\esreduce{\tte}{\ptrform{\ttmemid}{\ttC}}\nextclause \\ 
		\esmemlookup{\ttmemid}{\ttC}{(\ttv, \ttC_l)} }
	{\esreduce{\fetch{\tte}{\ttC_l}}{\ttv}}
\end{tabular}

\nextrulesmall
\irule
	{\esreduce{\tte}{\ptrform{\ttmemid}{\ttC}}\nextclause \\ 
		\esmemlookup{\ttmemid}{\ttC}{(\_, \ttC_m)}\nextclause \\
		\ttC_m \neq \ttC_l}
	{\esreduce{\fetch{\tte}{\ttC_l}}{\exprfail}}

\nextrulesmall
\begin{tabular}{cc}
\irule
	{\esreduce{\tte}{\ptrform{\ttmemid}{\ttC}}\nextclause \\ 
		\ptrform{\ttmemid}{\ttC}\notin\opsmenv}
	{\esreduce{\fetch{\tte}{\ttC_l}}{\exprfail}}
&
\irule
	{\esreduce{\tte}{\bitconst}}
	{\esreduce{\fetch{\tte}{\ttC_l}}{\exprfail}}
\end{tabular}

\nextrulesmall
\irule
	{\esreduce{\tte}{\ttC}}
	{\esreduce{\ptrform{\ttmemid}{\tte}}{\ptrform{\ttmemid}{\ttC}}}

\nextrulesmall
\begin{tabular}{cc}
\NEWW{
\irule
	{\venv(\texttt{EXTBRANCH})=\texttt{ext}}
	{\esreduce{\ebranchto{\ttextid}}{\tttrue}}
}
&
\NEWW{
\irule
	{\venv(\texttt{EXTBRANCH})=\cdot}
	{\esreduce{\ebranchto{\ttextid}}{\ttfalse}}
}
\end{tabular}

\nextrulesmall
\begin{tabular}{cc}
\irule
	{\forall \tti \in(1\ldots\ttk),\ \venv(\ttx_i) = \ttr_i}
	{\esreduce{\ttinbrace{\ttx_1, \ldots, \ttx_k}}{\ttinbrace{\ttr_1, \ldots, \ttr_k}}}
&
\irule
	{\esreduce{\tte}{\ttinbrace{\ttr_1, \ldots, \ttr_{\ttC}}}}
	{\esreduce{\sizeofset{\tte}}{\ttC}}
\end{tabular}

\nextrulesmall
\irule
	{\esreduce{\tte_1}{\ttr} \nextclause \\
		\esreduce{\tte_2}{\ttinbrace{\ttr_1, \ldots, \ttr_k}} \nextclause \\
		\exists \tti \in (1\ldots\ttk), \ttr_{\tti} = \ttr}
	{\esreduce{\setmemberof{\tte_1}{\tte_2}}{\tttrue}}

\nextrulesmall
\irule
	{\esreduce{\tte_1}{\ttr} \nextclause \\
		\esreduce{\tte_2}{\ttinbrace{\ttr_1, \ldots, \ttr_k}} \nextclause \\
		\forall \tti \in (1\ldots\ttk), \ttr_{\tti} \neq \ttr}
	{\esreduce{\setmemberof{\tte_1}{\tte_2}}{\ttfalse}}

\caption{\casp semantics for expressions.\label{fig:tr_casp_sem_expr}}
\end{figure*}

\mypara{Environment}
The execution environment $\venv$ maps \casp variables \ttx to values
\ttv.
\NNEW{For labels on memory regions, each label
maps to a pointer that points to the base (offset 0) of the region
associated with the label.}
However, we take advantage of the polymorphism and dynamic
typing of paper rules to also store the following in the same
environment:
\begin{itemize}
\item \ttfuncid (function names) map to pairs $\{\sseq{\ttx_i},
{\tte}\}$, which give the list of argument names and the body for
functions.
\item \ttprocid (procedure names) map to pairs $\{\sseq{\ttx_i},
{\ttS}\}$, which give the list of argument names and the body for
procedures.
\item \ttxop (operation/instruction names) map to triples $\{\sseq{\ttx_i},
\tte, {\ttS}\}$, which give the list of argument names, the expression
for the text form, and the body for operations.
\item \strof{\ttr} (the form for the text version of a register) maps
to a value.
\item \NEWW{The word \texttt{EXTBRANCH} maps to a branch state, which
must be either \texttt{ext} or $\cdot$.}
This reports whether, after executing a program, it branched to the
external label or not.
\end{itemize}
Since identifiers are not allowed to overlap in well-typed programs,
and register identities are not strings at all, \NEWW{and \texttt{EXTBRANCH}
is
reserved,} this usage creates no
conflicts.

Note that \ttmemid, \tttypeid, and \ttxmoduleid do not appear in
$\venv$ as these require no translation/lookup at runtime.

\mypara{Machine State}
In addition to the execution environment, we also need a representation
of machine state.
We define two stores, one for registers and one for memory.
The register store $\opsrenv$ maps registers \ttr to values \ttv.
The memory store $\opsmenv$ is more complicated: it maps pairs
$\ptrform{\ttmemid}{\ttC}$ (that is, pointer literals) to pairs
$(\ttv, \ttC_l)$, where \ttv is the value stored at that location and
$\ttC_l$ is the bit width.
The bit widths of memory regions are invariant, both across the region
when they are declared and also over time.
They are used to check the access widths appearing in fetch and store
operations.
Also note that new entries cannot be created in either the register
store or the memory store, as real hardware does not permit such
actions.
The values stored in registers and memory regions are restricted by
the typing rules to bitvectors (whether constants or pointers) of the
appropriate width.

Notice that stepping through the declarations does \emph{not}
initialize the machine state.
We want to reason about executions over ranges of possible starting
machine states; so instead we provide a judgment that uses the typing
environments to restricts the stores to forms consistent with the
declarations.
This is discussed further below.

\mypara{Expressions}
We describe expressions with a large-step operational semantics, shown in \autoref{fig:tr_casp_sem_expr}.
The form of an expression semantic judgment is:
$\esreduce{\tte}{\ttv}$,
which means that given the environment $\venv$ and the machine
state $\opsenv$, the expression \tte evaluates to the value \ttv.
Expressions may read the machine state, but not modify it.
Expressions can fail; in addition to the explicit failure cases seen,
some of the operators and built-in functions can fail.
For example, as mentioned earlier, attempting bitvector arithmetic
other than addition and subtraction on pointers will fail.
Furthermore, division by zero fails.

Note that we currently do not statically check (in the typing rules)
that the \strof{} form is present for every register or that it is
defined for registers on which it is used.
Thus we have an explicit failure rule for when no matching declaration
has been seen.
We also have failure rules for bad fetch operations: if the length
annotation is wrong, if the pointer is not in the machine state
(this covers both unaligned accesses and out of bounds accesses),
or if the value used is not a pointer.
Similarly, we have failure rules for when bit indexing/slicing a pointer. 
We do not, conversely, need explicit failure checks or rules for the
bit indexes in the bit extraction/slicing constructs as they are
statically checked.

Also note that we include in the semantics the obvious failure
propagation rules for when subexpressions fail.
We do not show these explicitly as they are not particularly
interesting or informative; however, note that the $\&\&$ and $||$
logical operators short-cut left to right.


\begin{figure*}
\textbf{(Statement Semantics)}\\
\nextrulesmall
\irule
	{\osreduce{\ttS_1}{\opsenv}{\ttskip}{\opsenvB}\nextclause \\
		\osreduce{\ttS_2}{\opsenvB}{\ttS_2'}{\opsenvC}}
	{\osreduce{\ttS_1\ttsemi\ \ttS_2}{\opsenv}{\ttS_2'}{\opsenvC}}

\nextrulesmall
\irule
{\forall \tti, \esreduce{\tte_i}{\ttv_i} \nextclause \\  	
	\venv\ttinparen{\ttprocid} = \{\sseq{\ttx_i},\ttS\}\nextclause \\ 
	\osreducetenv{\venv[\forall \tti, \ttx_i \envgoesto \ttv_i]}{\ttS}{\opsenv}{\ttS'}{\opsenvB}}
{\osreduce{\papp{\ttprocid}{\sseq{\tte_i}}}{\opsenv}{\ttS'}{\opsenvB}}

\nextrulesmall
\irule
	{\esreduce{\tte}{\ttv} \nextclause \\  
		\osreducetenv{\venv[\ttx \envgoesto \ttv]}{\ttS}{\opsenv}{\ttS'}{\opsenvB}}
	{\osreduce{\sLet{\ttx}{\cbasetyp}{\tte}{\ttS}}{\opsenv}{\ttS'}{\opsenvB}}

\nextrulesmall
\irule
	{\forall \tti \in (\ttC_1, \ttC_1+1,\ \ldots, \ttC_2),\ 
		\osreducetenv{\venv[\ttx \envgoesto \tti]}{\ttS}{\opsenvafter{i}{i}}{\ttskip}{\opsenvafter{i+1}{i+1}}}
	{\osreduce{\sFor{\ttx}{\ttC_1}{\ttC_2}{\ttS}}{\opsenvafter{\ttC_1}{\\tC_1}}{\ttskip}{\opsenvafter{\ttC_2+1}{\ttC_2+1}}}

\nextrulesmall
\irule
	{\esreduce{\tte}{\tttrue}\nextclause \\
		\osreduce{\ttS}{\opsenv}{\ttS_t}{\opsenvafter{t}{t}}}
	{\osreduce{\sITE{\tte}{\ttS}{\_}}{\opsenv}{\ttS_t}{\opsenvafter{t}{t}}}
	
\nextrulesmall
\irule
	{\esreduce{\tte}{\ttfalse}\nextclause \\
		\osreduce{\ttS}{\opsenv}{\ttS_f}{\opsenvafter{f}{f}}}
	{\osreduce{\sITE{\tte}{\_}{\ttS}}{\opsenv}{\ttS_f}{\opsenvafter{f}{f}}}

\nextrulesmall
\irule
	{\esreduce{\tte_1}{\ttr} \nextclause \\ 
		\ttr \in \opsrenv \nextclause \\
		\esreduce{\tte_2}{\ttv} \nextclause \\ 
		\opsrenv' = \opsrenv[\ttr\envgoesto\ttv]}
	{\osreduce{\slAssign{\tte_1}{\tte_2}}{\opsenv}{\ttskip}{\opsrenv',\opsmenv}}

\nextrulesmall
\irule
	{\esreduce{\tte_1}{\ptrform{\ttmemid}{\ttC}}\nextclause \\ 
		\esmemlookup{\ttmemid}{\ttC}{(\_, \ttC_l)} \nextclause \\
		\esreduce{\tte_2}{\ttv} \nextclause \\
		\opsmenv' = \opsmenv[ {\ptrform{\ttmemid}{\ttC}} \envgoesto(\ttv, \ttC_l)]}
	{\osreduce{\smAssign{\tte_1, \ttC_l}{\tte_2}}{\opsenv}{\ttskip}{\opsrenv,\opsmenv'}}

\nextrulesmall
\irule
	{\esreduce{\tte_1}{\ptrform{\ttmemid}{\ttC}}\nextclause \\ 
		\esmemlookup{\ttmemid}{\ttC}{(\_, \ttC_{m})} \nextclause \\
		\ttC_{m} \neq \ttC_{l}}
	{\osreduce{\smAssign{\tte_1, \ttC_l}{\tte_2}}{\opsenv}{\bferr}{\opsrenv,\opsmenv}}

\nextrulesmall
\begin{tabular}{cc}
\irule
	{\esreduce{\tte_1}{\ptrform{\ttmemid}{\ttC}}\nextclause \\ 
		{\ptrform{\ttmemid}{\ttC}} \notin \opsmenv}
	{\osreduce{\smAssign{\tte_1, \ttC_l}{\tte_2}}{\opsenv}{\bferr}{\opsrenv,\opsmenv}}
&
\irule
	{\esreduce{\tte_1}{\bitconst}}
	{\osreduce{\smAssign{\tte_1, \ttC_l}{\tte_2}}{\opsenv}{\bferr}{\opsrenv,\opsmenv}}
\end{tabular}

\nextrulesmall
\begin{tabular}{cc}
\irule
	{\esreduce{\tte}{\tttrue}}
	{\osreduce{\ttassert(\tte)}{\opsenv}{\ttskip}{\opsenv}}
&
\irule
	{\esreduce{\tte}{\ttfalse}}
	{\osreduce{\ttassert(\tte)}{\opsenv}{\bferr}{\opsenv}}
\end{tabular}

\nextrulesmall
\begin{tabular}{cc}
\NEW{
\irule
	{\esreduce{\tte}{0}}
	{\osreduce{\sBRANCH{\tte}}{\opsenv}{\ttskip}{\opsenv}}
}
&
\NEW{
\irule
	{\esreduce{\tte}{\bitconst} \nextclause \\
	 \texttt{0x00} < \bitconst < \texttt{0xff}}
	{\osreduceBR{\sBRANCH{\tte}}{\opsenv}{\ttskip}{\opsenv}{\bitconst}}
}
\end{tabular}

\nextrulesmall
\NEW{
\irule
	{\esreduce{\tte}{\texttt{0xff}}}
	{\osreduceBR{\sBRANCH{\tte}}{\opsenv}{\ttskip}{\opsenv}{\texttt{ext}}}
}
\caption{\casp semantics for statements.\label{fig:tr_casp_sem_stmt}}
\end{figure*}

\mypara{Statements}
Unlike expressions, statements can change machine state.
Thus, the form of a machine state semantics judgment (also large step)
is
$\osreduceBR{\ttS}{\opsenv}{\ttS'}{\opsenvB}{\xi}$, shown in \autoref{fig:tr_casp_sem_stmt}.
This means that the statement \ttS evaluates to the irreducible
statement \ttS' (which must be either \ttskip or \bferr) and in the
course of doing so changes the machine state from \opsenv to \opsenvB,
\NEW{and produces a branching state $\xi$, which can be either an 8-bit
bitvector, the reserved value \texttt{ext}, or a dot ($\cdot$).}
As with expressions, statements can fail.
Explicit failure rules are shown for bad stores (corresponding to the
cases for bad fetches) and for a failed assertions.
We also similarly include, but do not show, the obvious failure
propagation rules for cases where sub-statements, or expressions
within statements, fail.

\begin{figure*}
\textbf{(Declaration Semantics)}\\
\nextrulesmall
\begin{tabular}{cc}
\irule
   {~}
   {\SEMdeclAA{\eps}}
&
\irule
   {\SEMdeclAB{\ttdecl} \nextclause \\
      \SEMdeclBC{\ttdecls}}
   {\SEMdeclAC{\ttdecl \ttsemi\ \ttdecls}}
\end{tabular}

\nextrulesmall
\begin{tabular}{cc}
\irule
   {~}
   {\SEMdeclAA{\tttype\ \tttypeid = \cbasetyp}}
&
\irule
   {\esreduce{\tte}{\ttv}}
   {\SEMdeclAx{\ttlet\ \ttx\ \ttcol\ \cbasetyp = \tte}{\ttx \envgoesto \ttv}}
\end{tabular}

\nextrulesmall
\irule
   {\esreduce{\ttx}{\ttr} \nextclause \\ 
   		\esreduce{\tte}{\ttv}}
   {\SEMdeclAx{\ttlet\ \strof{\ttx} = e}{\strof{\ttr} \envgoesto \ttv}}

\nextrulesmall
\begin{tabular}{cc}
\irule
   {\venv' = \venv[\ttfuncid \envgoesto \{\sseq{\ttx_i}, \tte\}]}
   {\SEMdeclAB{\ttdef\ \ttfuncid\ \funtyp{\sseq{\ttx_i\ \ttcol\ \cbasetypi}}{\cbasetyp} = \tte}}
&
\irule
{\venv' = \venv[\ttprocid \envgoesto \{\sseq{\ttx_i}, \ttS\}]}
{\SEMdeclAB{\ttproc\ \ttprocid\ \proctyp{\sseq{\ttx_i\ \ttcol\ \cbasetypi}} = \ttS}}
\end{tabular}

\nextrulesmall
\begin{tabular}{cc}
\irule
   {\venv' = \venv[\ttx \envgoesto \ttr] \nextclause \\	
   		\ttr\ \mathrm{fresh}}
   {\SEMdeclAB{\ttlet\ttstate\ \ttx\ \ttcol\ \cregtyp}}
&
\irule
{\venv' = \venv[\ttx \envgoesto \ttr] \nextclause \\	
	\ttr\ \mathrm{fresh}}
{\SEMdeclAB{\ttlet\ttstate\ \ttcontrol\ \ttx\ \ttcol\ \cregtyp}}
\end{tabular}

\nextrulesmall
\irule
{\venv' = \venv[\ttx \envgoesto \ttr] \nextclause \\	
	\ttr\ \mathrm{fresh}}
{\SEMdeclAB{\ttlet\ttstate\ \ttcontrol\ \ttdontgate\ \ttx\ \ttcol\ \cregtyp}}

\nextrulesmall
\begin{tabular}{cc}
\irule
   {~}
   {\SEMdeclAA{\ttlet\ttstate\ \ttmemid\ \ttcol\ \cmemtyp}}
&
\irule
   {\venv' = \venv[\NNEW{\ttx} \envgoesto \ptrform{\ttmemid}{0}] \nextclause \\	}
   {\SEMdeclAB{\ttlet\ttstate\ \ttmemid\ \ttcol\ \cmemtyp\ \ttwith\ \NNEW{\ttx}}}
\end{tabular}

\nextrule
\begin{tabular}{cc}
\irule
   {~}
   {\SEMdeclAA{\eps}}
&
\irule
   {\SEMdeclAB{\defop} \nextclause \\
      \SEMdeclBC{\defops}}
   {\SEMdeclAB{\defop \ttsemi\ \defops}}
\end{tabular}
   
\nextrulesmall
\irule
{ \venv' = \venv[\ttxop \envgoesto \{[~], \tte, S\}]}
{\SEMdeclAB{\ttdefop\ \ttxop\ \ttinbrace{\tttxt=\tte\ttcomma\ttsem=\ttS}}}

\nextrulesmall
\irule
   { \venv' = \venv[\ttxop \envgoesto \{\sseq{x_i}, \tte, S\}]}
   {\SEMdeclAB{\ttdefop\ \ttxop\ \sseq{\ttx_i\ \ttcol\ \cbasetypi}\ \ttinbrace{\tttxt=\tte\ttcomma\ttsem=\ttS}}}
\caption{\casp semantics for declarations.\label{fig:tr_casp_sem_decl}}
\end{figure*}

\mypara{Declarations}
The semantics for declarations have judgments of the form
$\SEMdeclAB{\ttdecl}$, shown in \autoref{fig:tr_casp_sem_decl}.
This means that the given declaration updates $\venv$ as shown.
As stated above, we do not \emph{initialize} the machine state while
handling declarations; this instead allows us to work with arbitrary
(or universally quantified) machine states afterwards.
However, because the let-binding declaration evaluates an expression,
it potentially needs \emph{access} to a machine state.
Consequently we write the rules so they accept a machine state as input,
but do not update it.
In the case of machine descriptions, where there is no machine state,
we pass empty environments; let-binding declarations in machine descriptions
are not allowed to reference machine state.
In the case of the additional declarations that accompany a
specification, we pass in the initial machine state; this allows
values from the initial machine state to be globally bound so they can
be referred to in the postcondition.

We give first the rules for a list of declarations, then the rules for the various declarations, then the
rules for a list of operation definitions and a rule for a single operation definition.
Note that several of the declarations do not update $\venv$, and
nothing is placed in $\venv$ for memory regions.
For registers, only the mapping of the identifier to its
underlying register \ttr is entered; nothing for \ttr is inserted.

\begin{figure}
\textbf{(Machine Semantics)}\\
\nextrule
\irule
   {\SEMZZA{\decls} \nextclause \\
      \SEMmachAB{\defops}}
   {\SEMmachB{\decls \ttsemi\ \defops}}
\caption{\casp semantics for machines.\label{fig:tr_casp_sem_mach}}
\end{figure}

\mypara{Machines}
As with the typing rules, the semantics rule for a whole machine
description integrates the initial environment and gives a judgment of
the form $\SEMmachB{\machine}$, shown in \autoref{fig:tr_casp_sem_mach}.
We also include a comparable form that includes additional
declarations, as it will be used below.

\mypara{Programs}
Instructions (or more precisely, \casp operations) update the machine state, and we chose to represent
programs as lists of instructions rather than having dummy
instruction forms for skip and sequence.
Consequently the form of the judgments is slightly different, and
there are several of them, shown in \autoref{fig:tr_casp_sem_prog}.

First, we can execute an individual instruction using the form
$\psreduce{\itinst}{\opsenv}{\opsenvB, \NEW{\xi}}$,
meaning that the instruction executes and updates the machine state
\opsenv to \opsenvB, \NEW{producing the branching state $\xi$}.
Then, a list of instructions executes using the form
\psreduce{\itinsts, \NEW{\xi}}{\opsenv}{\opsenvB, \itinsts', \NEW{\xi'}},
which means that the list steps to a new list and updates \NEW{both} the
machine state \NEW{and the branching state}.
When the instruction list runs out, these reduce to a shorter form
via a judgment of the form
\psreduce{\itinsts, \NEW{\xi}}{\opsenv}{\opsenvB},
which discards the instructions and branching state and produces an
output machine state.
That in turn allows us to draw conclusions of the form
$\ttmach \oholds(\opsenv,\program)\goesto(\opsrenv', \opsmenv')$,
which means that a machine with the initial state $\opsenv$ executes
the program to produce the new machine state $(\opsrenv', \opsmenv')$.


Note that there are two versions of the judgment for instructions,
one specialized for no arguments/operands.
Instructions with no operands are declared as taking unit, but
invoked with empty operands (not with unit) to correspond to the way
assembly languages normally work.

We include a final judgment of the form 
$\ttmach \oholds(\opsenv,\program)\goesto(\opsrenv', \opsmenv')$
that puts the machine on the left-hand side of the turnstile.
It means that under a given machine the program maps \opsenv to
\opsenvB.
There is a limitation in the way we have formulated programs and the
rules for programs, which is that there is no easy way to represent
failure.
(Failure in this might represent triggering an exception and stopping
execution, which we do not model, or invoking ``unpredictable'' or
``undefined'' behavior in the processor and transitioning to an
arbitrary unknown machine state.)

The intended behavior is that a program that fails during execution
(that is, the body of one of its instructions steps to \bferr) enters
a state where no postcondition can evaluate to \tttrue.
We have decided for the moment that working this explicitly into the
formalism would result in a lot of complication and obscuration
without providing any useful information.

\begin{figure*}
\textbf{(Program Semantics)}\\
\nextrulesmall
\irule
   {\venv(\ttxop) = \{[~], \_, S)\} \nextclause \\
   	\osreduceBR{\ttS}{\opsenv}{\ttskip}{\opsenvB}{\xi}}
   {\psreduce{\ttxop}{\opsenv}{\opsenvB, \NEW{\xi}}}

\nextrulesmall
\irule
   {\forall \tti, \esreduce{\tte_i}{\ttv_i} \nextclause \\
      \venv(\ttxop) = \{\sseq{\ttx_i}, \_, S\}\nextclause \\
      \venv' = \venv[\forall i, \ttx_i \envgoesto \ttv_i] \nextclause \\
      \osreduceafterBR{\ttS}{\opsenv}{\ttskip}{\opsenvB}{\xi}}
  {\psreduce{\ttxop\ \sseq{\tte_i}}{\opsenv}{\opsenvB, \NEW{\xi}}}

\nextrulesmall
\begin{tabular}{cc}
\irule
	{\psreduce{\itinst}{\opsenv}{\opsenvB, \NEW{\xi}}}
	{\psreduce{\itinst \ttsemi\ \itinsts, \NEW{\cdot}}{\opsenv}{\opsenvB,
	\NEW{\itinsts, \xi}}}
&
\NEW{
	\irule{~}
	{\psreduce{\itinst \ttsemi\ \itinsts, \texttt{0x01}}{\opsenv}{\opsenv,
			\itinsts, \cdot}}
}
\end{tabular}
\nextrulesmall
\begin{tabular}{cc}
\NEW{
\irule
	{\bitconst > \texttt{0x01}}
	{\psreduce{\itinst \ttsemi\ \itinsts, \bitconst}{\opsenv}{\opsenv,
	\itinsts, \bitconst\ \bitsub\ \texttt{0x01}}}
}
&
\NEW{
\irule
  {~}
  {\psreduce{\_, \texttt{ext}}{\opsenv}{\opsenv,\NEWW{\eps,\texttt{ext}}}}
}
\end{tabular}

\nextrulesmall
\irule
   {\SEMmachA{\ttmach} \nextclause \\	
      \psreducemulti{\program, \NEW{\cdot}}{\opsenv}{\opsrenv',\opsmenv',\NEWW{\eps,\xi}}}
   {\ttmach \oholds(\opsenv,\program)\goesto(\opsrenv', \opsmenv',\NEWW{\xi})}

\caption{\casp semantics for programs.\label{fig:tr_casp_sem_prog}}
\end{figure*}

\begin{figure*}
\textbf{(Specification Semantics)}\\
\nextrulesmall 
\irule
	{\forall \tti, \esreduce{\ttregidi}{\ttr_i} \nextclause \\
		\forall \ttr \notin \{\sseq{\ttr_i}\}, \opsrenv (\ttr) = \opsrenv' (\ttr)}
	{\venv, \opsrenv, \opsmenv, \opsrenv', \opsmenv' \holds \ttfwrites\ \ttcol\ \sseq{\ttregidi} }

\nextrulesmall 
\irule
{\forall \tti, \esreduce{\ptrform{\ttmemidi}{\tte_i}}{\ptrform{\ttmemidi}{\ttC_i}} \nextclause \\
	\forall \ttmemid, \ttC, \ptrform{\ttmemid}{\ttC} \notin \{\sseq{\ptrform{\ttmemidi}{\ttC_i}}\}, \\
	\opsmenv (\ptrform{\ttmemid}{\ttC}) = \opsmenv' (\ptrform{\ttmemid}{\ttC})}
{\venv, \opsrenv, \opsmenv, \opsrenv', \opsmenv' \holds \ttfmwrites\ \ttcol\ \sseq{\ptrform{\ttmemidi}{\tte_i}} }

\nextrulesmall
\begin{tabular}{cc}
	\irule{~}
	{\venv, \opsrenv, \opsmenv, \opsrenv', \opsmenv' \holds \eps}
	&
	\irule
	{\venv, \opsrenv, \opsmenv, \opsrenv', \opsmenv' \holds \ttframe
		\nextclause \\
		\venv, \opsrenv, \opsmenv, \opsrenv', \opsmenv' \holds
		\ttframes
	}
	{
		\venv, \opsrenv, \opsmenv, \opsrenv', \opsmenv' \holds
		\ttframe\ \ttframes
	}
\end{tabular}

\nextrulesmall
\irule
{	\tholds \machine \ttsemi\ \ttdecls \ttthen \genv, \tenv \nextclause \\ 
  \SEMmachA{\machine} \nextclause \\ 
  \forall \opsenv,\ \left( \genv, \tenv, \venv \tholds \opsenv \right) \implies \\
  \SEMAAB{\decls} \implies \\
  \esreduceB{\ttpre}{\tttrue} \implies \\
  \forall \opsrenv', \opsmenv',
  \left( \psreducemultiB{\program,\NEWW{\cdot}}{\opsrenv,\opsmenv}
      {\opsrenv',\opsmenv',\NEWW{\eps,\xi}} \right) \implies \\
  ( \ensm{\venv'\NEWW{[\texttt{EXTBRANCH} \envgoesto \xi]}
      \oholds(\opsrenv', \opsmenv',\ttpost) \egoesto \tttrue} \ \wedge \ 
  \venv', \opsrenv, \opsmenv, \opsrenv', \opsmenv' \holds \ttframes )
}
{ \machine, (\ttdecls\ttsemi\ \ttframe\ttsemi\ \ttpre\ttsemi\ \ttpost) \holds \program}

\caption{\casp semantics for specifications.\label{fig:tr_casp_sem_spec}}
\end{figure*}

\mypara{Specifications}
For specifications we need three judgments, shown in \autoref{fig:tr_casp_sem_spec}: 
the first two state what the
\ttfwrites and \ttfmwrites clauses mean, respectively 
(they are properties on initial and final
register and memory states),  
and the last one says what it means for a program to
satisfy a specification.
Note that the \ttfwrites and \ttfmwrites rules as shown are slightly misleading, because
the register and pointer list provided in the input specification is implicitly
augmented with all registers and pointers mentioned in the postcondition before it
gets to this point.

\begin{figure*}  
\textbf{(Machine State Validity)}
\nextrulesmall
\irule 
{
	\left( \forall \ttx,\ \genv, \tenv \tholds \ttx\ \ttcol\ \bvltyp{\ttC} \wedge \venv (\ttx) = \ttr \right) \Leftrightarrow 
	\left( \exists \ttv,\ \opsrenv (\ttr) = \ttv \wedge \genv, \tenv \tholds \ttv\ \ttcol\ \bvaltyp{\ttC} \right) }
{\typenv, \venv \tholds \opsrenv}

\nextrulesmall
\irule 
{
	\forall \ttmemid,\ \genv, \tenv \tholds \ttmemid\ \ttcol\ \memtyp{\ttC_1}{\ttC_2}{\ttC_3} \Leftrightarrow \\
	( \forall\tti \in \{ 0, \ttC_1 / 8, \dots, (\ttC_2 - 1) * \ttC_1 / 8 \}, \ 
	\exists \ttv,\ \opsmenv (\ttmemid, \tti) = (\ttv, \ttC_1) \wedge \genv, \tenv \tholds \ttv\ \ttcol\ \bvaltyp{\ttC_1} )\ \wedge\\
	( \forall\tti \notin \{ 0, \ttC_1 / 8, \dots, (\ttC_2 - 1) * \ttC_1 / 8 \}, \ 
	\esreduce{\ptrform{\ttmemid}{\tti}}{\exprfail}) }
{\typenv \tholds \opsmenv}

\nextrulesmall
\irule 
{
	\typenv, \venv  \tholds \opsrenv \nextclause \\
	\typenv \tholds \opsmenv \nextclause
}
{\typenv, \venv  \tholds \opsenv}
\caption{\casp machine state validity.\label{fig:tr_casp_mach_valid}}
\end{figure*}

\mypara{Machine State Validity}
As discussed above, we do not initialize the machine state while
processing declarations.
Instead we treat the starting machine state as an input (e.g., in the
final judgment about programs) or quantify it universally as in the
specification judgment.
This requires that we have a predicate to reject
machine states that do not match the machine description.
The validity judgment has the form $\typenv, \venv \tholds
\opsrenv$, shown in \autoref{fig:tr_casp_mach_valid}, 
and correspondingly for $\opsmenv$ (except without $\venv$)
and then for $\opsenv$ (both stores at once).
This means that the given stores match the given environments.

We use this with the typing environments that come from
both the machine description and the
additional declarations arising from a specification.
In the case of registers we need access to $\venv$ to
handle the names of registers.
We do \emph{not} use the $\venv$ generated from the additional
declarations in a specification; this avoids circularity.
This is acceptable, because specifications are not allowed to
define new registers.
For memory regions we need to enumerate the valid offsets for the
region (note the literal 8 that hardwires 8-bit bytes) and check the
cell width.

\mypara{Branching}
\NEW{
To handle branching, we allow statements to produce a branching state,
which indicates the number of instructions to skip over.
Normally this is $\cdot$, which means none and has no effect; however,
when an instruction body produces something else we use it to branch.
A nonzero bitvector results in skipping the indicated number of
instructions; the out-of-band additional value \texttt{ext} causes a
branch to the external label.
The magic number used to select the external label appears only in one
of the statement rules; beyond that point we use \texttt{ext}
explicitly.
}

\NEWW{
The program rule in Figure \ref{fig:tr_casp_sem_spec} inserts the
branch state produced by the program into $\venv$, where it provides a
value for the \texttt{branchto} predicate found in the postcondition.
This allows specifications to enforce the external branching behavior.
}

\subsection{\ale Overview}
\label{sec:ale_lang}
This section describes \ale, our specification language for
writing abstracted machine-independent specifications of low-level
code.

\ale specifications are abstractions of machine-level \casp
specifications; we say that \casp constructs are \emph{lifted} into
\ale and \ale constructs are \emph{lowered} into \casp.
\ale is only for specifications, so there are no statements, no
updates, and no notion of instructions or programs.
We refer to the single synthesis problem in one \ale specification as a \emph{block}.

\mypara{Notation}
We use the following metavariables:

\begin{tabular}{llr}
\ttx, \tty, \ttz & Program variables (binders) \\
\ttr            & Registers (abstract) \\
\ttC            & Integer constants (written in decimal) \\
\bitconst       & Bitvector constants (written in binary) \\
\ttN            & Symbolic integer constants \\
\atyp           & Types \\
\ttv            & Values \\
\tte            & Expressions \\
\tti, \ttj      & Rule-level integers \\
\end{tabular}

(Other constructions are referred to with longer names.)

As noted previously, \ale types and expressions should be
considered distinct from \casp ones (even where they correspond
directly).
We use the same letters in the hopes that this will cause less
confusion (even in the definition of the translation) than picking an
entirely different set of letters for \ale.

\mypara{Identifiers and Variables}
In \ale there are \NNEW{five} syntactic categories of identifiers:
As in \casp, \ttmemid name memory regions.
\ttfuncid name functions, and \tttypeid are type aliases.
\ttxmoduleid name \casp \lowering modules, 
which are used to instantiate the abstract elements and conduct \ale{}--\casp translation for synthesis.
Other \ttx are ordinary variables that range over other things, and
may be presumed to not range over the above reserved categories.
All variables are immutable, in the sense that their values do not change once
bound.

\begin{figure*}
\centering
\begin{minipage}{0.48\textwidth}
\begin{align*}
	& & \textbf{(\ale Symbolic Constants)} \\
	& \ttN \bnfeq & \ttC \bnfor \ttx \\
	& & \\
	& & \textbf{(\ale Types)} \\
	& \atyp \bnfeq & \aprimitivetyp \bnfor \amemorytyp \bnfor \afunctiontyp \\
	& & \\
	& \aprimitivetyp \bnfeq & \ttint \bnfor \ttbool \bnfor \tttypeid \\
	& \bnfor & \ttN\ \ttwvec \bnfor \ttN\ \ttwptr \bnfor \ttN\ \abvltyp \\
	& \bnfor & \ttN\ \ttlabel \bnfor \aregsettyp\\
	& \aregsettyp \bnfeq & \ttN\ \abvstyp\\
	& & \\
	& \amemorytyp \bnfeq & \memtyp{\ttN_1}{\ttN_2}{\ttN_3} \\
	& \afunctiontyp \bnfeq & \afuntyp \\
	& & \\
& & \textbf{(\ale Values)} \\
& \ttv \bnfeq & \tttrue \bnfor \ttfalse \bnfor \ttC \bnfor \bitconst \bnfor \ptrform{\ttmemid}{\ttC} \\
& \bnfor & \ttassertfalse
\end{align*}
\end{minipage}
\hfill\vline\hfill
\begin{minipage}{0.49\textwidth}
\begin{align*}
	& & \textbf{(\ale Operators)} \\
	& \itunop \bnfeq & -  \bnfor \bitsub\ \bnfor \neg\ \bnfor \bitnot\\ 
	& \itbinop \bnfeq &  =\ \bnfor\ \neq\ \bnfor\ + \bnfor - \bnfor * \bnfor / \ \bnfor\ <\ \bnfor\ <=\ \bnfor\ >\ \bnfor\ >= \\
	& \bnfor & \&\& \bnfor || \bnfor \xor \\
	& \bnfor & >> \bnfor {>>}_S \bnfor << \bnfor \bitand \bnfor \bitor \bnfor \bitxor \\
	& \bnfor & \bitadd \bnfor \bitsub \bnfor \bitmul \bnfor \bitdiv  \\
	& \bnfor & \bitlt \bnfor \bitle \bnfor \bitgt \bnfor \bitge  \\
	& \bnfor & \bitslt \bnfor \bitsle \bnfor \bitsgt \bnfor \bitsge\\
	& \bnfor & \cup \bnfor \cap \bnfor \subseteq \bnfor \setminus \\
	& & \\
	& & \textbf{(\ale Expressions)} \\
	& \ttae \bnfeq & \ttv \bnfor \ttx \\
	& \bnfor & \fapp{\ttfuncid}{\sseq{\ttae}} \\ 
	& \bnfor & \itunop\ \ttae \\
	& \bnfor & \ttae_1\ \itbinop\ \ttae_2 \\
	& \bnfor & \bitidx{\ttae}{\ttC} \bnfor \bitfield{\ttae}{\ttC_1}{\ttC_2} \\ 
	& \bnfor & \eLet{\ttx}{\cbasetyp}{\ttae_1}{\ttae_2} \\ 
	& \bnfor & \sITE{\ttae_1}{\ttae_2}{\ttae_3} \\ 
	& \bnfor & \ptrform{\ttmemid}{\ttae} \\ 
	& \bnfor &  \regderef{\ttae}  \bnfor \fetch{\ttae}{\ttN} \\
	& \bnfor & \NEW{\texttt{\ebranchto{sym}}} \\
	& \bnfor & \setlit \\
	& \bnfor & \sizeofset{\ttae} \bnfor \setmemberof{\ttae_1}{\ttae_2} 
\end{align*}
\end{minipage}
\caption{\ale symbolic constants, types, values, operators and expressions.\label{fig:tr_ale_types1}}
\end{figure*}

\mypara{Symbolic Constants}
In \ale symbolic constants \ttN are permitted to
occur in some places where only integer constants are allowed in the
corresponding \casp constructions.
In particular, the bit sizes associated with types (and the lengths of
memory regions, which are functionally similar) may be given as
symbolic values \ttx instead of integer constants.
These must be bound to integer constants either directly in the
\ale spec, in the \casp \lowering, or by the \casp machine
description.
This allows the concrete sizes of bitvectors to vary depending on the
machine architecture.


\mypara{Types}
As in \casp, \ale types are divided syntactically into base types
and others, shown in \autoref{fig:tr_ale_types1}.
The chief difference from \casp is that bit widths (and the lengths of
memory regions) can be symbolic constants.
However, an additional difference is that pointers (\ttwptr) are
distinguished from plain bitvectors (\ttwvec).
This is reasonably possible in \ale, because it need not reason
about the progression of values through machine registers, only
pre- and post-block machine states.
Strings and unit are also absent, as they are not needed for
specifications.



\mypara{Values and Expressions}
The values in \ale correspond directly to the values in \casp as do
operators and most expressions, shown in \autoref{fig:tr_ale_types1}.
Note that the width argument of fetch can be a symbolic size.


\begin{figure*}
\centering
\begin{minipage}{0.54\textwidth}
\begin{align*}
	& & \textbf{(\ale Declarations)} \\
	& \ttdecls \bnfeq & \eps \bnfor \ttdecl\ttsemi\ \ttdecls \\
	& \ttdecl \bnfeq & \ttareq\ \ttatype\ \tttypeid \\ 
	& \bnfor & \ttareq\ \ttaval\ \hastyp{\ttx}{\aprimitivetyp} \\ 
	& \bnfor & \ttareq\ \ttafunc\ \hastyp{\ttfuncid}{\afunctiontyp} \\ 
	& \bnfor & \ttaprov\tsp\ttatype\tsp\tttypeid = \atyp\\
	& \bnfor & \ttaprov\tsp\ttaval\tsp\hastyp{\ttx}{\aprimitivetyp} = \ttae \\
	& \bnfor & \ttaprov\ \ttafunc\ \hastyp{\ttfuncid}{\funtyp{\sseq{\ttx_i\ \ttcol\ \cbasetypi}}{\cbasetyp}} = \ttae \\
	& \bnfor & \ttaregion\ \hastyp{\ttmemid}{\amemorytyp}   \\
	& \bnfor & \ttaregion\ \hastyp{\ttmemid}{\amemorytyp}\ \ttwith\ \NNEW{\ttx} \\
	& \bnfor & \ttalower\ \ttxmoduleid \\
	& \bnfor & \ttfwrites\ \ttcol\ \sseq{\ttx_i}  \\
	& \bnfor & \ttfmwrites\ \ttcol\ \sseq{\ptrform{\ttmemidi}{\tte_i}}
\end{align*}
\end{minipage}
\hfill\vline\hfill
\begin{minipage}{0.44\textwidth}
\begin{align*}
	& & \textbf{(Initial State Bindings)} \\
	& \ttblockbinds \bnfeq & \eps \bnfor \ttblockbind\ttsemi\ \ttblockbinds \\
	& \ttblockbind \bnfeq & \tLet{\ttx}{\aprimitivetyp}{\ttae}
\end{align*}
\begin{align*}
	& & \textbf{(\ale Specifications)} \\
	& \ttpre \bnfeq & \ttae \\ 
	& \ttpost \bnfeq & \ttae \\
	& \itspec \bnfeq & \ttdecls\ttsemi\ \ttblockbinds\ttsemi\ \ttpre\ttsemi\tsp \ttpost
\end{align*}
\end{minipage}
\caption{\ale declarations, block-lets, and specifications.\label{fig:tr_ale_types2}}
\end{figure*}

\mypara{Declarations and Frames}
\ale declarations come in two forms: \ttareq and \ttaprov, shown in \autoref{fig:tr_ale_types2}.
The second form declares elements in the ordinary way, while
the first form declares an element that must be provided
by the \casp \lowerings or the \casp machine description.
In this case, the type is given, but not the value.
This functions as a form of import, and allows an \ale file to be
checked on its own separately from any particular machine description
or \casp \lowerings.
However, we do not currently define or implement such a check.
Note that it is possible to \ttareq functions that implicitly depend
on machine state or that depend on machine state on some machines and
not others.
Such functions can also depend on constants or other elements that are
not visible in the \ale specification at all.
The \ttalower declarations specify all \lowering modules that 
are used to compile this \ale specification into a \casp specification.
The module name \ttxmoduleid is used to look up the \casp \lowering module to apply.
The \ttaregion declarations declare memory regions, like the
memory-typed \ttletstate declarations in \casp.
(These are implicitly always \ttaprov, because, for memory regions, the
corresponding \ttareq declaration would be entirely equivalent, requiring
duplication in the \casp \lowering.)
Note that the parameters of the region can be symbolic constants if
abstraction is needed.

Frame declarations in \ale, annotated with \ttfwrites and \ttfmwrites, are exactly the same as in \casp.
Because \ale files are machine-independent, the registers mentioned
must be abstract and concretized via the \casp \lowerings.


\mypara{Block-lets}
While \ale expressions include let-bindings, the scope of those
let-bindings is conventional: it lasts until the end of the
expression.
To refer to values taken from the initial state (that is, the
machine state of which the precondition must be true), we need a way to
bind these values so their scope extends to the postcondition.
The block-lets serve this purpose in \ale, shown in \autoref{fig:tr_ale_types2}, 
much like the additional declarations seen in \casp specs can.
These are found within a block (because a block corresponds to a
synthesis problem, it is meaningful to associate pre- and post-block
machine states with it), and the scope is the entire block.

\mypara{Specifications}
A full specification starts with a preamble of declarations, shown in \autoref{fig:tr_ale_types2}.
It also includes block-lets and the pre- and postconditions for the block.
Common declarations can be shared with \ttinclude.

\subsection{\ale Typing and Semantics}
\label{sec:ale_check}

\begin{figure*}
\textbf{(\casp Integer Constant Extraction)}
\nextrulesmall
\begin{tabular}{ccc}
\irule
	{~}
	{\KdeclAA{\eps}}
&
\irule
	{\KdeclAB{\ttdecl} \nextclause \\
		\KdeclBC{\ttdecls}}
	{\KdeclAC{\ttdecl\ttsemi\ \ttdecls}}
&
\irule
	{\KdeclAB{\ttdecls}}
	{\KmachineB{\ttdecls\ttsemi\ \defops}}
\end{tabular}

\nextrulesmall
\begin{tabular}{ccc}
\irule
	{~}
	{\KdeclAA{\tttype\ \tttypeid = \cbasetyp}}
&
\irule
	{\kenv' = \kenv[\ttx \envgoesto \ttC]}
	{\KdeclAB{\ttlet\ \ttx\ \ttcol\ \ttint = \ttC}}
&
\irule
	{\tte \neq \ttC}
	{\KdeclAA{\ttlet\ \ttx\ \ttcol\ \cbasetyp = \tte}}
\end{tabular}

\begin{tabular}{cc}
\irule
	{~}
	{\KdeclAA{\ttdef\ \ttfuncid\ \funtyp{\sseq{\ttx_i\ \ttcol\ \cbasetypi}}{\cbasetyp} = \tte}}
&
\irule
	{~}
	{\KdeclAA{\ttproc\ \ttprocid\ \proctyp{\sseq{\ttx_i\ \ttcol\ \cbasetypi}} = \ttS}}
\end{tabular}

\begin{tabular}{cc}
\irule
	{~}
	{\KdeclAA{\ttlet\ttstate\ \ttx\ \ttcol\ \cregtyp}}
&
\irule
	{~}
	{\KdeclAA{\ttlet\ttstate\ \ttmemid\ \ttcol\ \cmemtyp}}
\end{tabular}

\begin{tabular}{cc}
\irule
	{~}
	{\KdeclAA{\ttlet\ttstate\ \ttmemid\ \ttcol\ \cmemtyp\ \ttwith\ \NNEW{\ttx}}}
&
\irule
{~}
{\KdeclAA{\ttlet\ \strof{\ttx} = \text{\tte}}}
\end{tabular}

\caption{\casp integer constant extraction.\label{fig:tr_ale_const}}
\end{figure*}

We do not provide (or implement) a full typechecking pass for \ale.
Instead, when we lower to \casp, we allow the \casp typechecker to
reject invalid results, which might be caused by invalid \ale input or by
bad/mismatched \casp \lowering definitions.
The rules provided here are for doing scans over the declarations
sufficient to make the translation to \casp work and no more.

\mypara{Environments}
We retain the \casp typing environments, $\typenv$.
We add an additional environment \kenv, which maps identifiers to
integer constants.
This is a projection of the \casp execution environment $\venv$: it
holds mappings only for variables defined as integer constants and
excludes everything else.
We include a separate set of rules for extracting these integer
constants without doing a full \casp execution.
(Among other things, this avoids involving machine state or the
machine state stores.)

\mypara{Translation}
The translation (\lowering) from \ale to \casp, defined in the next
section, appears cross-recursively in the rules in this section.
Because $\typenv$ are \casp environments, they map identifiers to
\casp types, not \ale ones.
This means \ale types must be lowered on the fly to update
them correctly.

\mypara{Integer Constant Extraction}
The integer constant extraction rules do a simple pass over \casp
declarations to extract the variables defined as integer constants, shown in \autoref{fig:tr_ale_const}.
These populate a substitution environment \kenv that we use for
lowering \ale types containing symbolic constants.
These rules are judgments of the form \KdeclAB{\ttdecl} or
\KdeclAB{\ttdecls}, plus one of the form \KmachineA{\machine} for a
whole machine description.

\begin{figure*}
	\textbf{(\ale Declaration Typing)}
\nextrulesmall
\irule 
{\typenv, \kenv \tholds \ttdecl \ttthen \typenvBB, \kenv' \nextclause \\
	\typenvBB, \kenv \tholds \ttdecls \ttthen \typenvCC, \kenv''}
{\typenv, \kenv  \tholds \ttdecl\ttsemi\ \ttdecls \ttthen \typenvCC, \kenv''}

\nextrulesmall
\begin{tabular}{cc}
\irule 
{	
	\genv \twfholds \tttypeid}
{\typenv, \kenv \tholds \ttareq\ \ttatype\ \tttypeid \ttthen \typenv, \kenv}
&
\irule 
{	\acrule{\aprimitivetyp} = \atyp\nextclause \\
	\tenv(\ttx) = \atyp }
{\typenv, \kenv \tholds \ttareq\ \ttaval\ \hastyp{\ttx}{\aprimitivetyp} \ttthen \typenv, \kenv}
\end{tabular}
\nextrulesmall
\irule 
{   \acrule{\afunctiontyp} = \atyp\nextclause \\
	\tenv (\ttfuncid) = \atyp }
{\typenv, \kenv \tholds \ttareq\ \ttafunc\ \hastyp{\ttfuncid}{\afunctiontyp} \ttthen \typenv, \kenv}
	
\nextrulesmall
\irule 
	{ 
		\acrule{\atyp} = \atyp' \\
		\genv \twfholds \atyp' \nextclause \\
		\genv' = \genv[\tttypeid \goesto \atyp'] }
	{\typenv, \kenv \tholds \ttaprov\tsp\ttatype\tsp\tttypeid = \atyp \ttthen \typenvBA, \kenv }

\nextrulesmall
\irule 
	{ 
		\genv, \env \tholds\hastyp{\ttC}{\ttint} \nextclause \\ 
		\tenv' = \tenv[\ttx  \envgoesto \ttint] \nextclause \\
		\kenv' = \kenv[\ttx  \envgoesto \ttC] }
	{\typenv, \kenv \tholds \ttaprov\tsp\ttaval\tsp\hastyp{\ttx}{\ttint} = \ttC \ttthen \typenvAB, \kenv' }

\nextrulesmall
\irule 
	{ 
		\acrule{\aprimitivetyp} = \atyp\nextclause \\
		\genv \twfholds \atyp \nextclause \\
		\tte \neq \ttC \\
		\acrule{\tte} = \tte' \nextclause \\
		\genv, \env \tholds\hastyp{\tte'}{\atyp} \nextclause \\ 
		\tenv' = \tenv[\ttx  \envgoesto \atyp] }
	{\typenv, \kenv \tholds \ttaprov\tsp\ttaval\tsp\hastyp{\ttx}{\aprimitivetyp} = \ttae \ttthen \typenvAB, \kenv }

\nextrulesmall
\irule 
	{	
		\forall \tti,\ \acrule{\cbasetypi} = {\atyp_i} \wedge \genv \twfholds \atyp_i \nextclause \\
		\acrule{\cbasetyp} = \atyp \nextclause \\
		\genv \twfholds \atyp \nextclause \\ \\
		\acrule{\tte} = \tte' \nextclause \\
		\tenv' = \tenv[\forall \tti, \ttx_i\envgoesto {\atyp_i}] \nextclause \\ 
		\genv, \tenv'\tholds\hastyp{\tte'}{\atyp} \nextclause \\ 
		\tenv'' = \tenv[\ttfuncid \envgoesto \left( \funtyp{\sseq{\ttx_i\ \ttcol\ {\atyp_i}}}{\atyp} \right)] \nextclause \\
		}
	{\typenv, \kenv \tholds \ttaprov\ \ttafunc\ \hastyp{\ttfuncid}{\funtyp{\sseq{\ttx_i\ \ttcol\ \cbasetypi}}{\cbasetyp}} \\
		= \ttae \ttthen \typenvAC, \kenv}

\nextrulesmall
\irule 
	{	
		\acrule{\memtyp{\ttN_1}{\ttN_2}{\ttN_3}} =
			\memtyp{\ttC_1}{\ttC_2}{\ttC_3} \nextclause \\ \\
		\genv \twfholds \memtyp{\ttC_1}{\ttC_2}{\ttC_3} \nextclause \\
		\tenv' = \tenv[\ttmemid  \envgoesto \memtyp{\ttC_1}{\ttC_2}{\ttC_3}]  }
	{\typenv, \kenv \tholds \ttaregion\ \hastyp{\ttmemid}{\memtyp{\ttN_1}{\ttN_2}{\ttN_3}} \ttthen\ \typenvAB, \kenv }

\nextrulesmall
\irule 
 	{   
 		\acrule{\memtyp{\ttN_1}{\ttN_2}{\ttN_3}} =
 			\memtyp{\ttC_1}{\ttC_2}{\ttC_3} \nextclause \\ \\
 		\genv \twfholds \memtyp{\ttC_1}{\ttC_2}{\ttC_3} \nextclause \\
 		\genv \twfholds \labeltyp{\ttC_3} \nextclause \\ \\
 		\acrule{\labeltyp{\ttN_3}} = \labeltyp{\ttC_3} \nextclause \\
		\tenv' = \tenv[\ttmemid  \envgoesto \memtyp{\ttC_1}{\ttC_2}{\ttC_3} \ttsemi\ \NNEW{\ttx} \envgoesto \labeltyp{\ttC_3}] }
	{\typenv, \kenv \tholds \ttaregion\ \hastyp{\ttmemid}{\memtyp{\ttN_1}{\ttN_2}{\ttN_3}}
		\ \ttwith\ \NNEW{\ttx} \ttthen \typenvAB, \kenv }

\irule{~}
	{\typenv, \kenv \tholds \ttalower\ \ttxmoduleid \ttthen \typenv, \kenv}

\nextrulesmall
\irule
{
	\acrule{\ttfwrites\ \ttcol\ \sseq{\ttx_i}} =
		\ttfwrites\ \ttcol\ \sseq{\ttx_i}
	\nextclause \\
	\typenv \tholds
		\ttfwrites\ \ttcol\ \sseq{\ttx_i}
}
{
	\typenv, \kenv \tholds
		\ttfwrites\ \ttcol\ \sseq{\ttx_i}
		\ttthen \typenv, \kenv
}

\nextrulesmall
\irule
{
	\acrule{\ttfmwrites\ \ttcol\ \sseq{\ptrform{\ttmemidi}{\tte_i}}} =
		\ttfmwrites\ \ttcol\ \sseq{\ptrform{\ttmemidi}{\tte_i'}}
	\nextclause \\
	\typenv \tholds
		\ttfmwrites\ \ttcol\ \sseq{\ptrform{\ttmemidi}{\tte_i'}}
}
{
	\typenv, \kenv \tholds
		\ttfmwrites\ \ttcol\ \sseq{\ptrform{\ttmemidi}{\tte_i}}
		\ttthen \typenv, \kenv
}

\caption{\ale typing rules for declaration.\label{fig:tr_ale_type_decl}}
\end{figure*}

\mypara{Typing}
The declaration typing rules are intended to accumulate types for all
the declarations in an \ale specification.
They are applied concurrently with the \casp declaration rules to the
\ale specification, the \casp machine description, and the \casp
\lowering.
The declaration typing rules have judgments of the form
$\typenv, \kenv \tholds \ttdecl \ttthen \typenvBB, \kenv'$ and
$\typenv, \kenv \tholds \ttdecls \ttthen \typenvBB, \kenv'$, shown in \autoref{fig:tr_ale_type_decl}.
These mean that the declaration or declarations update the type
environment (and integer constant environment) as shown.
Note that there is a special-case rule for \ttaprov \ttaval for when
the value is an integer constant; this enters the constant into \kenv.
The integer constants are in turn used when lowering the types of
memory regions, which can be seen in the last two rules.

\begin{figure*}
	\textbf{(\ale Specification Typing)}

\nextrulesmall
\begin{tabular}{cc}
\irule 
{~}
{\typenv, \kenv \tholds \eps \ttthen \typenv, \kenv}
&
\irule 
	{\typenv, \kenv \tholds \ttblockbind \ttthen \typenvBB, \kenv' \ \ \ \ \ 
		\typenvBB, \kenv  \tholds \ttblockbinds \ttthen \typenvCC, \kenv'' }
	{\typenv, \kenv \tholds \ttblockbind\ttsemi\ \ttblockbinds \ttthen \typenvCC, \kenv'' }
\end{tabular}

\nextrulesmall
\irule 
{	
	\acrule{\cbasetyp} = \atyp \ \ \ \ \ 
	\genv \twfholds \atyp  \ \ \ \ \ 
	\acrule{\tte} = \tte' \ \ \ \ \ 
	\genv, \env \tholds\hastyp{\tte'}{\atyp} \ \ \ \ \ 
	\tenv' = \tenv[\ttx \envgoesto \atyp] }
{\typenv, \kenv \tholds \tLet{\ttx}{\aprimitivetyp}{\ttae}  \ttthen \typenvAB, \kenv }

\nextrulesmall
\irule
{
	\typenv, \kenv \tholds \ttdecls \ttthen \typenvBB, \kenv'
		\nextclause \\
	\typenvBB, \kenv' \tholds \ttblockbinds \ttthen  \typenvCC, \kenv''
		\nextclause \\
	\acruleCC{\ttpre} = \ttpre'
		\nextclause \\
	\acruleCC{\ttpost} = \ttpost'
		\nextclause \\
	\typenvCC \tholds \hastyp{\ttpre'}{\ttbool}
		\nextclause \\
	\typenvCC \tholds \hastyp{\ttpost'}{\ttbool}
}
{
	\typenv, \kenv \tholds
	    \ttdecls\ttsemi\ \ttblockbinds\ttsemi\ \ttpre\ttsemi\tsp \ttpost
	\ttthen \typenvCC, \kenv''
}

\caption{\ale typing rules for specifications.\label{fig:tr_ale_type_spec}}
\end{figure*}

\mypara{Block-lets}
The rules for block-lets are effectively the same as the rules for
declarations, shown in \autoref{fig:tr_ale_type_spec}.
The ways in which block-lets are special mostly do not apply here.
Note however that even though we pass through \kenv (for consistency
of the form of the rules) there is no rule for loading integer
constants into \kenv from block-lets.
Integer constants used in types and defined in the \ale specification should
be defined with \ttaprov \ttaval; block-lets are intended to provide
access to machine state.

\begin{figure*}
	\textbf{(\ale Specification Semantics)}
\nextrulesmall
\irule 
	{
		\itspec = 
			\ttdecls_{\mathit{ale}}\ttsemi\
			\ttblockbinds\ttsemi
			\ \ttpre\ttsemi\tsp \ttpost
		\nextclause \\\\
		\tholds \machine \ttthen \genv_0, \tenv_0 \ \ \ \ \ \ 
		\KmachineZ{\machine} \ \ \ \ \ \ 
		\left( \genv_0 \subseteq \genv \right) \wedge \left( \tenv_0 \subseteq \tenv \right) \wedge \left( \kenv_0 \subseteq \kenv \right)\nextclause \\
		\genv, \tenv, \kenv \tholds \ttdecls_{\mathit{ale}} \ttthen \genv, \tenv, \kenv \ \ \ \ \ \ 
		\acrule{\ttdecls_{\mathit{ale}}} \subseteq \ttdecls\ttsemi\ \ttframes \nextclause \\
		( \forall \ttxmoduleid,
			\ttalower\ \ttxmoduleid \in \ttdecls_{\mathit{ale}}
		\implies \nextclause \\
			\ttmodule\ \ttxmoduleid
			\ \{\ \ttdecls_{\mathit{lower}}\ttsemi
				\ \ttframes_{\mathit{lower}}\ \}
			\in \itmodules
		\wedge
			\ttdecls_{\mathit{lower}} \subseteq \ttdecls
		\wedge
			\ttframes_{\mathit{lower}} \subseteq \ttframes
		) \nextclause \\
		\genv, \tenv \tholds \decls \ttthen \genv, \tenv \ \ \ \ \ \ 
		\kenv \tholds \decls \ttthen \kenv \nextclause \\
		\genv, \tenv, \kenv \tholds \ttblockbinds \ttthen \genv, \tenv, \kenv \ \ \ \ \ \ 
		\acrule{\ttblockbinds} \subseteq \ttdecls \nextclause \\
		\acrule{\ttpre} = \ttpre' \nextclause \\
		\acrule{\ttpost} = \ttpost' \nextclause \\
		\Omega = \ttdecls\ttsemi\ \ttframes\ttsemi\ \ttpre'\ttsemi\tsp \ttpost'
	}
	{\machine, \itmodules, \itspec \ttthen \Omega}
\caption{\ale semantics for specifications.\label{fig:tr_ale_sem_spec}}
\end{figure*}

\mypara{Specifications}
The rule for the semantics of an entire specification is large and
complex.
The conclusion is that a given machine, \lowering module, and \ale
specification produce a final translation output $\Omega$.
The rules work by nondeterministically
taking fixpoints over all the material included.
$\ttdecls$ is the combination of all declarations found both in the
initial specification and all the included lowering modules, and
$\ttframes$ is the combination of all frame information (part of
the declarations in \ale; separated in \casp).
Similarly, the final set of environments $\tenv, \genv, \kenv$
represent fixpoints produced by processing all the declarations.

In \autoref{fig:tr_ale_sem_spec}, 
the first premise expands the \ale
specification as we will need to work with the components.
The next two premises generate initial environments: the \casp typing
environments induced by the machine description, and its integer
constants, and then we require that these are included in the
final environments.
In the fifth and sixth premises, we then require that 
the result of processing the declarations from
the specification appears in the final environments, and that the
translation of these to \casp is included in the final declarations
and frame rules.
Then for every lowering module requested by the specification, we
require that it be provided in the input modules list and that its
components appear in the final declarations and frame rules.
This is followed by two more rules to ensure that these results
are represented in the final environments.
Later, we include the block-let material in the final environments,
include its lowered form in the final declaration list (block-lets
lower to declarations), bind
the lowerings of the pre- and postconditions, and define
the output.

The fixpoint-based evaluation strategy for declarations is required,
because the
\ale declarations rely on the \casp \lowering file (most notably for
resolving symbolic constants), but the \casp \lowering file is in turn
also specifically allowed to refer to objects declared by the \ale
specification, such as memory regions.
In the implementation this circularity is resolved by lifting both the
\casp and \ale declarations (and block-lets) into a common
representation and topologically sorting them based on identifier
references.
(Genuinely circular references among identifiers are prohibited.)
From this point, they can be handled in order in a more conventional
way.

\mypara{Complete Output}
Note that the output includes the declarations
from the \casp \lowering modules (each $\ttdecls_{\mathit{lower}}$).
Apart from symbolic constants, we do not substitute the definitions of
the \lowering elements,
as that would greatly complicate things,
especially with functions; instead we include the definitions and let
the translation refer to them.
In fact, because of the declaration ordering issues, in the
implementation the \lowering declarations and
translated \ale declarations can be arbitrarily interleaved in the
output.

Note furthermore that it would not be sufficient to include
\emph{only} the \lowering declarations explicitly imported with \ttareq
declarations, as those may refer freely to other things declared in
the \lowering module that the \ale specification itself may have no cognizance
of whatsoever.

\subsection{Lowering \ale}
\label{sec:ale_exec}
\begin{figure*}
	\textbf{\ale{} -- \casp{} Type Translation}
	\begin{align*}
\actyprule{\ttN} = &
	\begin{cases} 
	\ttC & \ttN = \ttC \\
	\acsigma{\ttx} & \ttN = \ttx \wedge \ttx \in \kenv \\
	\bot & \ttN = \ttx \wedge \ttx \notin \kenv\\
	\end{cases}
&
\actyprule{\tttypeid} = &
	\begin{cases} 
		\acdelta{\tttypeid} & \tttypeid \in \genv \\
		\bot & \tttypeid \notin \genv
		\end{cases} \\
\actyptrans{\ttint} = &~ \ttint
&
\actyptrans{\ttbool} = &~ \ttbool \\
\actyptrans{\ttN\ \ttwvec} = &~ \bvaltyp{ \actyptrans{\ttN} } &
\actyptrans{\ttN\ \ttwptr} = &~ \bvaltyp{ \actyptrans{\ttN} }\\
\actyptrans{\ttN\ \abvltyp} = &~ \bvltyp{ \actyptrans{\ttN} } &
\actyptrans{\ttN\ \abvstyp} = &~ \bvstyp{ \actyptrans{\ttN} }
\end{align*}
\begin{align*}
\actyptrans{\memtyp{\ttN_1}{\ttN_2}{\ttN_3}} = &~ \ensm{\actyptrans{\ttN_1}\ \ttbit}\ \ensm{\actyptrans{\ttN_2} \ \ttmlen\ \actyptrans{\ttN_3}\ \texttt{ref}} \\
\actyptrans{\afuntyp} = &~ \funtyp{\sseq{\actyptrans{\cbasetypi}}}{\actyptrans{\cbasetyp}}
\end{align*}
\caption{\ale{} -- \casp{} type translation.\label{fig:tr_lower_type}}
\end{figure*}

The semantics of an \ale specification depend on material taken
from a \casp mapping and machine description.
This does not preclude defining a semantics for \ale in terms of
that material or even some abstracted concept of what any such \casp
material might be.
However, doing so is complicated (as can be seen from the material in the
previous section, which does not even attempt to handle expression
evaluation) and not perhaps very illuminating or rewarding.

So instead, we write only enough typing
rules to prepare material for writing a translation
(lowering) to \casp, and then apply the \casp typing to the
lowered material.
This gives \ale a semantics in terms of the \casp semantics.
The translated material goes into the \casp typing environments $\typenv$, and as
discussed in the previous section, we also maintain an additional
environment \kenv of integer constants used for substituting symbolic
constants in types.

This section defines the translation.
$\actrans{a}$ defines the \casp lowering of an \ale element $a$.
We make the translation polymorphic over the various kinds of element;
that is, $\actrans{\atyp}$ is the translation of a type (shown in \autoref{fig:tr_lower_type}),
$\actrans{\tte}$ is the translation of an expression, etc.
Some of the translation rules rely on the environments; these are
written $\acrule{a}$ (shown in \autoref{fig:tr_lower_all}).

Some of the translation rules produce $\bot$.
If these are reached, the translation fails; this can happen if the
\ale spec was malformed and, potentially, if the mapping module
failed to declare elements that were expected of it or declared them
in an incompatible or inconsistent way.
The rules in the previous section exclude some of these cases, but we
are not (yet) prepared to argue that they rule out all
translation-time failures.

Notice that the translations for \ttareq declarations are empty,
because the declarations from the mapping module are output along with
the translated \ale specification.

\begin{figure*}
\textbf{\ale{} -- \casp{} Expression Translation}
\begin{align*}
\acrule{\ttx} = &
	\begin{cases}
 	\ttx & \ttx \in \tenv \\
	\bot & \ttx \notin \tenv
 	\end{cases} 
&
\acrule{\fapp{\ttfuncid}{\sseq{\ttae}}} = &~
\begin{cases} 
	\fapp{\ttfuncid}{\sseq{\actrans{\ttae}}} & \ttfuncid \in \tenv \\
	\bot & \ttfuncid \notin \tenv
\end{cases} \\
\actrans{\tttrue} = &~ \tttrue &
\actrans{\ttfalse} = &~ \ttfalse \\
\actrans{\ttC} = &~ \ttC &
\actrans{\bitconst} = &~ \bitconst \\
\actrans{\itunop\ \ttae} = &~ \itunop\ \actrans{\ttae} &
\actrans{\ttae_1\ \itbinop\ \ttae_2} = &~ \actrans{\ttae_1}\ \itbinop\ \actrans{\ttae_2} \\
\actrans{\bitidx{\ttae}{\ttC}} = &~ \bitidx{\actrans{\ttae}}{\ttC} &
\actrans{\bitfield{\ttae}{\ttC_1}{\ttC_2}} = &~ \bitfield{\actrans{\ttae}}{\ttC_1}{\ttC_2}
\end{align*}
\vspace{-.6cm}
\begin{align*}
\actrans{\eLet{\ttx}{\cbasetyp}{\ttae_1}{\ttae_2}} = &~ \ensm{\texttt{let}\tsp\oftyp{\ttx}{\actyptrans{\cbasetyp}} } \ensm{ = \actrans{\ttae_1}\ \ensm{\texttt{in}\tsp{\actrans{\ttae_2}}}} \\
\actrans{\sITE{\ttae_1}{\ttae_2}{\ttae_3}} = &~ \ensm{\texttt{if}\tsp{\actrans{\ttae_1}}}\ \ensm{\texttt{then}\tsp{\actrans{\ttae_2}}}\ \ensm{\ensm{\texttt{else}\tsp{\actrans{\ttae_3}}}} \\
\acrule{\ptrform{\ttmemid}{\ttae}} = &
	\begin{cases} 
	\ptrform{\ttmemid}{\actrans{\ttae}} & \ttmemid \in \tenv \\
	\bot & \ttmemid \notin \tenv
	\end{cases}
\end{align*}
\vspace{-.3cm}
\begin{align*}
\actrans{\ttassertfalse} = &~ \ttassertfalse \\
\actrans{\regderef{\ttae}} = &~ \regderef{\actrans{\ttae}} &
\actrans{\fetch{\ttae}{\ttN}} = &~ \fetch{\actrans{\ttae}}{\actyptrans{\ttN}} \\
\NEW{\actrans{\ebranchto{sym}} = }&~\NEW{ \ebranchto{sym}} &
\actrans{\setlit} = &~ \ensm{\ttinbrace{\actrans{\ttx_1}, \ldots, \actrans{\ttx_k}}} \\
\actrans{\sizeofset{\ttae}} = &~ \sizeofset{\actrans{\ttae}} &
\actrans{\setmemberof{\ttae_1}{\ttae_2}} = &~ \setmemberof{\actrans{\ttae_1}}{\actrans{\ttae_2}}
\end{align*}
\vspace{-.5cm}
\nextrulesmall
\textbf{\ale{} -- \casp{} Block-Lets Translation}
\begin{align*}
	\actrans{\tLet{\ttx}{\aprimitivetyp}{\ttae}} = &~ \tLet{\ttx}{\actrans{\aprimitivetyp}}{\actrans{\ttae}} 
\end{align*}
\vspace{-.5cm}
\nextrulesmall
\textbf{\ale{} -- \casp{} Declaration Translation}
\begin{align*}
\acrule{\ttareq\ \ttatype\ \tttypeid} = &
	\begin{cases} 
		\eps & \tttypeid \in \genv \\
		\bot & \tttypeid \notin \genv
		\end{cases} \\
\acrule{\ttareq\ \ttaval\ \hastyp{\ttx}{\aprimitivetyp}} = &
	\begin{cases} 
	\eps & \ttx \in \tenv \\
	\bot & \ttx \notin \tenv
	\end{cases} \\
\acrule{\ttareq\ \ttafunc\ \hastyp{\ttfuncid}{\afunctiontyp} } = &
	\begin{cases} 
	\eps & \ttfuncid \in \tenv \\
	\bot & \ttfuncid \notin \Sigma
	\end{cases} \\
\actrans{\ttaprov\tsp\ttatype\tsp\tttypeid = \atyp} = &~ \tttype\ \tttypeid = \actyptrans{\atyp}\\
\actrans{\ttaprov\tsp\ttaval\tsp\hastyp{\ttx}{\aprimitivetyp} = \ttae} = &~ \ttlet\ \ttx\ \ttcol\ \actyptrans{\cbasetyp} = \actrans{\tte}\\
\actrans{\ttaprov\ \ttafunc\ \hastyp{\ttfuncid}{\funtyp{\sseq{\ttx_i\ \ttcol\ \cbasetypi}}{\cbasetyp}} = \ttae} = &~ \ttdef\ \ttfuncid\ \funtyp{\sseq{\ttx_i\ \ttcol\ \actyptrans{\cbasetypi}}}{\actyptrans{\cbasetyp}} \\
	& ~ = \actrans{\tte}\\
\actrans{\ttaregion\ \hastyp{\ttmemid}{\amemorytyp}} = &~ \ttlet\ttstate\ \ttmemid\ \ttcol\ \actyptrans{\cmemtyp}\\
\actrans{\ttaregion\ \hastyp{\ttmemid}{\amemorytyp}\ \ttwith\ \NNEW{\ttx}} = &~ \ttlet\ttstate\ \ttmemid\ \ttcol\ \actyptrans{\cmemtyp}\ \ttwith\ \NNEW{\ttx} \\
\actrans{\ttalower\ \ttxmoduleid} = &~ \epsilon \\
\actrans{\ttfwrites\ \ttcol\ \sseq{\ttx_i}} = &~ \ttfwrites\ \ttcol\ \sseq{\actrans{\ttx_i}} \\
\actrans{\ttfmwrites\ \ttcol\ \sseq{\ptrform{\ttmemidi}{\tte_i}}} = &~ \ttfmwrites\ \ttcol\ \sseq{\ptrform{\actrans{\ttmemidi}}{\actrans{\tte_i}} }
\end{align*}
\caption{\ale{} -- \casp translation.\label{fig:tr_lower_all}}
\end{figure*}

\end{document}